\documentclass[twocolumn, pra, nofootinbib, aps]{revtex4}
\usepackage{CJK}
\usepackage{graphicx}
\usepackage{dcolumn}
\usepackage{bm}
\usepackage{amsmath}
\usepackage{amsthm}
\usepackage{amssymb}
\usepackage{physics}
\usepackage{diagbox}
\usepackage{dsfont}
\usepackage[normalem]{ulem}
\usepackage{multirow}
\usepackage[shortlabels]{enumitem}
\usepackage[colorlinks]{hyperref}
\usepackage[toc,page]{appendix}
\usepackage{hyperref}
\usepackage{adjustbox}

\newcommand*{\MyIncludeGraphics}[2][]{%
\begin{adjustbox}{max size={\textwidth}{\textheight}}
    \includegraphics[#1]{#2}%
\end{adjustbox}
}

\hypersetup{colorlinks, citecolor=blue, filecolor=blue, linkcolor=blue, urlcolor=magenta}

\def\setelem#1{\expandafter\def\csname myarray(#1)\endcsname}
\def\dispsymbol#1{\csname myarray(#1)\endcsname}

\newcolumntype{?}{!{\color{black}\vrule width 1pt}}

\newcolumntype{C}[1]{>{\centering\let\newline\\\arraybackslash\hspace{0pt}}m{#1}}
\newcolumntype{N}{@{}m{0pt}@{}}

\def\squareforqed{\hbox{\rlap{$\sqcap$}$\sqcup$}}
\def\qed{\ifmmode\squareforqed\else{\unskip\nobreak\hfil
		\penalty50\hskip1em\null\nobreak\hfil\squareforqed
		\parfillskip=0pt\finalhyphendemerits=0\endgraf}\fi}

\def\duzomniejsze{<\kern-.7mm<}
\def\duzowieksze{>\kern-.7mm>}

\def\textbf#1{{\bf #1}}
\def\beq{\begin{equation}}
\def\eeq{\end{equation}}
\def\be{\begin{equation}}
\def\ee{\end{equation}}

\def\bal{\begin{align}}
\def\eal{\end{align}}

\def\ben{\begin{eqnarray}}
\def\een{\end{eqnarray}}
\def\beqa{\begin{eqnarray}}
\def\eeqa{\end{eqnarray}}
\def\eea{\end{array}}
\def\bea{\begin{array}}

\newcommand{\bei}{\begin{itemize}}
	\newcommand{\eei}{\end{itemize}}
\newcommand{\bee}{\begin{enumerate}}
	\newcommand{\eee}{\end{enumerate}}
\newcommand{\nc}{\newcommand}
\nc{\1}{{\openone}}

\def\tr{{\rm Tr}}

\def\>{\rangle}
\def\<{\langle}

\def\ot{\otimes}

\def\mP{{\mathrm{P}}}

\newcommand{\overbar}[1]{\mkern 1.5mu\overline{\mkern-1.5mu#1\mkern-1.5mu}\mkern 1.5mu}

\newtheorem{lemma}{Lemma}

\newtheorem{proposition}{Proposition}
\newtheorem{theorem}{Theorem}
\newtheorem{definition}{Definition}
\newtheorem{observation}{Observation}

\newtheorem{rem}{Remark}

\newtheorem{fact}{Fact}
\newtheorem{notation}{Notation}

\theoremstyle{plain}
\newtheorem*{corone}{Corollary 1}
\newtheorem*{cortwo}{Corollary 2}
\newtheorem*{corthree}{Corollary 3}
\newtheorem*{corfour}{Corollary 4}
\newtheorem*{corfive}{Corollary 5}

\newtheorem*{prop2}{Proposition 2}
\newtheorem*{thm1}{Theorem 1}
\newtheorem*{thm2}{Theorem 2}
\newtheorem*{thm3}{Theorem 3}

\def\bed{\begin{definition}}
	\def\eed{\end{definition}}
\def\bel{\begin{lemma}}
	\def\eel{\end{lemma}}
\def\bet{\begin{theorem}}
	\def\eet{\end{theorem}}
\def\bet{\begin{notation}}
	\def\eet{\end{notation}}
\def\bet{\begin{remark}}
	\def\eet{\end{remark}}

\begin{document}


	
	\title{Limitations on device--independent key secure against non--signaling adversary\\ via the squashed nonlocality}

\author{Marek Winczewski$^{(1,2)}$, Tamoghna Das$^{(2)}$, Karol Horodecki$^{(2,3)}$}


\affiliation{$^{1}$Institute of Theoretical Physics and Astrophysics, National Quantum Information Centre, Faculty of Mathematics, Physics and Informatics, University of Gda\'nsk, Wita Stwosza 57, 80-308 Gda\'nsk, Poland}
\affiliation{$^{2}$International Centre for Theory of Quantum Technologies, University of Gda\'nsk, Wita Stwosza 63, 80-308 Gda\'nsk, Poland}
\affiliation{$^{3}$Institute of Informatics, National Quantum Information Centre, Faculty of Mathematics, Physics and Informatics, University of Gda\'nsk, Wita Stwosza 57, 80-308 Gda\'nsk, Poland}

	\begin{abstract}
We initiate a systematic study to provide upper bounds on device-independent key, secure against a non-signaling adversary (NSDI). We employ the idea of “squashing” on the secrecy monotones and show that squashed secrecy monotones are the upper bounds on the NSDI key. Our technique for obtaining upper bounds is based on the non-signaling analog of quantum purification: the complete extension. As an important instance of an upper bound, we construct a measure of nonlocality called “squashed nonlocality”. Using this bound, we identify numerically a certain domain of two binary inputs and two binary outputs non-local devices for which the squashed nonlocality is zero. Therefore one can not distill secure-key from these non-local devices via a considered (standard) class of operations. Showing a connection of our approach to [New J. Phys., 8:126, 2006] we provide, to our knowledge, the tightest known upper bound in the (3,2,2,2) scenario. Moreover, we formulate a security condition, equivalent to known ones, for the considered class of protocols. To achieve this, we introduce a non-signaling norm that constitutes an analogy to the trace norm used in the security condition of the quantum key distribution.
\end{abstract}

\maketitle


\section{Introduction}
 Secure key distribution is a process of generation of secret key bits between two distant parties, in presence of an eavesdropper
\cite{CsisarKorner_key_agreement,Maurer93,Gisin-crypto}. 
There are four major cryptographic security paradigms developed in the last several decades that provide a background for our investigation. These are:  (i) \emph{secret key agreement scenario} (SKA) \cite{CsisarKorner_key_agreement,Maurer93}, (ii) \emph{device-dependent security against a quantum adversary} (QDD) \cite{BB84,Ekert1991,B92,Gisin-crypto,AcinBBBMM2004-key}, (iii) \emph{device-independent security against a quantum adversary} (QDI) \cite{Ekert1991,Bell-nonlocality,Mayers-Yao,acin-2007-98, Masanes2011, lit12} 
and (iv) \emph{device-independent security against a non-signaling adversary} (NSDI) \cite{Kent,AcinGM-bellqkd, masanes-2006, acin-2006-8, hanggi-2009}.
We have enumerated them in order of increasing power of the eavesdropper. In what follows, we 
are going to use concepts of SKA paradigm to place upper bounds on the secret key rate in the NSDI scenario in a manner that is known from the QDD paradigm. Let us then begin with a short reminder of the main ideas behind the aforementioned cryptographic setups.

In the SKA scenario, the parties share marginals of a classical probability distribution $P(ABE)$, respectively. The honest parties (often called  Alice and Bob) can process their data by the so-called {\it Local Operations and Public Communication} (LOPC). At the same time,  the eavesdropper Eve can listen to public communication and can apply any stochastic map on her data \cite{Maurer93,CsisarKorner_key_agreement}. This paradigm is of special interest in context of security of the wireless communication.

The QDD scenario, originating conceptually from the SKA, was introduced at the early stage of quantum cryptography \cite{BB84,Ekert1991}. In this paradigm, the three parties share (in the worst case) a subsystem of a joined pure quantum state $|\Psi_{ABE}\>$. Alice and Bob can process this state by {\it Local quantum Operations and Classical Communication} (LOCC).  At the same time, Eve obtains any system which is discarded by Alice and Bob  and can perform any quantum operation on her subsystem \cite{DevetakWinter-hash-prl,keyhuge,pptkey}. This scenario has a drawback that Alice and Bob have to trust the inner working of their device:
the dimensionality of the state and operations of measurement performed by the device. This problem has been resolved in a much more sophisticated 
approach of QDI, {\it quantum device-independent scenario}. 
In this paradigm, the honest parties share an untrusted device,  described by a joint conditional probability distribution $P(AB|XY)$ originating from a measurement on a quantum state $\rho_{AB}$: $P(AB|XY)=\tr (M_{A|X}\otimes M_{B|Y}\rho_{AB})$. Security in this scenario is based solely on statistics of the inputs $X, Y$, and outputs $A,B$ of the device. Eavesdropper is assumed to be restricted by the laws of quantum mechanics. She is therefore bound to hold a purifying system of a $\rho_{AB}$
i.e., the system $E$ of such a pure state $|\psi_{ABE}\>$, that $\tr_{E}|\psi\>\<\psi|_{ABE}=\rho_{AB}$. 




\subsection{Non-signaling adversary scenario} In this manuscript, we focus on another  branch of key distribution that
has emerged in the last two decades, which is the non-signaling device-independent  (NSDI) scenario \cite{ Kent, Kent-Colbeck, Scarani2006, acin-2006-8,AcinGM-bellqkd, masanes-2009-102, hanggi-2009,masanes-2006}. This scenario has even more relaxed assumptions than QDI. Here, the eavesdropper is restricted only by the non-signaling condition, i.e., she can not influence statistics of the honest parties in a faster than light manner. Similarly the honest parties can share a possibly supra-quantum correlation only constrained by the non-signaling condition.
The advantage of NSDI approach over SKA, QDD, and QDI scenarios is the fact that it assures security even if a new theory replacing quantum mechanics became established, as long as it is non-signaling. The object shared by Alice, Bob, and Eve is a tripartite non-signaling device, $P(ABE|XYZ)$, with $Z$ and $E$ being the input and output respectively of  Eve's part of the device. On this device, the parties perform some measurements $(X,Y)$ and post-process their output data $(A,B)$ by some LOPC operations, to produce the secure key.  
This device is assumed to be (in a worst-case) created by the eavesdropper who can listen to the public communication, and perform certain operations on her subsystem.
  
 The first NSDI protocol, whose security was proven, was given by Barrett, Hardy, and Kent \cite{Kent}. The protocol results in a single key bit in the noiseless scenario.
Later, lower bounds on the key rate have been derived in \cite{AcinGM-bellqkd,acin-2006-8, Scarani2006},  via several key distillation protocols, under the assumption that eavesdropper attacks each of the subsystems separately.
In the presence of a collective eavesdropping attack, it was shown in Refs. \cite{hanggi-2009, masanes-2009-102,masanes-2006},  that one can obtain a non-zero key rate under the fully non-signaling constraint. By fully non-signaling, we mean that none of the subsystems of a device can signal to each other. More precisely, a device with $2N+1$ inputs and $2N+1$ outputs ($N$ for each of the honest parties and one for the eavesdropper) is understood to have $2N+1$ subsystems none subset $k$ of which can signal to the remaining $2N+1-k$.\footnote{In what follows, by "device" we mean a single-use device. A device can be used by measuring its input. A single-use device can not be measured more than once. If there is a need to perform multiple measurements on a device,  then it will be assumed as a composite device consists of multiple single-use devices.}
 This assumptions is vital, because if the device can perform signaling between its subsystems (of one party) \cite{hanggi-2009b}, then no hash function is known to achieve privacy amplification against the non-signaling eavesdropper. Moreover, if the device has a memory \cite{Rotem-Sha,Salwey-Wolf},  or can signal forward (from one run\footnote{By one single run of the protocol, we mean one use of a  particular single use device.}  of the protocol to the next one) \cite{Rotem12},  then a  wide class of hash functions can be attacked by a non-signaling Eve. The assumption of full non-signaling can be achieved by performing measurements in parallel on all of the $2N$ subsystems. We refer to this approach as to parallel measurement model.

 The non-signaling paradigm that allows defining the NSDI scenario became an active field of research since the seminal papers of Rastall \cite{Rastall1985-RASLBT}, Khalfin, and Tsirelson \cite{Tsirelson_Khalfin} as well as Popescu and Rohrlich \cite{PR} (for a recent review on Bell nonlocality see \cite{Bell-nonlocality}). Our findings will
contribute not only to the aforementioned cryptographic scenarios (NSDI and SKA) but also to  the  domain of Bell nonlocality. This is because some of the functions that serve as upper bounds on the key rate that we establish in the NSDI scenario, are novel measures of nonlocality.

\subsection{Motivation:}


In the NSDI scenario described above, mainly the lower bounds on the key rate has been considered \cite{Kent,AcinGM-bellqkd, acin-2006-8, Scarani2006, masanes-2009-102, hanggi-2009, Kent-Colbeck, masanes-2006}.
For the upper after seminal result given in \cite{acin-2006-8} based on intrinsic information, upper bounds were not studied systematically until recently (an upper bound based on intrinsic information has been proposed in parallel to the approach presented in this work in \cite{Kaur-Wilde}). In contrast, if one considers the QDD scenario, both lower bounds \cite{DevetakWinter-hash-prl, RGKinfo_sec_proof_short,DevetakWinter-hash,Christandl12}, and upper bounds 
on the secure key rate are well known. Indeed, the upper bounds in this scenario where studied both in the context of quantum states \cite{pptkey,AugusiakH2008-multi,keyhuge,Christandl_2002,Christandl12,multi-sqent,Wil16} and quantum channels \cite{TGW14,Takeoka_2014,Pirandola2017} (see also \cite{Takeoka_2016,Wilde_2017,Laurenza_2017,Pirandola_2018,Pirandola_2020} in this context). Similarly in the case of QDI scenario, after seminal upper bound of \cite{Kaur-Wilde,Eneetthesis}, a sequence of other proposals were provided recently \cite{CFH21,AFL21,Farkas_2021,KaurHorodeckiDas,Horodecki_2022}.
Some of the upper bounds in QDD and QDI scenario \cite{Christandl_2002,Christandl12,multi-sqent,Wil16,TGW14,Takeoka_2014,CFH21,AFL21,Farkas_2021}  are based on the entanglement measure called ``squashed entanglement" \cite{Christandl12}. A welcome feature of this measure is that it is an additive function, i.e., one avoids regularization like it is the case for the relative entropy of entanglement \cite{keyhuge,pptkey,AEJPVM2001,Pirandola_2018,Pirandola_2020}. We aim at both constructing upper bounds in the NSDI scenario and introducing novel measures of nonlocality. Although the analog of relative entropy - the ``strength of nonlocality proof" \cite{vanDamGrunwaldGill} (also called relative entropy of nonlocality \cite{Grudka_contextuality}) has been constructed, no analog of squashed entanglement was known in the realm of nonlocality (for the parallel, and different approach see \cite{Kaur-Wilde}). In our approach to the problem, we are guided by an analogy between entanglement and nonlocality. Interestingly the measure which we construct is, up to maximization over the inputs of the honest parties, equal to the one implicitly considered in \cite{acin-2006-8}. It is however differently formulated, as we use the notion of a complete extension \cite{CE} to formalize it. Moreover, we prove that our measure is a convex function of the devices of the honest parties, what allows for the use of the convexification technique (that we formulate) for finding the numerical upper bound. We will see that this reformulation is fruitful for studying
properties of this upper bound, which we call here "the squashed nonlocality".

\section{Main results}

In this manuscript, we construct upper bounds on the NSDI  key rate, distillable via (i) direct measurement, changing device into a distribution followed by (ii) Local Operations and Public Communication (denoted together as MDLOPC operations). Aiming at upper bounds, we study the scenario in which the shared device consists of $N$ independent and identically distributed (iid) copies of a non-signaling device $P(AB|XY)$. We define a wide class of secrecy quantifiers taken from the so-called SKA (secure key agreement) model \cite{Maurer93}. One such quantifier, we call the \textit{squashed nonlocality}, as we define it in analogy to squashed entanglement \cite{Christandl12}, however, in the realm of non-signaling devices. We then show that the squashed nonlocality serves as an upper bound on the key distilled by MDLOPC operations. 
It is important to note that almost all of the secure key distillation protocols in QDI and NSDI, proposed so far, belong to the MDLOPC class of operations (see however recent proposal \cite{Scarani_2021}).
Therefore, our bounds, on the amount of key, bound from above the
key rate achieved by a wide class of practical protocols.

\subsection{Family of novel nonlocality measures as upper bounds}

One of our achievements is a construction of upper bound on the secret-key in the NSDI scenario that is in an addition a (non-faithful) measure of nonlocality. Informally, the squashed nonlocality $ {\cal N}_{sq}(P)$, of a bipartite non-signaling device $P:= P(AB|XY)$ is given by
    \begin{align}\label{eq:Nsq0}
    &{\cal N}_\mathrm{sq}(P) := \widehat{\mathrm{I}}\left(A:B\downarrow E\right)_\mathrm{\mathcal{E}\left(P\right)(ABE|XYZ)}, \nonumber \\
    &= \max_{x,y} \min_z \mathrm{I}\left(A:B\downarrow E\right)_{(\mathcal{M}^F_{x,y} \otimes \mathcal{M}^G_{z}) \mathrm{\mathcal{E}\left(P\right)(ABE|XYZ)}},
    \end{align}
where $\mathcal{E}\left(P\right)(ABE|XYZ)$ is the complete extension of the device $P$~\cite{CE}, and $I(A:B\downarrow E)_{P(ABE)}$ is the intrinsic information of a distribution $P(ABE)$ \cite{Intrinsic-Maurer,MauWol97c-intr}. Furthermore, the honest parties choose inputs $x,y$ (for a full direct measurement $\mathcal{M}^F_{x,y}$), while the eavesdropper is allowed to perform a more general measurement $\mathcal{M}^G_{z}$ that contains in particular probabilistic mixing of input choices.

The squashed nonlocality, as we prove, possesses many properties of those desired for a measure of nonlocality such as convexity and additivity. As we show the above function is only an example of an upper bound that can be introduced using our approach. The other function that we study in this paper to be lifted from the SKA to the NSDI scenario are mutual information and conditional mutual information.

We note however, that the above function
can be equivalently formulated in a way
considered implicitly in \cite{acin-2006-8} by A. Acin, S. Massar and S. Pironio (AMP). Consider a function $\mathrm{I}_{\mathrm{AMP},(x,y)}$:
\begin{align}
    &\mathrm{I}_{\mathrm{AMP},(x,y)}(P(AB|X=x,Y=y):=\nonumber\\
    &\inf_{\{p(E=e), P(ABE=e|X=x,Y=y)\}} \mathrm{I}(A:B\downarrow E)_{P(ABE|XY)}
\end{align}
where $P(ABE|XY)=p(E=e)P(AB,E=e|XY)$, and the infimum is taken over all ensembles
$\{p(E=e),P(AB,E=e|XY)\}$ of the device $P(AB|X,Y)=\sum_e P(E=e)P(AB,E=e|X,Y)$.
The equivalence can be establish as follows for a device $P\equiv P(AB|XY)$:
\begin{equation}
\max_{(x,y)} \mathrm{I}_{\mathrm{AMP},(x,y)}(P) = {\cal N}_{sq}(P).
\label{eq:equivalence}
\end{equation}
This fact, along with our proof of the convexity of ${\cal N}_{sq}$ leads to the tightest known bound in the scenario (3,2,2,2) (see Fig. \ref{fig:numerical-ub-3222_main}) i.e. with three inputs for one party, binary inputs for the other and binary outputs for both (for the proof of Eq. (\ref{eq:equivalence}) and consequences of it see Section \ref{sec:quantitative}).

We provide a method of generating tighter (though possibly harder to compute) upper bounds. Indeed, in defining the squashed nonlocality, we used the secrecy monotone called  {\it intrinsic information}. The non-faithfulness\footnote{The property of non-faithfulness of a measure of nonlocality means that the measure is zero for some non-local behaviors.} of the squashed nonlocality is therefore due to the property inherited from the classical intrinsic information that can be zero for correlated distribution. One can, however, use some other quantifiers of secret correlations, e.g., the so-called {\it reduced intrinsic mutual information}, which also leads to an upper bound.  Due to an analogy between entanglement and nonlocality, the upper bounds we provide here are also measures of nonlocality, and as such, can be studied independently.

Furthermore, we notice that our approach can be readily modified in order to construct upper bounds for a wider class of protocols in which one of the inputs of the honest parties is not announced~\cite{AcinGM-bellqkd}. This can be done by changing $\max_{x,y} \min_z$ to $\max_{y} \min_z \max_x$ in equation~\eqref{eq:Nsq0}, what reflects  the action of the parties in the latter scenario (only Bob announces his inputs).

\begin{figure}[t]
    \begin{center}
        \includegraphics[width=0.5\textwidth]{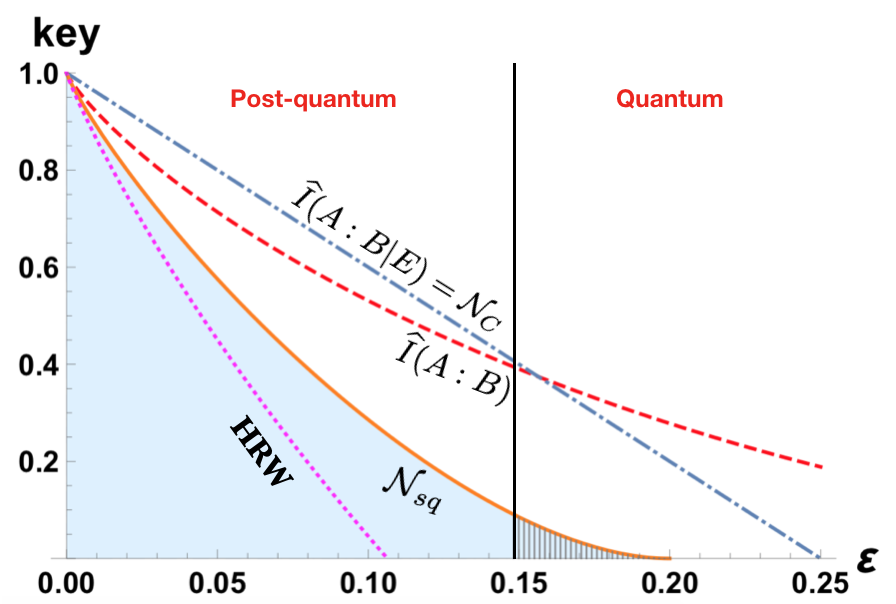}
    \end{center}
    \caption{Plot of several secrecy quantifiers $\widehat{\mathrm{M}}(A:B||E)$, as an upper bound on  $K_{DI}^{(iid)}$,  for a bipartite binary input-output device lying on the  isotropic line. The dashed red line represents  squashed mutual information $\widehat{\mathrm{I}}(A:B)_{P_{iso}}$. The straight blue  line represents the nonlocality cost, as well as the squashed conditional mutual entropy $\widehat{\mathrm{I}}(A:B|E)_{{\cal E}({P_{iso}})}$, over the complete extension ${\cal E}({P_{iso}})$ of the given device $P_{iso}$. The solid orange line represents the upper bound on the squashed nonlocality $N_{sq}$, which is the lower convex hull of the several other upper bounds on ${\cal N}_{sq}$. The dotted pink curve (HRW) corresponds to the lower bound achieved by H{\"a}nggi, Renner and Wolf's protocol~\cite{hanggi-2009}. }
    \label{fig:numerical-ub}
\end{figure}

\subsection{MDLOPC-Bound nonlocality}
Using the bound, we then obtain numerically a region of non-local two binary input and two binary output, $(2,2,2,2)$ devices, from which {\it no key} can be distilled via MDLOPC operations. These are the ``isotropic'' mixtures of the devices, namely the Popescu-Rohrlich (PR) box and the box complementarity to it, the anti-PR box when the admixture of the PR box is less than $80\%$. Notably, this result implies that in parallel measurement model, when the same measurement on each device is performed, nonlocality does not imply secrecy.  Indeed, quantum devices with mixture of PR box more than $75\%$ exhibit nonlocality, that is they violate the CHSH inequality \cite{CHSH}, while as we show, all the devices below $80\%$ have zero key distillable by MDLOPC protocols. We compare also the upper bound via non-signaling squashed nonlocality for isotropic devices with the lower bound
on the key rate taken from \cite{hanggi-2009} (see figure \ref{fig:numerical-ub}).  The lower bound and the upper come pretty close for the state close to PR box. 

We note here that in \cite{AcinGM-bellqkd} a protocol for distillation of private key from 
isotropic devices were given which is out of MDLOPC class: one of the parties do not announce the input 
from used to generate the key. There also it was shown that the so called {\it intrinsic information} is zero when both the parties announce their inputs after measurements. Our bound does not extends straightforwardly to this scenario, as in our case Eve knows that Alice and Bob draw key from {\it single} pair of inputs. However it indicates that
keeping one of the inputs used for generating key secret, is crucial for non-zero key rate in the non-signaling adversary scenario. 

This indication is confirmed by recent result given in \cite{Farkas_2021} for the case of device-independent quantum key distribution with quantum adversary. There, a broader notion of protocols is considered, also called "standard".  These are protocols during which for generation of the key each device is measured by a pair of inputs $(X=x,Y=y)$ with probability $p(x,y)$ drawn in i.i.d manner, an announced before post-processing the output key rate.
It is shown there, that such protocols admits an upper bound $\sum_{x=0,y=0}^{1,1}p(x,y)I(A:B\downarrow E,xy)$, i.e the intrinsic information~\cite{Intrinsic-Maurer,MauWol97c-intr} averaged over choices of the inputs. Moreover it is argued, that there exist non-local devices (violating CHSH inequality) for which the latter upper bound is zero.
This implies that no such "standard" protocol is able to achieve non-zero key rate in the case of quantum adversary.

In similar way, we show the MDLOPC-bound nonlocality in the (3,2,2,2) scenario \cite{Ekert1991,acin-2006-8}. In the latter one party has inputs $x\in\{0,1,2\}$ and the other $y\in\{0,1\}$. The inputs $x\neq 0, y$ are used for testing the value of the CHSH inequality \cite{CHSH}, while the pair $x=0,y=0$ is used for generation of the raw key. The fact that distributions with isotropic parameters $p \in [0.7071,0.8284]$ are non-local but no key can be distilled from them in the latter scenario was left open in \cite{acin-2006-8}. Showing the equivalence given in Eq. (\ref{eq:equivalence}) and the fact that ${\cal N}_{sq}$ upper bounds the distillable key, we close the mentioned open problem, by confirming that no key can be obtained by a protocol drawing key from a single pair of settings $x=0,y=0$.
The obtained results are shown in Fig. \ref{fig:numerical-ub-3222_main}.

\begin{figure}[t]
    \begin{center}
        \includegraphics[width=0.48\textwidth]{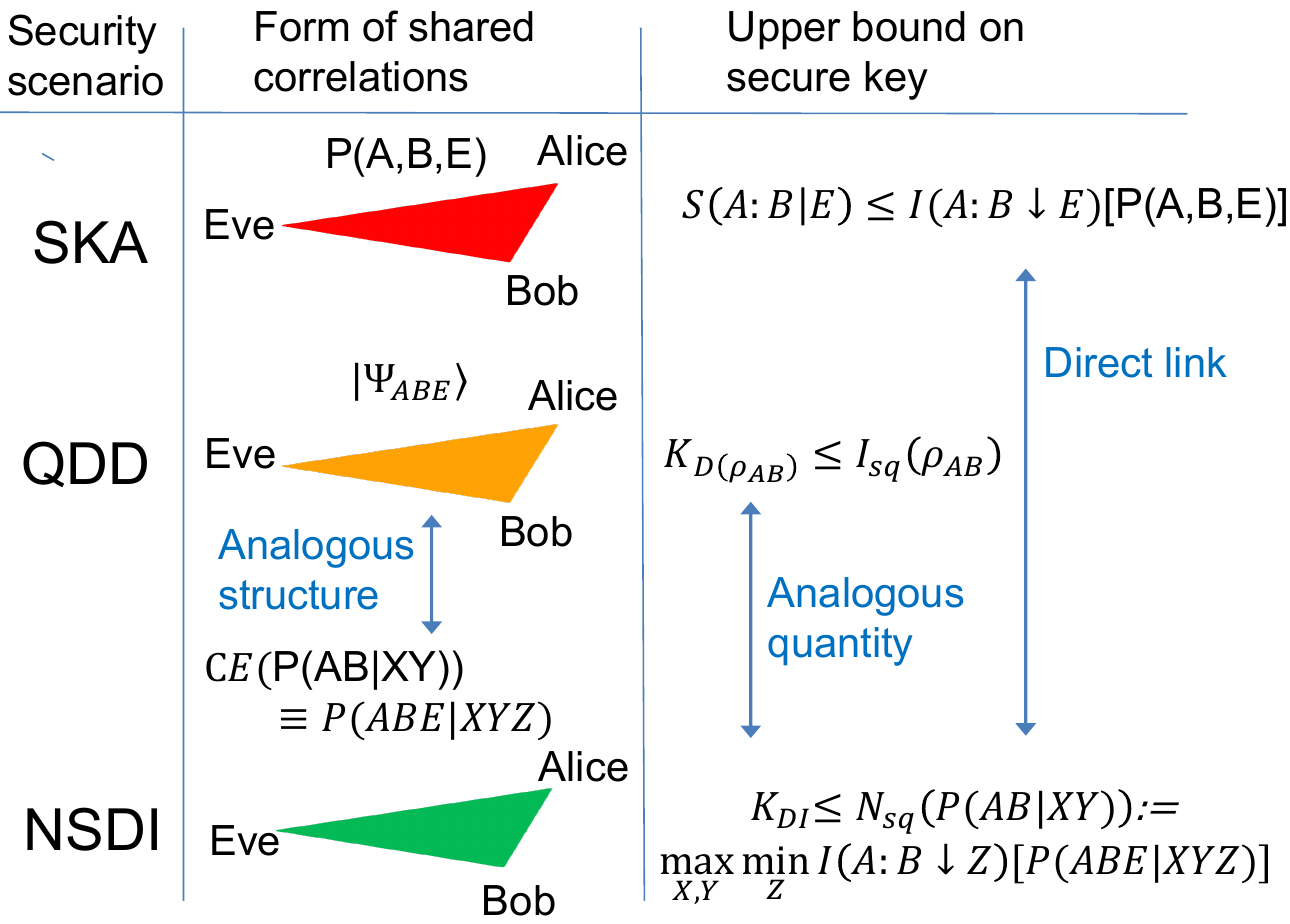}
    \end{center}
    \caption{Summary of part of the results which contribute to the analogy between security paradigms: SKA scenario where distributions are processed, and one of the upper bounds is the intrinsic information $I(A:B\downarrow E)$, QDD protocol, where the shared pure state is processed, and distillable key $K_D$ is upper bounded (among others) by the measure "squashed entanglement" $I_{sq}$ \cite{Tucci-squashed, Winter-squashed-ent}. We reformulate NSDI paradigm so that it bases on the complete extension, ${\cal E}(P(AB|XY))$,  of a device (conditional distribution) $P(AB|XY)$,
        introduce an analogue of intrinsic information and squashed entanglement called "squashed nonlocality". }
    \label{analogy}
\end{figure}

\subsection{Analogies between different cryptographic paradigms}

We finally compare the proposed security criteria with the previously known ones \cite{masanes-2009-102,masanes-2006,hanggi-2009,hanggi-2009b,Hanggi-phd,Renner-Hanggi}, and prove their equivalence. In the case of quantum mechanics, the power of eavesdropper is fully described by  system of the honest parties through the so-called {\it purification}. However, it is known that there is no analog of the quantum purification in the realm of devices \cite{Chiribella2010,Chiribella2011}. To overcome this problem, we have used a recently introduced notion of {\it complete extension}  \cite{CE}, to describe the eavesdropper's power. The complete extension, ${\cal E}(P)(ABE|XYZ)$, of the shared device $P(AB|XY)$,  is the worst-case extension that Eve can share with the honest parties. It is the worst case in the sense that it gives the 
eavesdropper an ultimate power as 
compared to quantum purification does in QDD and QDI scenarios. Indeed, the complete extension gives access to all possible ensembles of the device of the honest parties, when randomizing input and post-processing channel is applied on the extended part. It implies, as we show in detail, that this structural approach is equivalent to the one proposed in \cite{hanggi-2009}. 

 We have further introduced a novel criterion of security, based on
an operational distance measure between non-signaling devices - the non-signaling norm (\emph{NS norm})  analogous to the trace norm in quantum mechanics (related to the one given in \cite{ChristandlToner}).
We have also proved equivalence between our criteria and the two proposed so far in \cite{masanes-2006,masanes-2009-102,Masanes2011} and \cite{hanggi-2009,hanggi-2009b,Renner-Hanggi,Hanggi-phd}, respectively. As a byproduct,
we have shown that the latter two definitions are equivalent.
By proving equivalence of our definition based on $||.||_{NS}$ norm to the definition of \cite{hanggi-2009,hanggi-2009b,Renner-Hanggi,Hanggi-phd}, we have shown that the former is composable, in a sense given there\footnote{Naturally however, the device can not be reused in composing the protocols due to the threat of the memory attack \cite{memory-attack}}. 
A visualization of some of the main results that contribute to developing  a structural analogy between SKA, QDD, and NSDI are presented in Figure~\ref{analogy}.







\section{Security definition in the IID scenario}

In every DI secure key distillation protocol, the honest parties perform several numbers of test runs to estimate the non-local correlation present in the system and a (larger) number of key generation runs to generate the raw key. The raw key is further processed to yield the final key only if the device has passed the test run, i.e., the data are compatible with a sufficiently non-local device. Aiming at upper bounds, we study only the performance of the key generation runs. 
We, therefore, assume that, on the $N$ iid (identical, independently distributed)\footnote{For QDI, it is known that any arbitrary device can not be expressed in terms of the IID single use device, but the security proof for a broad range of cryptographic protocols can be performed via a reduction to IID \cite{Rotem-phd}. 
} copies of the shared device $P(AB|XY)^{\otimes N}$, the honest parties perform full direct measurement 
$[\mathcal{M}_{x,y}^F]^{\otimes N}$, by setting $X=x$ (Alice) and $Y=y$  (Bob) at their choice,  followed by any composition of classical post-processing of the distribution $P(AB|xy)$, and public communication (denoted as $Q$). These operations result in a pair of random variables $(S_A,S_B)$ that represents the {\it key}.  That is, on the outputs of the measured device, the honest parties perform an LOPC protocol. An operation performed on a device, that is a composition of the direct measurements and an LOPC operations we call {\it Measurement on Device Local Operation and Public Communication operation (MDLOPC)}.

In NSDI scenario Eve collects all the public communication $Q$, and then post-process her data
represented by $\tilde{P}(E|Z,Q)$. She can also perform a wider class of operations than the honest parties, including the general measurement $M_{z'}^G = \sum_z p(z|z')\mathcal{M}_{z}^F$. This is equivalent to a probabilistic choice of the inputs for direct measurements. She can do so by the general measurement $M_{z'}^G$, by wiring the output of her local auxiliary device (a dice), that generates a random conditional probability distribution $p(z|z')$, to the input of her part of the device, i.e., $Z$ of $\tilde{P}(E|Z,Q)$.
However, the ultimate power of eavesdropping in this scenario is fixed by definition of the class of operations that a hypothetical agent called {\it distinguisher}  could perform. It is assumed that distinghuisher has access to {\it both} the output of the protocol (i.e., the keys of the honest parties) {\it and} the Eve's device $\tilde{P}(E|Z,Q)$. By his operations, distinguisher should be almost not able to tell apart this so-called ``real'' device 
$P^\mathrm{real}(S_A,S_B,Q,E|Z)$
from an ``ideal'' one i.e., containing perfectly uniform and correlated keys, product with Eve's system.



We can specify now what the key distillation protocol is.
\emph{A protocol of key distillation is a sequence of MDLOPC operations $\Lambda=\left\{\Lambda_N\right\}$, performed by the honest parties on $N$ iid copies of the shared devices. Each of this $\Lambda_N$,  consists of a measurement stage $\{{\cal M}_N\}$, followed by post-processing $\{{\cal P}_N\}$, on $N$ iid copies of $P(AB|XY)$. Moreover, for each consecutive, complete extension of $N$ copies of shared devices $\mathcal{E}(P^{\ot N})(\bm{AB}E|\bm{XY}Z)$, the protocol outputs a probability distribution in part of Alice and Bob and a device in part of Eve, which is arbitrarily close to an ideal distribution,  satisfies
    \ben
    \label{eq:pdit}
    {||P_{\mathrm{out}}- P_\mathrm{ideal}^{(d_N)} ||}_\mathrm{NS} \le \varepsilon_N \stackrel{N\rightarrow \infty}{\longrightarrow} 0.
    \een
}
\noindent  Here $P_{\mathrm{out}} = \Lambda_N ~ \left({\mathcal{E}} \left(P^{\ot N}\right)\right)$. Moreover $\bm{A} = A_1A_2 \ldots A_N$, $\bm{B}$, $\bm{X}$ and $ \bm{Y}$ are similarly defined.

The definition of the secret key rate, based on the notion of the (i) complete extension and (ii) the key distillation protocol, satisfying the proximity in the NS norm security criterion according to the Eq. (\ref{eq:NS-norm-simplified}), is given below.
    \begin{definition}\label{def:key_rate}
    Given a bipartite device $P\equiv P(AB|XY)$ the secret key rate of the protocol of key distillation  $\Lambda_N$, on $N$ iid copies of the device, denoted by $\mathcal{R}\left(\left.\Lambda\right|_P\right)$ is a number $\limsup_{N\rightarrow \infty} \frac{\log d_N}{
        N}
    $, where $\log d_N$ is the length of a secret key shared between Alice and Bob, with $d_N=\mathrm{dim}_\mathrm{A}\left(\Lambda_N \left({\cal E} \left(P^{\otimes N}\right)\right)\right) \equiv |S_A|$. The  device independent key rate of the {\it iid} scenario is given by
    \be
    K_{DI}^{(iid)}(P)=\sup_{ \Lambda} {\cal R}\left(\left.\Lambda\right|_P\right),
    \ee
    where the supremum is taken over all MDLOPC  protocols~$\{\Lambda\}$.
\end{definition}
Later in this manuscript, we argue that the above definition is equivalent in terms of security
to the one adopted earlier \cite{masanes-2006,masanes-2009-102,Masanes2011,hanggi-2009b,Renner-Hanggi,hanggi-2009,Hanggi-phd}, which allows us to compare some of the existing lower bounds with
the upper bounds that we provide.

\section{Squashing procedure}

Let us suppose that $\mathrm{M}(A:B||E)$ is a real-valued and non-negative function, with domain in the set of tripartite probability distributions $P(ABE)$, which is an upper bound on secret key rate $\mathrm{S}(A:B||E)$ in SKA cryptographic paradigm \cite{Maurer93}, i.e., $\forall {P(ABE)},~~\mathrm{M}(A:B||E) \ge \mathrm{S}(A:B||E)$.  We will refer to $\mathrm{M}(A:B||E)$ as to {\it secrecy quantifier}. Additionally, if $\mathrm{M}(A:B||E)$ is monotonic with respect to LOPC and zero for product distributions, we call it a \emph{secrecy monotone}. Squashing a secrecy monotone will not yield an MDLOPC monotonic quantifier in general. 
The quantifiers of secret correlations in the NSDI model can be constructed by mapping the tripartite non-signaling device $R(ABE|XYZ)$ to a joint probability distribution, as given in the definition.
\begin{definition}\label{def:SQ} Corresponding to each secrecy quantifiers in SKA model $\mathrm{M}(A:B||E)$, we associate a non-signaling secrecy quantifier $\widehat{\mathrm{M}}(A:B||E)$ acting on the tripartite non-signaling devices:
    \begin{align}\label{eq:squashed-fn}
    &\widehat{\mathrm{M}} \left(A:B||E\right)_{R(ABE|XYZ)} :=\nonumber \\ &\max_{x,y} \min_z \mathrm{M}\left(A:B||E\right)_{(\mathcal{M}^F_{x,y} \otimes \mathcal{M}^G_{z}) R(ABE|XYZ)},
    \end{align}
where    
\ben 
(\mathcal{M}^F_{x,y} \otimes \mathcal{M}^G_{z'})R(ABE|XYZ)= \nonumber\\ 
  \sum_{z} p(z|z') R(ABE|X=x,Y=y,Z=z).
\een
If $R(ABE|XYZ) \equiv {\cal E}(P)(ABE|XYZ)$, is the complete extension of a bipartite device $P(AB|XY)$, we call $\widehat{\mathrm{M}} \left(A:B||E\right)_{\mathcal{E}(P)(ABE|XYZ)}$ the non-signaling squashed secrecy quantifier. If $\mathrm{M}(A:B||E)_{R(ABE|XYZ)}$ is a secrecy monotone, we call $\widehat{\mathrm{M}}(A:B||E)_{R(ABE|XYZ)}$ a non-signaling secrecy monotone. Additionally if $R(ABE|XYZ)$ is 
a complete extension, we call it a non-signaling squashed monotone.
\end{definition}
\begin{figure}[h]
	\begin{center}
		\includegraphics[width=0.5\textwidth]{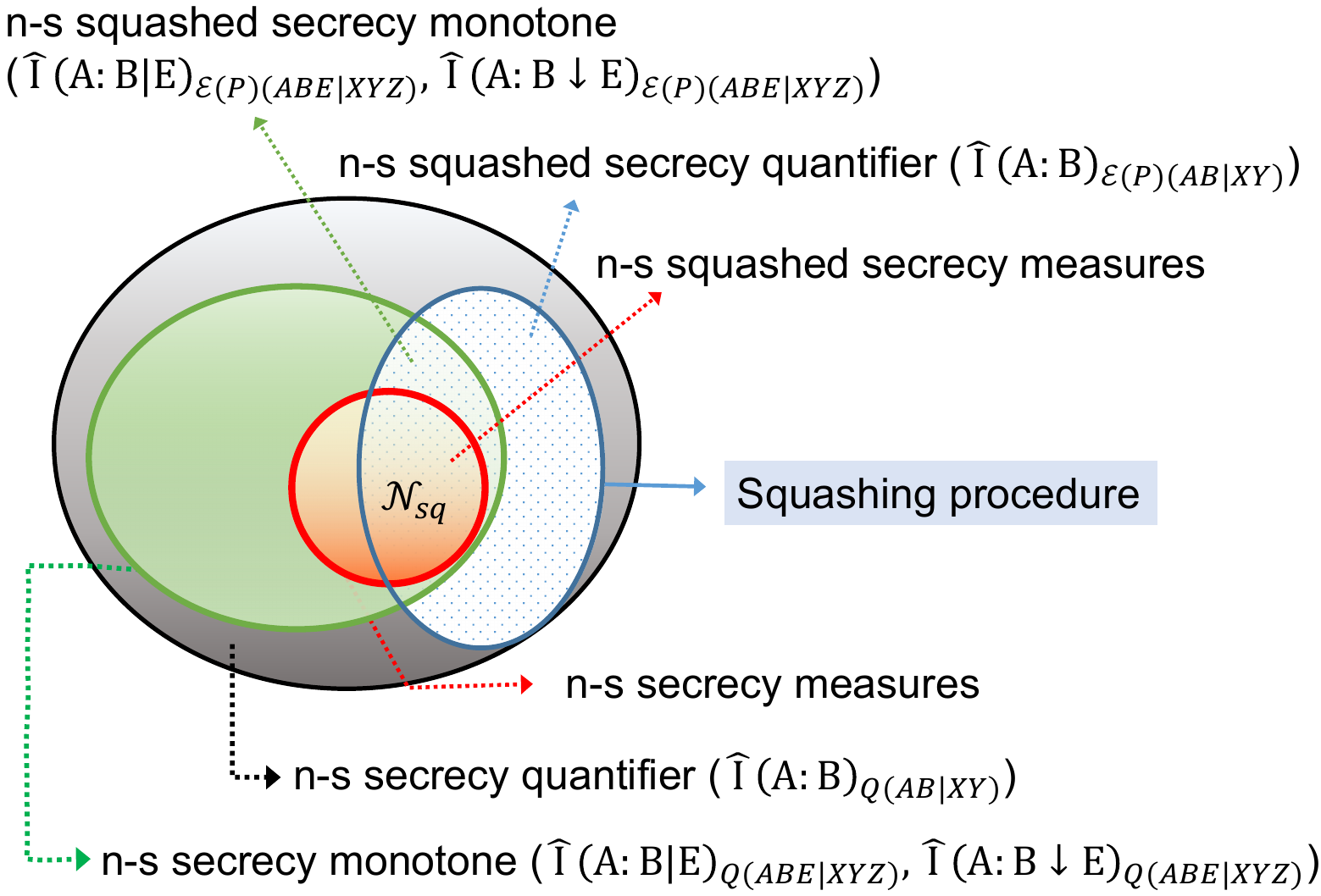}
	\end{center}
	\caption{The relative hierarchy of the squashed function $\widehat{\mathrm{M}} \left(A:B||E\right)$ of any bipartite device $P(AB|XY)$. Any $\widehat{\mathrm{M}} \left(A:B||E\right)$ functions which is positive semidefinite and vanishes for devices which is the product of two local devices, is called a non-signaling squashed secrecy quantifier, and the set of all such functions is the entire region inside the black ellipse.  The quantifiers, that are generated from the monotones of the SK paradigm are called 
	\emph{non-signaling squashed secrecy monotones}, and are represented by the green region inside the green ellipse, as a a subset of the secrecy quantifiers.
	 $\widehat{\mathrm{I}}(A:B|E)$, $\widehat{\mathrm{I}}(A:B\downarrow E)$ and $\widehat{\mathrm{I}}(A:B\downarrow \downarrow E)$ are the monotones belonging to this category. If a function is additionally  monotonic under MDLOPC operations for devices, vanishes for local ones, then we call it the \emph{non-signaling squashed secrecy measure}, and the set  represented by the red region in the figure. Any secrecy quantifier will be called squashed secrecy quantifier if the extension of the device $P(AB|XY)$ has been taken to be the CE ${\cal E}(P)(ABE|XYZ)$. The set of such functions are denoted by the dashed blue region.  The intersection of the dashed blue  region with the green region includes all squashed secrecy monotones, whereas its intersection with the red contains all squashed secrecy measures.   The squashed nonlocality ${\cal N}_{sq}$ is a particular function from that region, which is depicted as the black dot.}
	\label{fig:hierarchy}
\end{figure}	
Here, by ${\max_{x,y}}$, we mean the maximization over all possible direct measurements, $\mathcal{M}^F_{x,y} \equiv \mathcal{M}^F_{x}\otimes \mathcal{M}^F_{y}$  by the honest parties, whereas the $\min_z$ implies that the eavesdropper will try to minimize the function over all possible choices of measurements, direct and general. 
Optimization over direct measurements involves a fixed input choice, whereas for general measurement, one needs
to perform optimization over all possible conditional probability distributions $p(z|z')$.
In our MDLOPC key distillation protocol, the eavesdropper can choose her measurement adaptively, based on the public communication variable $Q$.
Hence the causal order of the optimization on the secrecy quantifier is that Alice and Bob first choose their optimal measurements, and then Eve performs her part. This gives her the maximal operational power to reduce the correlations between the honest parties 
\footnote{One can also consider the reverse order of optimization, but that opens up a different, uncommon paradigm of key distillation.}. 

The motivation to use the term ``squashed" in the above measures, comes from the fact that the definition of squashed entanglement, of an arbitrary quantum state $\rho_{AB}$, contains an optimization over all possible extensions $\rho_{ABE}$, where $\text{tr}_E(\rho_{ABE}) =  \rho_{AB}$. This arbitrary extension $\rho_{ABE}$ can be obtained from the purification $|\psi\rangle_{ABE}$ of the quantum state \cite{Winter-squashed-ent}. In the analogy of these, here we use the complete extension ${\cal E}(P)$, the non-signaling equivalent of quantum purification,  which is the key ingredient to perform an optimization over all possible non-signaling extensions \cite{CE} of a given device $P$. 
The secrecy quantifiers, we have used for squashing,  are  the mutual information  $\mathrm{I}(A:B)$, the conditional mutual information $\mathrm{I}(A:B|E)$, the intrinsic information $\mathrm{I}(A:B\downarrow E)$ \cite{Intrinsic-Maurer} and the reduced intrinsic information $\mathrm{I}(A:B\downarrow\downarrow E)$ \cite{reduced-intrinsic}.  Among them, $\mathrm{I}(A:B|E)$, $\mathrm{I}(A:B\downarrow E)$ and $\mathrm{I}(A:B\downarrow\downarrow E)$ are secrecy monotones. Hence  $\hat{\mathrm{I}}(A:B|E)$, $\hat{\mathrm{I}}(A:B\downarrow E)$ and $\hat{\mathrm{I}}(A:B\downarrow\downarrow E)$ are non-signaling squashed secrecy monotones while $\hat{\mathrm{I}}(A:B)$ is an example of a non-signaling squashed secrecy quantifier.

The inclusions between gray, green, and orange sets in Fig. \ref{fig:hierarchy} follow directly from the definition of different classes of functions. Namely, all n-s secrecy measures are necessarily n-s secrecy monotones, and all n-s secrecy monotones are necessarily n-s secrecy quantifiers, but not vice versa. The strictness of the inclusions follows from a trivial example of n-s mutual information (gray area), n-s intrinsic information (orange area) and n-s intrinsic information shifted by a non-zero constant (green area). Analogous relation is true for the squashed version of the aforementioned functions. Nevertheless, the squashing procedure does not imply that the resulting function is automatically a secrecy measure or a secrecy monotone; therefore, the representatives of squashed functions are present in all three sets.

\section{Generic upper bound and the squashed nonlocality}

Below, we use the aforementioned idea of squashing for upper-bounding the secret key in the NSDI scenario with MDLOPC operations.
~~\\
~~\\
\begin{theorem} \label{thm:main}
    The secret key rate, in the non-signaling device-independent {\it iid} scenario achieved with MDLOPC operations, $K_{DI}^{(iid)}$, from a device $P$, is upper bounded by any non-signaling squashed secrecy quantifier evaluated for the complete extension of $P$:
    \be
    \forall_{P}~~
    K_{DI}^{(iid)}(P) \le \widehat{\mathrm{M}} \left(A:B||E\right)_{\mathcal{E}\left({P}\right)} ,
    \ee
    where $P\equiv P(AB|XY)$ is a single copy of a bipartite non-signaling device shared by the honest parties, and $\mathcal{E}(P) \equiv \mathcal{E}(P)(ABE|XYZ)$ is its complete extension to the eavesdropper's system. 
\end{theorem}
\noindent\textit{Proof}. For the proof, see Sec. \ref{sec:upper-iid} of  the Appendix.

 Theorem \ref{thm:main}, together with Definition \ref{def:SQ}, establishes a connection between the secret key rate in the SKA and NSDI scenario. 
 The novelty of our approach is that not only it connects at least two major security paradigms, but it also opens up a new area of research - to study more tighter upper bounds on the key rate in the NSDI scenario (for parallel, different approach see \cite{Kaur-Wilde}). In this paper, we focus on the secrecy monotone called intrinsic information $I(A: B\downarrow E)$.  From this secrecy monotone via squashing we construct the so called {\it squashed nonlocality}, as an upper bound on the NSDI key. We then prove several important properties of squashed nonlocality, which promotes it as a measure of nonlocality. Secrecy monotone called the reduced intrinsic information $I(A:B\downarrow \downarrow E)$, provide a tighter bound on the key rate in the SKA scenario, as $I(A:B\downarrow \downarrow E) \leq I(A: B\downarrow E)$ for tripartite probability distribution $P(ABE)$ \cite{RennerK04-key}. Hence we open a possibility to study even tighter upper bound on the $K_{DI}^{(iid)}$, upon squashing the $I(A:B\downarrow \downarrow E)$. 
%
We focus now on the definition of the aforementioned  {\it squashed nonlocality}.
\begin{definition}\label{def:squashed_nonlocality}
    The squashed nonlocality $ {\cal N}_{sq}(P)$, of a bipartite non-signaling device $P:= P(AB|XY)$ is 
    \ben\label{eq:Nsq}
    &{\cal N}_\mathrm{sq}(P) := \widehat{\mathrm{I}}\left(A:B\downarrow E\right)_\mathrm{\mathcal{E}\left(P\right)(ABE|XYZ)}= \nonumber \\
    & \max_{x,y} \min_z \mathrm{I}\left(A:B\downarrow E\right)_{(\mathcal{M}^F_{x,y} \otimes \mathcal{M}^G_{z}) \mathrm{\mathcal{E}\left(P\right)(ABE|XYZ)}}, \nonumber 
    \een
    where $\mathcal{E}\left(P\right) := \mathcal{E}\left(P\right)(ABE|XYZ)$ is the complete extension of the device $P$ \cite{CE}.
\end{definition} 
We note here, that the above definition is tuned to the deifinition of $K_{DI}^{(iid)}$. The order of the $\max_{x,y}$ and $\min_z$ stems from the fact that we consider the scenario of key distillation in which
Eve knows $x,y$ beforehand. In our case the inputs $(x,y)$ are fixed before the beginning of the protocol, but in general it could be announced during the protocol's execution. This is important point,
as alternative protocols exist in which only one party announces the inputs, and the key is distilled from output of all the inputs \cite{AcinGM-bellqkd}. In the latter case a positive key rate can be obtained even from the quantum {\it isotropic devices} in the scenario of two binary inputs and two binary outputs, while in the scenario which we consider where both inputs are known to the eavesdropper no positive lower bound on the key rate is known. It is possible that the upper bounds on the protocols such as those from \cite{AcinGM-bellqkd} where $x$ is not announced, are provided in terms of the squashed nonlocality where however $\max_y\, \min_z\, \max_x$ appears in front instead of $\max_{xy}\,\min_z$.

 From the definition of a complete extension of a device (see Ref. \cite{CE}) we know that in order to construct it, one needs to identify all possible so-called {\it minimal ensembles} of the device. For example, in the polytope of two binary input and two binary output devices $(2,2,2,2)$, a device lying on the isotropic line between Popescu-Rohrlich and Tsirelson's one\footnote{By Tsirelson's device we mean a one attaining maximal value of violation of the CHSH inequality \cite{CHSH} among quantum $(2,2,2,2)$ devices \cite{Tsirelson-bound}.} has up to $354$ minimal ensembles (achieved for the Tsirelson's device). However, a priori, there are $880946$ of ensembles that can be potentially minimal \cite{CE}. Hence, obtaining all possible minimal ensembles, and therefore finding out the complete structure of the CE may be an arduous task. However, we observe that to obtain a non-trivial upper bound on the ${\cal N}_{sq}$, not the whole complete extension has to be even known. 

We collect below certain properties of the above measure. Some of them are used in what follows, and some of  them are of independent interest in the context of Bell nonlocality.
\begin{proposition} \label{pro:properties}
	Besides being non-faithful, the squashed nonlocality satisfies the following properties:
\begin{enumerate}
        \item \label{pro:prop1}\textbf{Positive.} It is a non-negative real function of bipartite non-signaling devices, and equal to zero for local devices.\footnote{By local we mean devices which possess a local hidden variable model \cite{Bell-nonlocality}.}.
        \item \label{pro:prop2}\textbf{Monotonic} with respect to MDLOPC class of operations.
    \item \label{pro:prop3}\textbf{Convex} with respect to  the mixture of devices.
        \item \label{pro:prop4}\textbf{Superadditive}  over joint non-signaling devices.
        \item \label{pro:prop4a}\textbf{Additive}  for  product devices.
        \item \label{pro:prop6}\textbf{Subextensive.}  $\mathcal{N}_\mathrm{sq}(P) \le \log\left(\min\left\{d_A,d_B\right\}\right)$.
    \end{enumerate}
\end{proposition}
\noindent\textit{Proof}. For the proof, see Sec. \ref{appen:proofproperties} of the Appendix. See also the discussion in Section \ref{sec:quantitative}. 

Note: On the completion of the main results (preliminary version of this paper) contained in Sections \ref{appen:feature-boxnorm}-\ref{sec:upper-iid}, \ref{sec:nsq} and \ref{sec:numerical} in the Appendix, we have noticed the preprint of the paper by E. Kaur, M. Wilde and A. Winter \cite{Kaur-Wilde} also related to upper bounds on device independent key. The proofs of monotonicity, subadditivity and additivity over tensor
product devices (see Sections \ref{sec:monotonicity} and  \ref{sec:superadditivity} of the Appendix), were inspired by the analogous result for the squashed intrinsic nonlocality presented there. 

Calculating ${\cal N}_{sq}$ for an arbitrary bipartite device $P$ is a non-trivial task, but we can use the convexity of this measure to simplify the procedure of finding an upper bound of it. 
Positivity, monotonicity, and additivity of squashed nonlocality 
 lead to the following Corollary.
\begin{cortwo}\label{cor:measure}
    The squashed nonlocality is a measure of non-local correlation  of the bipartite device $P$.
\end{cortwo}

We describe now, how to use the convexity of the squashed nonlocality (this technique proposed in this manuscript proved already useful in context of upper bounds on the secure key in QDI scenario \cite{KaurHorodeckiDas}). Consider 
any set of functions  ${\cal F} = \{F_i(P)\}$, that are convex w.r.t. the mixture of devices, each of which upper bounds the squashed nonlocality $F_i(P) \geq {\cal N}_{sq}(P), \forall i$. Then the lower convex hull (LCH) of ${\cal F}$ denoted as  $F(P) (\equiv \mathrm{LCH}({\cal F}))$ upper bounds ${\cal N}_{sq}(P)$,  i.e., ${\cal N}_{sq}(P) \leq F(P)$, as a consequence of property \ref{pro:prop3}. To exemplify the above convexification process, let ${\cal F} = \{\widehat{\mathrm{I}}(A:B)_{P(AB|XY)}, \widehat{\mathrm{I}}(A:B|E)_{{\cal E}(P)(ABE|XYZ)}\}$, then ${\cal N}_{sq}(P) \leq F(P) \equiv \mathrm{LCH}(\widehat{\mathrm{I}}(A:B)_{P(AB|XY)}, \widehat{\mathrm{I}}(A:B|E)_{{\cal E}(P)(ABE|XYZ)}).$ This fact
is used in order to construct  figure \ref{fig:numerical-ub}: the orange curve is, in fact, a convex hull of several
upper bounds that are incomparable with each other.

\section{Quantitative results}\label{sec:quantitative}

In Figure \ref{fig:numerical-ub}, we construct numerically an upper bound on the ${\cal N}_{sq}$, with the help of above specified convexification procedure. We also draw several other squashed quantifiers for the set of $(2,2,2,2)$ devices, lying in the isotropic line, i.e., $P_{iso} = (1 -  \varepsilon) PR +  \varepsilon \overline{PR} $. Where $PR$ is the famous Popescu-Rohrlich box \cite{PR}, and $\overline{PR}$ is the anti-PR box\footnote{Anti-PR box is a binary input output device, satisfy $\overline{PR} (ab|xy) = \frac 12 \delta_{a\oplus b, \overline{x.y}}, ~\forall a,b,x,y \in \{0,1\}$ \cite{Barrett-info-proc}}.
The non-faithfulness of our measure,  ${\cal N}_{sq}$ is visible from the numerical results. The orange curve is the upper bound on ${\cal N}_{sq}$, and we have found that the bound reaches $0$ for $\varepsilon = 0.2$ (it remains equal to $0$ for $\varepsilon \in [0.2, 0.25]$ due to the convexity of the measure). 
This is since, in MDLOPC protocol, Eve can perform adaptive general measurements and post-process her output through a classical post-processing channel to reduce the correlations between Alice and Bob. In the range $\varepsilon \in [0.2, 0.25]$, corresponding to each input $(x,y)$ of the honest parties, we have found a measurement and a post-processing channel on Eve, which partitioned the device into an ensemble of product distributions.
This proves that there exists nonlocality which can not be turned into security via MDLOPC protocols. 
Interestingly, these devices are quantum realizable ones. One can conjecture that even the general operation, including the so-called ``wirings''\footnote{Operations of feeding input of one device with the output of the other.}
can not help in distilling key out of these isotropic devices. Indeed, using wirings that is necessary for the key to be non-zero, which implies that we enter to some extent the general scenario of key distillation for which there is a wide class of attacks by employing the forward signaling attacks found in \cite{Rotem-Sha,Salwey-Wolf}.

\begin{figure}[t]
    \begin{center}
        \includegraphics[trim={1.2cm 0 0 0},clip,width=0.515\textwidth]{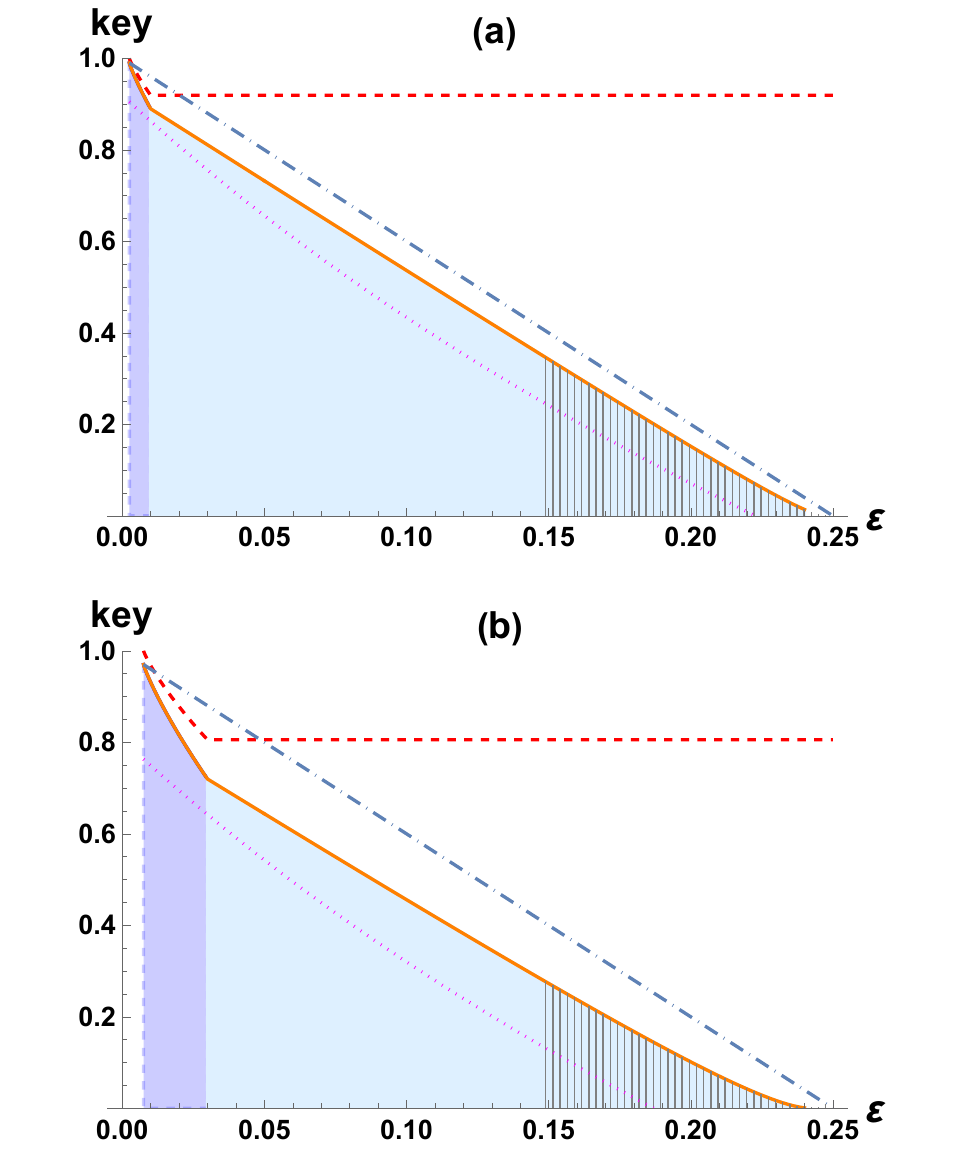}
    \end{center}
    \caption{
     Plot of several non-signaling secrecy quantifiers $\widehat{\mathrm{M}}(A:B||E)$, as an upper bound on secure key rate $K_{DI}^{(iid)}$,  for the bipartite binary input output device $P_{HRW}$ given in Eq. (\ref{eqn:RH-box_0}) (also in Ref. \cite{hanggi-2009}). The parameters used to draw plot (a) are $\delta=0.01$, $\epsilon=\frac{1}{16} \left(3.04+12 \varepsilon\right)$, and for plot (b) we used $\delta=0.03$, $\epsilon=\frac{1}{16} \left(3.12+12 \varepsilon\right)$. The dashed red line corresponds to the non-signaling squashed mutual information $\widehat{\mathrm{I}}(A:B)_{P_{HRW}}$. The blue straight line represents the nonlocality cost, as well as the non-signaling squashed conditional mutual information $\widehat{\mathrm{I}}(A:B|E)_{{\cal E}({P_{HRW}})}$ over the complete extension ${\cal E}({P_{HRW}})$ of the given device $P$. The solid orange line represents the upper bound on the non-signaling squashed nonlocality ${\cal N}_{sq}$ which is in fact  the lower convex hull of the several other upper bounds on ${\cal N}_{sq}$. The magenta dotted line is the key rate  $\mathcal{R}\left(\left.\mathcal{P}\right|_{\mathrm{P_{HRW}}}\right)$ of the protocol design by H{\"a}nggi, Renner and Wolf \cite{hanggi-2009}.
	The region with black stripes corresponds to the devices that are quantum realizable ones.		}
    \label{fig:numerical-ub-quantititve}
\end{figure}

In Figure \ref{fig:numerical-ub-quantititve} (a) and (b), we plot upper bounds on ${\cal N}_{sq}$ for several other sets of $(2,2,2,2)$ devices (non-isotropic), parameterized as in equation \eqref{eqn:RH-box_0}. In fact, the parametrization that we use is the same as in Ref. \cite{hanggi-2009} as we want to compare our upper bound with the lower bound obtained therein. One can see that there exists some region of non-local correlation (Figure \ref{fig:numerical-ub-quantititve} (a) and (b)), which can be simulated by a quantum device and for which the lower bound obtained by \cite{hanggi-2009,Hanggi-phd} is positive, and therefore the secret-key can be generated. As we observe and ${\cal N}_{sq}$ is also non-trivial and close to the lower bound in the case considered here. We address the interested reader to Section \ref{sec:numerical} of the Appendix, where more plots are provided.

\begin{align} 
 		&\mathrm{P}_\mathrm{HRW}\left(ab|xy\right)= \nonumber\\
		&\begin{array}{cc|cc|cc}
			&\multicolumn{1}{c}{x} & \multicolumn{2}{c}{0}	& \multicolumn{2}{c}{1} \\
			y &$\diagbox[width=1.8em, height=1.8em, innerrightsep=0pt]{$b$}{$a$}$  & 0	& 1	&0  & 1 \\
			\hline \\ [-0.9em]
			\multirow{2}{*}{0 }  & 0 & \frac{1}{2}	- \frac{\delta}{2}	& \frac{\delta}{2}		& \frac{3}{8}	-\frac{\epsilon}{2}	& \frac{1 }{8} +  \frac{\epsilon}{2}\\[0.3em]
			& 1	& \frac{\delta}{2}	& \frac{1}{2}	- \frac{\delta}{2}	& \frac{1 }{8} +  \frac{\epsilon}{2}	& \frac{3}{8}	-\frac{\epsilon}{2}  \\ [0.2em]
			\hline  \\ [-0.9em]
			\multirow{2}{*}{1 }   &0 & \frac{3}{8}	-\frac{\epsilon}{2}	   &  \frac{1}{8}	+\frac{\epsilon}{2}	  &   \frac{1}{8}	+\frac{\epsilon}{2}& \frac{3}{8}	-\frac{\epsilon}{2} \\ [0.3em]
			& 1 & \frac{1}{8}	+ \frac{\epsilon}{2}		&       \frac{3}{8}	-\frac{\epsilon}{2}	  &  \frac{3}{8}	-\frac{\epsilon}{2}	 	& \frac{1}{8}	+\frac{\epsilon}{2}
		\end{array}. ~~~~~~
		\label{eqn:RH-box_0}
\end{align}

We note here, that the result presented in Figure \ref{fig:numerical-ub} exhibits that in our approach the nonlocality measure based on the intrinsic information can be non-faithful i.e. zero for some non-local devices. This is inherited after the intrinsic information, which is known to be zero for some tripartite distributions in spite of the fact that the latter are not of the product form $P(A|E)P(B|E)$. We note here, that \cite{Nonsig_theories} claimed, that the intrinsic information is non-zero for all devices violating Bell inequality (cf \cite{acin-2006-8}). We reformulate the result of \cite{Nonsig_theories} as follows:
\begin{align}
    &\forall_{P(ABE)}\,\forall_{\Lambda:E\rightarrow E'} \exists_{(x,y)}\, \nonumber\\
    &P(ABE')\neq P(AE')P(BE') \Leftrightarrow
    I(A:B|E') >0.
\end{align}
The above implies that if we can adjust the inputs {\it after} the attack by Eve represented by the map $\Lambda$ is performed, we will obtain non-zero conditional information. This implies also non-zero intrinsic information as the map can realize the infimum over such maps in the definition of the latter. However this approach does not fit the usual cryptographic scenario: it is that Eve is listening to Alice and Bob and adjusts her measurement to their announcement and not vice versa. 
Owing to that observation, one should consider the inputs $(x,y)$ to be chosen {\it before} the map $\Lambda$ of the attack is performed. This happens e.g. whenever the input is fixed from advanced as we assume, or when it is announced right after has been made. This change in the paradigm has important consequences. What both the \cite{Farkas_2021} and our result implies goes with no contradiction with the above, as is based on the following fact:
\begin{equation}
\exists_{P(ABE)\neq P(AE)P(BE)}\,\forall_{(x,y)}\exists_{\Lambda:E\rightarrow E'} \,
    I(A:B|E') =0.
\end{equation}
Indeed, in the case of the above mentioned quantitative results we adjust the measurement and post-processing of Eve to the inputs of the honest parties.

Finally we note, that a more common approach to key distribution in device independent scenarios is such that, following A. Ekert \cite{Ekert1991}, one of the honest parties has one more input, which is use to key generation. This so called (3,2,2,2) scenario has been considered in \cite{acin-2006-8} in context of a non-signaling adversary, along with a protocol of key distillation and an upper bound on it in terms of the intrinsic information. 
To see the relation between our results with
bthat of \cite{acin-2006-8}, we show
the Eq. (\ref{eq:equivalence}), that is $\max_{(x,y)}\mathrm{I}_{\mathrm{AMP},(x,y)}={\cal N}_{sq}$ (see Sec. \ref{sec:app:RelationAcin} of the Appendix).
 We note here, that by this fact we show that the bound given in $\mathrm{I}_{\mathrm{AMP},(x,y)}$ hold for any MDLOPC protocol using inputs (x,y) for generating key, closes the problem left open in \cite{acin-2006-8} concerning possibility of key distillation from states that violate CHSH inequality but have zero $\mathrm{I}_{\mathrm{AMP},(x,y)}$ bound.

\begin{figure}[h]
    \begin{center}
        \includegraphics[width=0.45\textwidth]{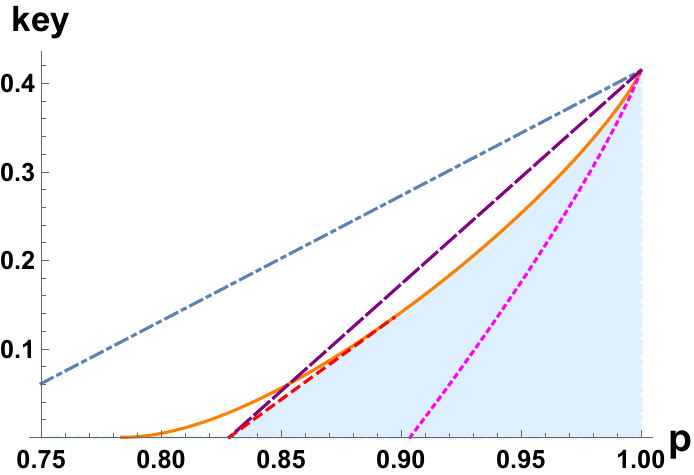}
    \end{center}
    \caption{ Plot of non-trivial upper bound on the secret-key rate $K_{DI}^{(iid)}$ given by ${\cal N}_{sq}$, of $\mathrm{P}_\mathrm{AMP}\left(ab|xy\right)$ given in Eq. (\ref{eqn:AMP-box}) (see Appendix), by the blue shaded region under the orange solid line and a red dashed line. The red dashed line is the (segment of) lower convex hull of the orange solid curve and the purple ``big-dashed" straight line. The solid orange line is obtained by the lower convex hull of several upper bounds of ${\cal N}_{sq}$, with the help of Eq. (339). Blue dashed-dotted line is the squashed conditional mutual information $\widehat{\mathrm{I}}(A:B|E)_{{\cal E}(\mathrm{P}_\mathrm{AMP})}$. The magenta dotted line is the lower bound on the key rate, whereas the purple big-dashed line is the upper bound on intrinsic information of the eavesdropping strategy used in \cite{acin-2006-8}. We observe that the convexification technique resulting in the convex-hull bound allows to obtain tighter upper bound on ${\cal N}_{sq}$, and therefore the tightest known upper bound on the secret key-rate in the non-signaling scenario.  }
    \label{fig:numerical-ub-3222_main}
\end{figure}

As we will see this fact proves useful, since we have shown that ${\cal N}_{sq}$ is convex. This will enable us to use the convexification method to obtain tighter upper bounds. Following \cite{acin-2006-8}, 
\color{black} as a noise model, we consider the isotropic state $p|\psi_+\>\<\psi_+|_{AB} + \frac{(1-p)}{4}\mathds{1}_{AB}$ with $|\psi_+\>=\frac{1}{\sqrt{2}}(|00\> +|11\>)$ with $p\in[0,1]$.
The bound outperforms existing one \cite{acin-2006-8} in a wide range of a parameter $p$ (see the orange curve in Fig. \ref{fig:numerical-ub-3222_main}). In general however it is  incomparable (for the whole range of parameters) with the one given in \cite{acin-2006-8}. It is possible that a more refined optimization procedure, involving all the extremal points of the non-signaling polytope in $(3,2,2,2)$ scenario, would provide a tighter bound. It is however computationally involved.

Moreover the convex hull of the bound given in \cite{acin-2006-8} and ours (which we got by convexification of two upper bounds, given in Eq. (339)), is also an upper bound on the distillable key. This is because ${\cal N}_{sq}$ is 
a lower bound to both upper bounds, and is convex. Hence is less than the convex hull of the latter two bounds. This gives to our knowledge the tightest bound known so far in this scenario.



\section{Rephrasing the key rate of the SKA model}
 
In the SKA model of key distillation, the honest parties and the eavesdropper share a joint probability distribution $P(ABE)$. The task of the honest parties is to perform LOPC operation to distill a secret-key, in such a manner that the eavesdropper's knowledge about the key remains negligibly small. 
In the following lines, we propose an alternative definition of the key rate in the aforementioned scenario and prove that it is equivalent to the definition of the secret key rate introduced in the literature \cite{CsisarKorner_key_agreement,Maurer93,MaurerWolf00CK,Christandl12}. Rephrasing, the definition of the secret-key in the SKA model to the form similar to the one used in quantum cryptography serves not only as a connection between different cryptographic paradigms. Indeed, the Theorem below, besides being interesting on its own, is a crucial ingredient used to prove Theorem \ref{thm:main}, i.e., our main result.

\begin{theorem}[Informal]\label{thm:SK_key0} The secret key rate ${S} ( A : B||E)$ of SKA cryptographic model \cite{CsisarKorner_key_agreement,Maurer93,MaurerWolf00CK,Christandl12} 
{is equivalent to the following asymptotic expression:}
	\be\label{eq:equality}
	\mathrm{S} ( A : B||E) =
	\sup_{\cal P} \limsup_{N\rightarrow \infty} \frac{\log \mathrm{dim}_\mathrm{A} \left( \mathcal{P}_N\left({P}^{\ot N}\left(ABE\right)\right)\right)}{N},
	\ee
	{with security condition}
	\be
	\norm{\mathcal{P}_N\left({P}^{\ot N}\left(ABE\right)\right) - P_N^\mathrm{ideal}}_1 \le \delta_N \stackrel{N\to \infty}{\longrightarrow}0,
	\label{eqn:normAP}
	\ee
	where $\mathcal{P}=\cup_{N=1}^\infty \{{\mathcal{P}}_N\}$ is a cryptographic protocol consisting of LOPC operations, acting on $N$ iid copies of the classical probability distribution $P(ABE)$ {, and $P_N^\mathrm{ideal}$ is the distribution containing ideal secret key, with adequate dimensions}. 
\end{theorem}
\textit{Proof}. For the proof, see Sec. \ref{sec:Maurer-rate} of the Appendix.

The aim of this rephrasing is to show and utilize a connection between the definition of secret key rate in SKA and NSDI scenarios, as it was done {in the case of quantum cryptography} \cite{Christandl12}.

The link we have made in the above Theorem, is technical, however important in our method for obtaining the upper bound on the key rate in NSDI scenario.
We rephrase the security definition of SKA proposed by U. Maurer \cite{Maurer93},  with the one based on the trace norm $||.||_1$. What is crucial in the choice of the latter criterion is the fact that it is equivalent to the NS norm $||.||_{NS}$ for tripartite probability distributions\footnote{Note however that $||.||_{NS}$ norm applies also to conditional distributions i.e. devices. Only for devices with unary input i.e. distributions, it is equivalent to $||.||_1$.}. We recall here that the security definition in SKA is based on the control of the
correlations (in terms of the mutual information) of the random variable of the honest parties with a random variable representing
Eve's knowledge. These correlations should tend to zero for a large number of copies, $N$. Thus, in other words, in the above Theorem, we have modified the security criterion of the SKA to an equivalent form, which is more useful for our purpose. We have done so by demanding that the output distribution of the protocol should be close to an ideal one. The ideal is the distribution representing perfectly correlated uniform random variables, of the honest parties close to being product with the variable of Eve, in trace norm distance $||.||_1$. As it will appear later, this 
technical change turns to be useful when we pass to the case of devices because the NS norm of a device is in fact a trace norm of a distribution coming from this device after measurement. 


\section{Equivalence of the security conditions}

In this Section we show the equivalence between two different known definitions of the security of the secret-key in the NSDI scenario via showing that each of them is equivalent to the one proposed by us. Indeed, we show that the security definition proposed by us that bases on the NS norm is equivalent both to the definition that employs secrecy and correctness as well as the so-called distinguisher  \cite{hanggi-2009,hanggi-2009b,Renner-Hanggi,Hanggi-phd} and the other one given in  \cite{masanes-2006,masanes-2009-102,Masanes2011}. 

\subsection{The definition and the properties of the NS-norm}

In this Section we provide the explicit description of the NS-norm that is an important ingredient of our security criterion. The tensor product  should be understood as an algebraic tensor product in $\mathbb{R}^N$ space \cite{TensorProduct}. To measure the closeness between two devices $P$ and $P'$, we use the newly defined distance measure, the NS norm  which reads
\be\label{box-norm}
\left| \left|P-P'\right| \right|_\mathrm{NS}:= \sup_{ g\in\mathcal{G}} \frac 12  \left| \left|g(P)- g(P')\right| \right|_1,
\ee 
where $\mathcal{G}$, is a set of certain operations that map a device to probability distributions and $||.||_1$ is a variational distance between two distributions. 
More precisely, operations from $\mathcal{G}$ are convex combinations of operations that can be composed of the following basic ones (i) fetching an
auxiliary device that has single input and single output (a {\it dice}) (ii) connecting the output of a device/dice to the input of a dice/device respectively, called {\it wirings} (iii) pre-processing the inputs of device(s)  (iv) post-processing inputs and outputs of the devices. We call them {\it generating} operations\footnote{Name for these operations stems from the fact that they are proven in \cite{CE} to generate from the complete extension any possible other non-signaling extension of a conditional probability distribution.}, and refer to this norm as to {\it non-signaling norm}. The set of generating operations $\mathcal{G}$ is a subset of all linear operations $\mathcal{L}$ mapping device to distribution, that were considered in \cite{ChristandlToner}. Operational characterization of the set $\mathcal{L}$ is interesting, yet, to our knowledge, unresolved task. However, as we show (see Proposition \ref{cor:box_norm_exp}), the set $\mathcal{G}\subseteq \mathcal{L}$ has enough power in discriminating between devices, to be used in security definition in place of $\mathcal{L}$. Indeed, $NS$ norm via Eq. (\ref{box-norm}) 
leads to security definition, which is equivalent to the other two already present in literature (\cite{masanes-2006,masanes-2009-102,Masanes2011} and \cite{Hanggi-phd,Renner-Hanggi,hanggi-2009,hanggi-2009b}). For more detailed discussion, see Sec. \ref{appen:feature-boxnorm} of the Appendix.

After the MDLOPC key distribution protocol, the output of the honest parties reduces to a classical-classical-probability distribution, whereas the part shared by  Eve still remains a device,  of the form $\Lambda_N ~ \left({\mathcal{E}} \left(P^{\ot N}\right)\right)_{S_A,S_B, Q, E|Z}(s_A, s_B, Q,E|Z)$, where $s_A$ and $s_B$ are the instances of the key shared between Alice and Bob. 
We will denote random variables $S_A$, $S_B$  for the secret keys in possession of Alice and Bob, whereas $Q$ stands for all possible classical communications between Alice and Bob; $E$, $Z$ for Eve's output and input (and the lower case letters are for their values).
This distribution, which is, in part a probability distribution, and in part a conditional probability distribution, i.e., device. Hence we will refer to it as to   ``classical-classical-device" (cc-d) distribution throughout the paper.
The $\left(P_\mathrm{ideal}^{(d_N)}\right)_{S_A,S_B, Q, E|Z}(s_A, s_B, Q,E|Z) =\frac{1}{|S_A|}\delta_{s_A,s_B} \otimes \sum_{s'_A,s'_B} \Lambda_N ~ \left({\mathcal{E}} \left(P^{\ot N}\right)\right)(s'_A,s'_B,Q,E|Z)$,
is an {\it ideal} cc-d distribution which contains uniform and perfectly correlated outcomes shared between the honest parties. Eve is completely uncorrelated in case of this distribution, and it is assumed that Eve's system is  the same as she possesses at the end of the real protocol $\Lambda_N$.

For the cc-d distribution shared at the end of the MDLOPC protocol, the NS norm given in Eq. (\ref{box-norm}) takes a more simplified form, stated in the following Proposition:
    \begin{proposition}\label{cor:box_norm_exp} For the cc-d states $P$ and $R$ shared at the end of the MDLOPC protocol $\Lambda_N$, the NS norm can be rephrased with a simplified expression:
    \ben\label{eq:NS-norm-simplified}
    &&\left| \left| P_{S_A,S_B,Q,E|Z}-R_{S_A,S_B,Q,E|Z}\right| \right|_\mathrm{NS}= \nonumber \\
    &&\frac{1}{2}
    \sum_{s_A,s_B,q} \max_z \sum_{e}
    \left| P_{S_A,S_B,Q,E|Z}(s_A,s_B,q,e|z) \right. \nonumber \\ && \left.  -R_{S_A,S_B,Q,E|Z}(s_A,s_B,q,e|z) \right|, 
    \een
\noindent where $\max_z$, stands for the maximization over all possible direct measurements performed by the eavesdropper. 
\end{proposition}
\noindent\textit{Proof}. For the proof, see Sec. \ref{appen:feature-boxnorm} of  the Appendix. \\
 In the above equality, one can see that the adopted definition of security is equivalent to the one used in \cite{masanes-2009-102,masanes-2006,Masanes2011} in the case of the NSDI scenario (the latter is defined as in RHS of the (\ref{eq:NS-norm-simplified}). This justifies our security definition given in Eq. (\ref{eq:pdit}), in particular, the choice of the set of operations ${\mathcal G}$, that define the NS norm $||.||_{NS}$. However, in literature another definition of security is adopted, given in \cite{Hanggi-phd,Renner-Hanggi,hanggi-2009,hanggi-2009b}. This one is based on assuring high correlations between the parties and low correlations with the eavesdropper. In this approach, Eve can generate {\it ensembles} of the device of the honest parties i.e., representation of a device as   probabilistic mixtures of devices.
	In later part of this manuscript we show that the latter definition is also equivalent to the newly proposed one based on the NS norm. By doing so, as a byproduct, we have also
	proven that our, and the two definitions given in \cite{masanes-2006,masanes-2009-102,Masanes2011} and \cite{Hanggi-phd,Renner-Hanggi,hanggi-2009,hanggi-2009b} respectively, are equivalent.


\subsection{Equivalence of security criteria}

We show that in the NSDI scenario, in analogy to quantum cryptography \cite{Portmann-Renner,Beaudry}, there exist two different, however equivalent definitions of security. One connected to the notion of the so-called distinguisher and the other one based on the proximity in norm \cite{Ben-Or-Mayers-compos,Ben-OrHLMO05}. In the case of NSDI, Renner, H{\"a}nggi, and Wolf \cite{hanggi-2009} present the approach via the notion of distinguisher.  Recall here, that to develop the latter approach, we consider the non-signaling norm, which is a total
variational distance for two devices mapped into probability distribution with the so-called  {\it non-signaling operations}, over which we take a supremum (see \cite{hanggi-2009,ChristandlToner} in this context). We then focus on tripartite cc-d distributions (classical distribution is isomorphic to a device with unary input) as these are encountered at the end of an NSDI cryptographic protocol. The two classical parts are in the hands of the honest parties, while eavesdropper holds some device. We then show that the NS norm takes for such cc-d distribution a closed-form expression. In particular, we prove that the supremum over Eve's operations reduces to a maximization over direct measurements (for the proof, see Sec. \ref{appen:feature-boxnorm} of the Appendix). 

 We present below the Theorem, which states that our definition of NS norm security criterion is equivalent to the criteria used by Renner, H{\"a}nggi, and Wolf \cite{hanggi-2009}. We do it in analogy to the results of Refs. \cite{Portmann-Renner, Beaudry} related to quantum device-dependent security, but for non-signaling devices:
\begin{theorem}[Equivalence of the NSDI security criteria]
    \label{thm:equivalence}
    For an MDLOPC protocol $\Lambda$, the proximity in the NS norm security criterion is equivalent to the criterion based on secrecy and correctness of the protocol. That is for any $\varepsilon_\mathrm{sec}+\varepsilon_\mathrm{cor} \equiv \varepsilon  \ge \varepsilon_\mathrm{sec},\varepsilon_\mathrm{cor} \ge 0$ the following relation holds:
    \ben \label{eqn:security0}
        &&\left(1-\mathrm{p}_\mathrm{abort}\right) \norm{ P_{S_A,S_B,Q,E|Z}^\mathrm{real|pass}-P_{S_A,S_B,Q,E|Z}^\mathrm{ideal|pass}}_\mathrm{NS} \le O(\varepsilon)\nonumber\\
        &&\Longleftrightarrow~ 
         \left\{
        \left(1-\mathrm{p}_\mathrm{abort}\right)\mathrm{P}\left[S_A \neq S_B|\mathrm{pass} \right] \le O(\varepsilon_{\mathrm{cor}}) \right.
        \\&&\left.  \wedge~ 
        \left(1-\mathrm{p}_\mathrm{abort}\right) \norm{ P_{S_A,Q,E|Z}^\mathrm{real|pass}-P_{S_A,Q,E|Z}^\mathrm{ideal|pass}}_\mathrm{NS}
        \le O(\varepsilon_{\mathrm{sec}}) \right\} \nonumber,
        \een
    where $\mathrm{p}_\mathrm{abort}$ is the probability for the protocol to abort and the constant $O(\varepsilon)$ does not depend on any parameter of the protocol. 
\end{theorem}
\noindent\textit{Proof}. For the proof, see Sec. \ref{sec:equivalence-security} of the Appendix.

Following arguments in Ref. \cite{Portmann-Renner}, {as a consequence of the above Theorem, we can claim that our definition of security is restricted composable  \cite{Ben-Or-Mayers-compos,Ben-OrHLMO05,Can01} provided the device is not reused.
In that sense, our definition diverges from that of \cite{hanggi-2009} formally in two ways. First, we use the notion of the complete extension. This encapsulates the access of the eavesdropper to all ensembles of the
device shared by the honest parties - the fact used in \cite{hanggi-2009}. Furthermore, in our approach, the memory of Eve is finite and minimal without compromising her eavesdropping power. Second, as we have mentioned, we modify the security criterion,
without losing the effect of composability. 
We use the proximity in NS norm
to the ideal classical-classical-device distribution. We show that it is equivalent to the statement that (as it was used in \cite{hanggi-2009}) the distinguisher can not tell apart the real cc-d distribution from the ideal one.

\section{Discussion and open problems}
In this manuscript, we have contributed in three ways to the topics of  cryptographic security and Bell nonlocality.  We describe them below along with possible directions to follow that naturally appears in consequence.

Firstly, we have initiated a systematic study on the upper bounds on the secret key rate on the NSDI scenario and defined a computable function, the squashed nonlocality as one of the bounds. We have also demonstrated a direct link between the Secrete Key Agreement scenario and that of NSDI by systematic construction of the bounds in the latter case from the secrecy monotones of the former. Interestingly this method leads among others to a known measure of nonlocality, which is the nonlocality  fraction. However, our approach goes much beyond that by offering construction of novel nonlocality measures, which confirms the generality of our paradigm.  Looking for tighter upper bounds stemming from (or even going beyond) the relationship between SKA and NSDI scenarios is a new direction to study.

The numerical estimate of the upper bound suggests
that only a limited amount of key can be obtained from quantum devices with two binary inputs and two binary outputs via direct measurement followed by local operations and public communication. For the family of devices studied here, it is below $40\%$. Given characterization from \cite{Geometry-of-small-boxes} of the boundary of the quantum set, one can find limitations on the key rate obtained via quantum mechanics against a non-signaling adversary for the set of $(2,2,2,2)$ devices.
It appears plausible that employing similar idea to the contextual set of observables may also lead to a novel measure of contextuality which upper bounds their private randomness content \cite{ran-context}. 

One of the most important problems which arise here is a dual one - whether the isotropic devices in $(2,2,2,2)$ scenario with less than $80\%$ weight of Popescu-Rohrlich box are key undistillable in general. We have shown that one can not distill them by MDLOPC operations, i.e., by direct measurements on device and LOPC operations. However, one might consider that grouping several of such devices together and distilling one of them via the so-called ``wirings'', could lead to a positive key if followed by MDLOPC operations. Although one can not exclude this case, it is rather improbable, because an action of wiring, within a group of wired devices, opens a possibility of the forward-signaling attack, as discovered in \cite{Rotem-Sha} and developed in \cite{Salwey-Wolf} (the two-way signaling case was excluded already in \cite{hanggi-2009b}). This is the reason why the non-signaling between individual devices seems necessary precondition of security in NSDI. 
 In any case, extending presented results to a more general class of operations e.g., including {\it wirings}, is an important open problem. As a step in this direction, one can consider how the key rate changes if the honest parties have access to randomness private from Eve. Such randomness could be in principle used for performing general measurements. 
We have also demonstrated applicability of our bound in the $(3,2,2,2)$ scenario, giving a tighter bound to the one provided in \cite{acin-2006-8}. A more careful study, which takes into account all the extremal points of the non-signaling polytope in the $(3,2,2,2)$ scenario could be a basis for further tighter bounds.

As the second of the main contributions, we have provided a method of constructing novel measures of nonlocality and proved several important properties for one of them - the squashed nonlocality. Among these properties are the monotonicity, convexity, and additivity.  
One property which is not considered here, the asymptotic continuity of the squashed nonlocality, will be presented in the forthcoming contribution \cite{future}.

Comparing it with the other measure - the relative entropy of nonlocality \cite{AxiomContext,Errata,QuantContext} may lead to interesting results and possibly the proof that the latter is also an upper bound on the distillable device-independent key.
Exploring further the analogy between squashed entanglement and squashed nonlocality may lead to novel analogous results in the realm of quantum devices. We also notice that the squashing procedure can be naturally extended to an arbitrary number of parties.  This can be achieved by following Ref. \cite{multi-sqent}, where the multipartite version of the intrinsic information in SKA has been shown to upper bound the conference key in the latter scenario.

As the third contribution, we have realized a novel idea of incorporating the eavesdropper in the scenario by applying the newly introduced concept of the complete extension \cite{CE}. Eve controls the additional interfaces of the extended part.
This provides the NSDI protocol a structural definition like the quantum purification did for QDD and QDI. 
Although the security condition derived from this approach is equivalent to the former, it shows a direct structural analogy between NSDI and QDD paradigms. In consequence, the complete extension models an adversary with minimal memory required for ultimate eavesdropping power. The amount of memory needed for a given attack in a non-signaling scenario to best our knowledge has not been studied so far and deserves attention in the future.
 To formalize security, we considered the NS norm analogous to the trace norm in quantum mechanics. We have proven that this approach is equivalent to the two former ones \cite{masanes-2006,masanes-2009-102,Masanes2011,hanggi-2009,hanggi-2009b,Renner-Hanggi,Hanggi-phd}. We obtained that our definition of security is  composably secure if the same device is not reused in composing the protocols (restricted composable). The properties of this NS norm computed for classical-classical-devices may become useful also in the context of Generalized Probabilistic Theory \cite{Hardy2001, Chiribella2010, Chiribella2011}. In this context, it is an important open problem if the class of operations $\mathcal{G}$, over which supremum is taken in the definition of the NS device norm,  is equal to the set of all linear operations $\mathcal{L}$ considered in \cite{ChristandlToner}.
 Finding  an answer to this 
problem may lead to the full operational characterization of the set of maps that can be performed on devices.

\begin{acknowledgments}
    MW, TD and KH acknowledge grant Sonata Bis 5 (grant number: 2015/18/E/ST2/00327) from the National Science Center. M.W thanks Eneet Kaur and Mark Wilde for the discussion during QIP2019. MW, TD and KH acknowledge partial support by the Foundation for Polish Science through IRAP project co-financed by EU within Smart Growth Operational Programme (Contract No. 2018/MAB/5).
    The authors acknowledge Ryszard Pawe{\l} Kostecki for useful comments.
\end{acknowledgments}

\bibliographystyle{unsrt}

\bibliography{references}



	

\onecolumngrid

	\section*{List of Symbols:}
	\vspace{2em}
	\begin{itemize}
		\item[]{\makebox[4cm]{$\mathbf{P(AB|XY)}$\hfill}: Bipartite non-signaling device.}
		\item[]{\makebox[4cm]{$\mathbf{P(ABE|XYZ)}$\hfill}: Tripartite non-signaling device.}
		\item[]{\makebox[4cm]{$\mathbf{P(ABE)}$\hfill}: Tripartite probability distribution.}
		\item[]{\makebox[4cm]{$\mathbf{|\psi\rangle_{ABE}}$\hfill}: A pure tripartite quantum state.}
		\item[]{\makebox[4cm]{$\mathbf{  S(A:B||E)  }$\hfill}:    Secure key rate in SKA model. }
		\item[]{\makebox[4cm]{$\mathbf{ I(A:B)   }$\hfill}:  Mutual information.  }
		\item[]{\makebox[4cm]{$\mathbf{ I(A:B|E)   }$\hfill}:  Conditional mutual information.  }
		\item[]{\makebox[4cm]{$\mathbf{ I(A:B\downarrow E)   }$\hfill}:  Intrinsic information.  }
		\item[]{\makebox[4cm]{$\mathbf{  I(A:B\downarrow \downarrow E)  }$\hfill}:   Reduced intrinsic information.  }
		\item[]{\makebox[4cm]{$\mathbf{ K_D(\rho_{AB})   }$\hfill}: Key rate in QDD scenario.   }
		\item[]{\makebox[4cm]{$\mathbf{  I_{sq}(\rho_{AB})  }$\hfill}: Quantum squashed entanglement.   }
		\item[]{\makebox[4cm]{$\mathbf{  K_{DI}  }$\hfill}:  Non-signaling Device independent key rate  }
		\item[]{\makebox[4cm]{$\mathbf{  {\cal N}_{sq}(P)  }$\hfill}:  Non-signaling squashed nonlocality  }
		\item[]{\makebox[4cm]{$\mathbf{  {\cal E}(P)  }$\hfill}: Complete extension of a device $P$.   }
		\item[]{\makebox[4cm]{$\mathbf{  {\cal M}  }$\hfill}:  Measurements, maps devices to distributions.  }
		\item[]{\makebox[4cm]{$\mathbf{ {\cal M}^F   }$\hfill}: Full direct measurements.   }
		\item[]{\makebox[4cm]{$\mathbf{  {\cal M}^G  }$\hfill}: General measurements.   }
		\item[]{\makebox[4cm]{$\mathbf{ \Lambda_N   }$\hfill}: MDLOPC protocol of key distribution acting on $N$ iid copies of a device.   }
		\item[]{\makebox[4cm]{$\mathbf{ \Lambda   }$\hfill}: The set of all MDLOPC protocol $\{\Lambda_N\}$.   }
		\item[]{\makebox[4cm]{$\mathbf{  P(AB|XY)^{\otimes N}  }$\hfill}:  Tensor product of $N$ iid copies of the device $P$.  }
		\item[]{\makebox[4cm]{$\mathbf{  {\cal E}\left( P^{\otimes N}\right)  }$\hfill}:  Complete extension of $N$ iid copies of the device $P$.  }
		\item[]{\makebox[4cm]{$\mathbf{  P^{d_N}_{ideal}  }$\hfill}: Ideal cc-d distribution of dimension $d_N$.   }
		\item[]{\makebox[4cm]{$\mathbf{  || P - Q ||_{NS}  }$\hfill}:  Non-signaling device norm of two devices $P$ and $Q$.  }
		\item[]{\makebox[4cm]{$\mathbf{  {\cal O}  }$\hfill}:  All possible linear operations which map a device to a distribution.  }
		\item[]{\makebox[4cm]{$\mathbf{  S_A   }$\hfill}:  The set of all possible key string in part of Alice after the MDLOPC operation.  }
		\item[]{\makebox[4cm]{$\mathbf{  S_B  }$\hfill}:  The set of all possible key string in part of Bob after the MDLOPC operation.   }
		\item[]{\makebox[4cm]{$\mathbf{  Q  }$\hfill}: Classical communication variable.   }
		\item[]{\makebox[4cm]{$\mathbf{  {\cal R}(\Lambda|_P)  }$\hfill}: NSDI key rate for a particular MDLOPC protocol.   }
		\item[]{\makebox[4cm]{$\mathbf{  M(A:B||E)  }$\hfill}:  Secrecy quantifiers of probability distribution $P(ABE)$  }
		\item[]{\makebox[4cm]{$\mathbf{  \widehat{M}(A:B||E)   }$\hfill}: Non-signaling squashed secrecy quantifiers of the device $P$.  }
		\item[]{\makebox[4cm]{$\mathbf{ PR   }$\hfill}: Popescu Rohrlich box   }
		\item[]{\makebox[4cm]{$\mathbf{ \overline{PR}   }$\hfill}: Complementary box to Popescu Rohrlich box.   }
		\item[]{\makebox[4cm]{$\mathbf{  P_{iso}  }$\hfill}:  Device lying on the isotropic line connecting $PR$ and $\bar{PR}$ box. }
		\item[]{\makebox[4cm]{$\mathbf{  \varepsilon  }$\hfill}: Error in the CHSH game.   }
		\item[]{\makebox[4cm]{$\mathbf{  P_E  }$\hfill}:  Extremal device  in the polytope of all non-signaling devices.}
		\item[]{\makebox[4cm]{$\mathbf{  \{p_i, P^i\}  }$\hfill}:  An ensemble of a device $P$.  }
		\item[]{\makebox[4cm]{$\mathbf{  \{p_i, P^i_E\}  }$\hfill}: Pure members ensemble of the device $P$.   }
		\item[]{\makebox[4cm]{$\mathbf{ D   }$\hfill}: A dice, source of additional randomness.   }
		\item[]{\makebox[4cm]{$\mathbf{ {\cal W}   }$\hfill}: Variable designate  wirings between two devices.  }
		\item[]{\makebox[4cm]{$\mathbf{ {\cal P}_N   }$\hfill}:  LOPC operations on $N$ copies of the distribution.  }
		\item[]{\makebox[4cm]{$\mathbf{ {\cal P}   }$\hfill}: Class of LOPC operations $\{{\cal P}_N\}_{N = 1}^\infty$, also a protocol for SKA model.   }
		\item[]{\makebox[4cm]{$\mathbf{ P_{B,A_1|X_1}   }$\hfill}: A classical-device distribution.   }
		\item[]{\makebox[4cm]{$\mathbf{ S_{ABE}   }$\hfill}:  Total state of the system after the MDLOPC protocol.  }
		\item[]{\makebox[4cm]{$\mathbf{  P^{real}_{S_A, S_B, Q, E|Z}  }$\hfill}: Classical-classical-device distribution  after the execution of a real protocol.  }
		\item[]{\makebox[4cm]{$\mathbf{  p^{abort}  }$\hfill}: Probability of aborting the protocol.   }
		\item[]{\makebox[4cm]{$\mathbf{  P^{real|abort}_{S_A, S_B, Q, E|Z}  }$\hfill}: Classical-classical-device distribution  after the execution of a real protocol}
		\vspace{-1em}       
		\item[]{\makebox[4cm]{$\mathbf{    }$\hfill}~ conditioning of aborting.   }
		\item[]{\makebox[4cm]{$\mathbf{  P^{real|pass}_{S_A, S_B, Q, E|Z}  }$\hfill}: Classical-classical-device distribution  after the execution of a real protocol} 
		\vspace{-1em}    
		\item[]{\makebox[4cm]{$\mathbf{    }$\hfill}~   conditioning of not aborting. }  
		\item[]{\makebox[4cm]{$\mathbf{  P^{ideal|pass}_{S_A, S_B, Q, E|Z}  }$\hfill}: Classical-classical-device distribution  after the execution of an ideal protocol}
		\vspace{-1em}    
		\item[]{\makebox[4cm]{$\mathbf{    }$\hfill}~ conditioning of not aborting. }  
		\item[]{\makebox[4cm]{$\mathbf{  S_{AE}  }$\hfill}:  State of the system after the protocol in part of Alice and Eve.  }
		\item[]{\makebox[4cm]{$\mathbf{  {\cal D}(P,Q)  }$\hfill}: Distance of two devices $P$ and $Q$.   }
		\item[]{\makebox[4cm]{$\mathbf{ P[S_A \neq S_B]   }$\hfill}: Probability of not having the same key strings between Alice and Bob.   }
		\item[]{\makebox[4cm]{$\mathbf{  dim_A({\cal P}_N\left( (P(ABE))^N  \right)}$\hfill}: Dimension of part  $A$ after the LOPC operation on the $N$ copies of the probability}
		\vspace{-0.8em}    
		\item[]{\makebox[4cm]{$\mathbf{    }$\hfill}~ distribution.  }
		\item[]{\makebox[4cm]{$\mathbf{ C_i   }$\hfill}:  Message sent from Alice to Bob as part of SKA protocol or vice versa.  }
		\item[]{\makebox[4cm]{$\mathbf{ C^t   }$\hfill}:  Collection of all messages $C^t = C_1C_2\ldots C_t$ sent between Alice and Bob in the $t$th }
		\vspace{-1em}    
		\item[]{\makebox[4cm]{$\mathbf{    }$\hfill}~ step. }
		\item[]{\makebox[4cm]{$\mathbf{  I(S:C^tE^N)   }$\hfill}:   Mutual information between the final key string and Eve's information. }
		\item[]{\makebox[4cm]{$\mathbf{  H(S)  }$\hfill}: Entropy of the final key $S$.   }
		\item[]{\makebox[4cm]{$\mathbf{  \Lambda^\eta_N  }$\hfill}: $\eta$  optimal MDLOPC protocol on $N$ iid copies of the device.  }
		\item[]{\makebox[4cm]{$\mathbf{  {\cal P}^\eta_N  }$\hfill}:  $\eta$  optimal LOPC protocol on $N$ iid copies of the distribution.  }
		\item[]{\makebox[4cm]{$\mathbf{  \tilde{\cal E}(P)  }$\hfill}:  Overcomplete extension of the device $P$.  }
		\item[]{\makebox[4cm]{$\mathbf{  {}^{x,y} {\cal P}^\eta_N  }$\hfill}:  Measurement dependent $\eta$  optimal LOPC protocol on $N$ iid copies of the }
		\item[]{\makebox[4cm]{$\mathbf{    }$\hfill} ~ distribution.  }
		\item[]{\makebox[4cm]{$\mathbf{  \Omega_{\mathrm{GMDLOPC}}  }$\hfill}:  LOPC operations involve general measurements on the devices.  }
		\item[]{\makebox[4cm]{$\mathbf{  \Lambda_{\mathrm{MDLOPC}}  }$\hfill}:  LOPC operations involve direct measurements on the devices.  }
		\item[]{\makebox[4cm]{$\mathbf{  || P - Q||_{NS}^{res}  }$\hfill}:  Restricted NS norm of two devices.  }
		\item[]{\makebox[4cm]{$\mathbf{  {\cal N}_C  }$\hfill}:  Nonlocality cost of a non-signaling device. }
		\item[]{\makebox[4cm]{$\mathbf{ C(P)   }$\hfill}:   Nonlocality fraction of a non-signaling device $P$. }
	\end{itemize}

	\newpage
	
	\MyIncludeGraphics{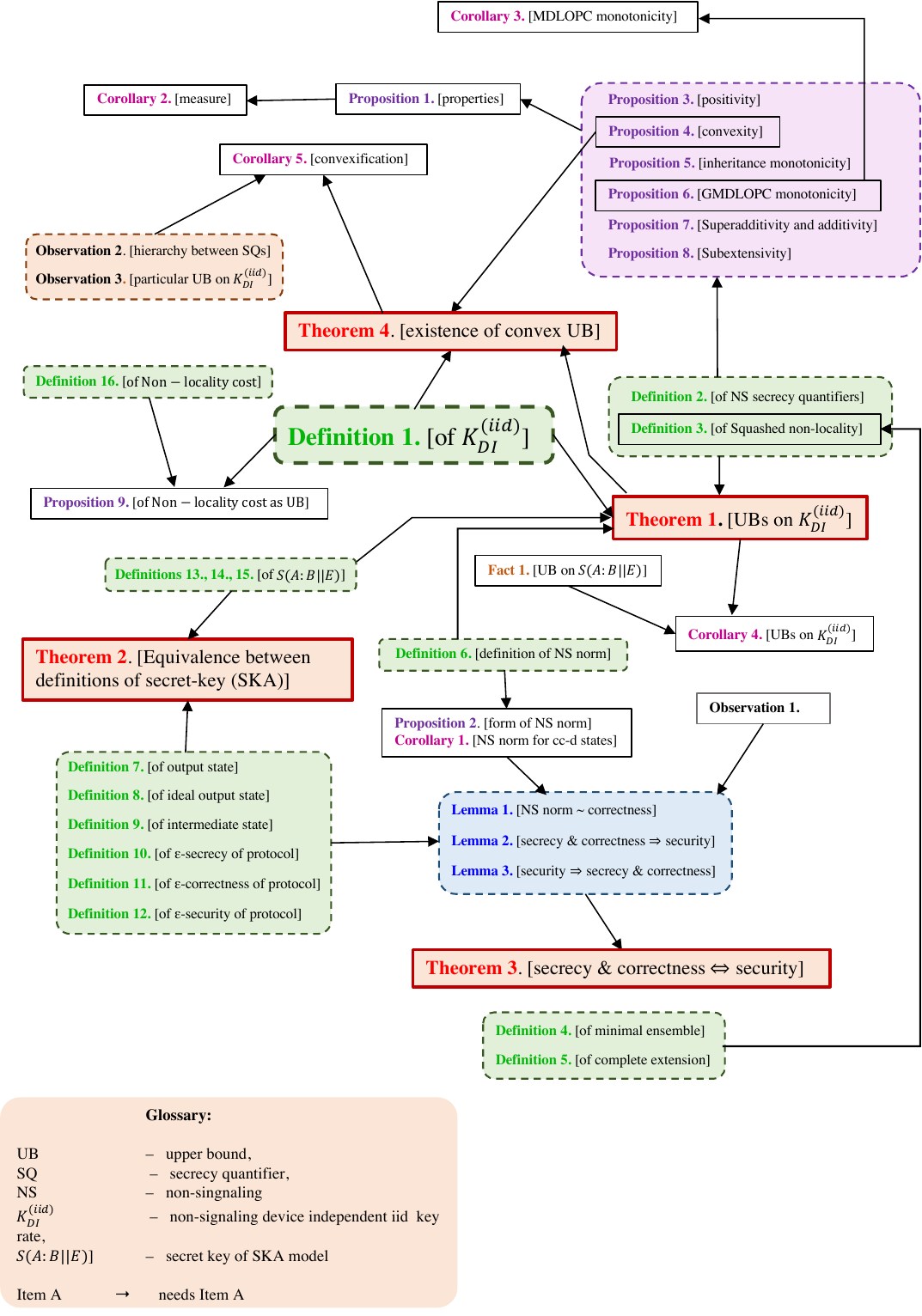}
	
	\newpage
	
	\begin{center}
		{\Large Appendix}
	\end{center}

	\textbf{Notation:} \textit{In the Appendix we adapt two different notations for conditional probability distributions (devices). We do this to avoid small fonts in multi-level mathematical expressions that appear in forthcoming parts of this work and hence to make them more readable.}

	\section{Definitions of Entropic functions}
	\label{entropies}
	{In this section, we recall definitions of basic quantities associated with random variables.} Suppose $A, B$ and $E$ are discrete random variables, with outcomes $a \in A,$ $b \in B$ and $e \in E$.  Let {$P(ABE)$} be the joint probability distribution {of random variables A, B, E}. {Similarly, let} {$P(A=a,B=b,E=e) \equiv p(abe)$} be  {the probability for obtaining the outcome $A=a, B=b $ and $E=e$. }
	\begin{itemize}
		\item The \textbf{Shannon entropy} of a random variable {(variables)} is defined as
		\ben
		H(A) &=& - \sum_a p(a) \log_2 p(a), \\
		H(AB) &=& - \sum_{ab} p(ab) \log_2 p(ab), \\
		H(ABE) &=& - \sum_{abe} p(abe) \log_2 p(abe),
		\een
		where, $p(ab) = \sum_e p(abe)$ and $p(a) = \sum_{b} p(ab)$ are the marginal {probabilities} of the joint probability distribution {$P(ABE)$}.
		\item The \textbf{conditional Shannon entropy} of any random variable  $A$ with respect to the random variable $B$, quantifying the {lack of knowledge about the outcome of} $A$ when one already knows {the value of} $B$, is given by
		\ben
		H(A|B) = \sum_b p(b) H(A|B = b) = H(AB) - H(B).
		\een
		\item \textbf{ {The m}utual information} $I(A:B)$, measuring the correlations between $A$ and $B$, is defined as 
		\ben
		I(A:B) = H(A) + H(B) - H(AB).
		\een
		\item The \textbf{conditional mutual information} $I(A:B|E)$, quantifying the correlation remaining between variables $A$ and $B$ conditioned upon the knowledge about value of third variable $E$,  is given by
		\ben
		I(A:B|E) &=& \sum_e I(A:B|E = e)\\& =& H(A|E) + H(B|E) - H(AB|E).
		\een

		\item The \textbf{intrinsic mutual information} \cite{Intrinsic-Maurer,MauWol97c-intr} $I(A:B \downarrow E)$ is 
		\ben
		I(A:B \downarrow E) = \inf_{\Theta_{E'|E}} I(A:B|E'),
		\een
		where $I(A:B|E')$ is the conditional mutual information of the probability distribution $P(ABE') =  \sum_e \Theta_{E'|E}(E'|E=e)P(AB,E=e)$, while the infimum is taken over all possible conditional channels $\Theta_{E'|E}$.
		\item The \textbf{reduced intrinsic information} \cite{reduced-intrinsic,lit3} of random variables $A$, $B$ and $E$, denoted by $I(A:B\downarrow \downarrow E)$ is defined as 
		\ben\label{eq:reduintrinsic}
		I(A:B\downarrow \downarrow E) = \inf_{\Theta_{U|ABE}} \left( I(A:B\downarrow EU) + H(U)\right),
		\een
		where the infimum is taken over all possible conditional channels $\Theta_{U|ABE}$.
	\end{itemize}

	\section{The world of non-signaling devices and the NSDI cryptographic scenario}\label{sec:notation}
	
	In the NSDI cryptographic scenario, we {consider}  that  the honest parties, Alice and Bob, share a cryptographic device of unknown internal structure, identified with a non-signaling conditional probability distribution $P(AB|XY)$ (we use also $P_{AB|XY}$ notation). We refer to $P(AB|XY)$, as to a non-signaling device throughout our paper.
	Here $A$, $B$, $X$, and $Y$ are random variables and $a \in A$, $b \in B$, $x \in X$, and $y \in Y$ are respectively their values. The indices $x$ and $y$ are considered to be choices of inputs {of} the honest parties, whereas the respective outcomes are denoted by $a$ and $b$. The non-signaling condition {for} $P_{AB|XY}(ab|xy)$, {that roughly speaking forbids faster than light communication between the two parties}, is defined as 
	\ben
	&& P_{A|X}(a|x) = \sum_b P_{AB|XY}(ab|xy) = \sum_b P_{AB|XY}(ab|xy') ~\forall ~a,x, y, y', \label{eq:nons:A}\\
	&& P_{B|Y}(b|y) = \sum_a P_{AB|XY}(ab|xy) = \sum_a P_{AB|XY}(ab|x'y) ~\forall ~b,x, x', y. \label{eq:nons:B}
	\een
	We incorporate the {no-signaling} eavesdropper (Eve) in the system by giving her the access to the additional interfaces of the {\it complete extension} (CE) \cite{CE}, of the shared {tripartite non-signaling} device {(see next Subsection \ref{subsec:CE} for reference on CE)}.
	{We denote the complete extension of a bipartite device $P(AB|XY)$ as ${\cal E}(P)(ABE|XYZ)$, where the additional input $z \in Z$ and the corresponding output $e \in E$, are controlled by Eve. Extending a bipartite device with CE ensures that the non-signaling constraints also hold  between Eve and Alice's and Bob's joint subsystem.}
	Additionally, Eve can also apply local randomness in both her input and output  to generate general measurements and to post-processing the output, which gives her the ultimate operational eavesdropping power, as {then by construction of CE}, she can access all possible ensembles of the extended device \cite{CE}.
	
	\subsection{The notion of the complete extension}\label{subsec:CE}
	\label{sec:0}
	For an arbitrary device $P(A|X)$, one can always find its extension $P(AE|XZ)$ in {the space of a larger dimension}, such that {the non-signaling constraints are satisfied (see equations (\ref{eq:nons:A}), (\ref{eq:nons:B}))}. 
	{Some extensions of bipartite non-signaling boxes have been studied in the past} \cite{acin-2006-8, AcinGM-bellqkd, hanggi-2009, lit13}. The complete extension defined in \cite{CE}, is {an extension of the lowest possible dimension}, that possesses all basic properties of quantum purification except extremality.
	
	Let us consider a polytope {(state space)} of non-signaling devices, with a fixed number of parties and fixed cardinalities of inputs and outputs. An arbitrary device $P$, in that polytope, can always be expanded as a convex combination of the extremal (pure) devices $\{P_E^i\}$, as $P = \sum_{i} p_i P_E^i$.
	The ensemble $\{p_i, P_E^i\}$ will be called a pure members ensemble (PME). {The} decomposition $\{p_i\}$ is not unique {in general.} \cite{CE}.
	
	\begin{definition}[Minimal ensemble]
		A pure members ensemble,  $\{p_i, P_E^i\}_{i \in \mathcal{I}}$ will be called a {\it minimal ensemble} of  $P$, if all the members are {\it pure} and if any proper subset of  $\{P^i_E\}_{i \in \mathcal{I}}$ for any new choices of the corresponding probabilities $\{p'_i\}_{i \in \mathcal{I}}$ is not an ensemble of the device P.
		\label{def:minimal-ensemble}
	\end{definition}
	
	We can now invoke the definition of a complete extension. Qualitatively, it is such an extension of a device, which enables to produce {all} minimal ensembles of it, {with the choice of input in the extending part resolving which minimal ensemble will be generated}.
	The complete extension is, by {its} definition a non-signaling extension, which makes it {a perfect tool for} the NSDI cryptography (see \cite{hanggi-2009} in this context).  
	
	\begin{definition}[Complete extension \cite{CE}] 
		Given a device $P_{\cal A}(A|X)$, we say that a device ${\cal E}(P)_{\cal AX}(AE|XZ)$ is 
		its complete extension to system $\cal X$ if for any $z \in Z$ and $e \in E$ there holds
		\ben
		{\cal E}(P)_{\cal AX}(A,E =e|X,Z=z) = 
		p(e|z) P^{e,z}_{\cal A}(A|X),
		\een
		such that the ensemble $\{p(e|z), P^{e,z}_{\cal A}(A|X)\}$ is a minimal ensemble of the device $P_{\cal A}(A|X)$, and corresponding to each minimal ensemble of $P_{\cal A}(A|X)$, there is exactly one $z \in Z$ which generates it\footnote{The calligraphic $\mathcal{X}$ stands here for the extending system, and should not be confused with the input of the system $\mathcal{A}$. }. 
		\label{def:complete-extension}
	\end{definition}
	
	Here we slightly abuse the notation, so by $P_{\cal A}(A|X)$, we mean the device $P(A|X)$ with random variables $A$ and $X$. The subscript ${\cal A}$ denotes that the device is in possession of party ${\cal A}$. Similarly, the subscript ${\cal X}$, for the complete extension ${\cal E}(P)_{\cal AX}(AE|XZ)$,   stands for the extending party ${\cal X}$, who controls the additional interfaces $Z$ and $E$.
	
	\pagebreak
	
	The complete extension satisfies the following properties alike the quantum purification, what makes CE its counterpart \cite{CE}.
	\begin{enumerate}
		\item ACCESS: A complete extension of a device $P$, together with access to arbitrary randomness, gives access to any ensemble of a device $P$.
		\item GENERATION: The complete extension can be transformed to any other extension.
	\end{enumerate}

	\subsection{Possible eavesdropping actions }\label{sec:Evesdrpping-action}
	
	In this section, we define the building blocks of the set of allowed operations that the non-signaling eavesdropper can perform.
	In every device-independent key distribution protocol, the honest parties hold a  device, the internal structure of which is completely unknown to them.
	Their task is to share {at the end of the protocol} a cryptographically secure key, which is perfectly correlated between the honest parties and completely secret with respect to the eavesdropper \cite{Beaudry}, {by use of several copies of the device $P(AB|XY)$. As we are interested in finding the upper bound on the key rate, we consider the attacks by the eavesdropper as an independent and identically distributed (iid) attack} {as a choice of particular eavesdropping strategy}. In this attack, the eavesdropper prepares $N$ iid devices  $\big(P(AB|XY)\big)^{\otimes N} \equiv P^{\otimes N}(\bm{AB}|\bm{XY})$ for Alice and Bob and holds the {extending} part of the CE ${\cal E}( P^{\otimes N})(\bm{AB}E|\bm{XY}Z)$, where $\bm{A} = A_1A_2 \cdots A_N$, and similarly for $\bm{B}$, $\bm{X}$, and $\bm{Y}$. 
	{At this point we are ready to describe the possible actions of Eve on input and output of the extending system }
	\begin{enumerate}
		\item Full direct measurement, $\{{\cal M}^F_z\}$ defined by choice of input $Z = z$. The inputs correspond to the choices of different minimal ensembles. In a cryptographic sense, some inputs are in favour of Eve, and some are not.
		\item General  measurement, $\{{\cal M}^G_z\}$, defined by a probabilistic choice of direct measurements ${\cal M}^G_{z} = \sum_{z'} p(z'|z){\cal M}_{z'}^F $. Upon each choice of general measurement on the CE of the shared device, Eve can generate any pure members ensemble of the device. Here $\{p(z'|z)\}$ represents  the dice, an external randomness.
		\item Classical post-processing channel $\Theta_{E'|E}$ on the output of the {extending subsystem that can also be  conditioned upon values of inputs and outputs of the dice}. {These operations when considered together with a general measurement} gives  access to all ensembles (possibly mixed) {of the} part of the device shared by the honest parties.
		\item Eve can also {monitor the communication}, i.e., collect the classical information exchanged between the honest parties. 
	\end{enumerate}
	
	The most general strategy of the eavesdropper is to utilize both the {general measurement} and the post-processing channel. Any other strategy is a specific case of the general one { described above}. For example, the full direct measurement can be considered as a combination of deterministic dice and an identity  post-processing channel. 

	\subsection{Cryptographic protocol}
	In this section, we {describe} the building blocks of the set of operations that the honest parties can perform  to generate a cryptographically secure key. In {the} case of non-signaling device-independent protocol, the honest parties can perform the following operations {on} their shared devices: 
	\begin{enumerate}
		\item Full direct measurements on the input, i.e., {setting certain values $x$, $y$ of their inputs $X$, $Y$,}  followed by any composition of operations 2 and 3 below:
		\item Classical post-processing of the distribution 
		\item Public communication.
	\end{enumerate}
	
	We call this class of operations as  {\it Measurement on Devices followed by Local Operations and Public Communications} ({\bf MDLOPC}) \cite{masanes-2009-102}. {Here we do not allow the honest parties to perform wirings}  between their subsystems {because} the forward signaling between the subsystems has been proved to be an insecure procedure for many important examples of post-processing \cite{Rotem-Sha, Salwey-Wolf}. Limitation from a general measurement to a direct one is because,  {in the former case,} Eve does not have access to correlation with the whole system of Alice and Bob. 
	
	In our cryptographic protocol, we prove the security when the Eve's attacking strategy is to prepare $N$
	iid copies of a non-signaling device $P(AB|XY)$ and hands them over to the honest parties. Eve controls the CE of the full system, i.e., $P^{\otimes N}(AB|XY)$.
	It is important to note that  CE of a tensor product of devices is not a tensor product of CE's of these  devices. This is the most general eavesdropping strategy (in the iid case) since it gives Eve access to all possible statistical ensembles of the shared device. {Incorporating CE} in this NSDI scenario  encompasses a structural way to access to all ensembles of the extended device, which is the key point in all NSDI security protocol \cite{Kent,AcinGM-bellqkd, acin-2006-8, Scarani2006, masanes-2009-102, hanggi-2009, Kent-Colbeck, lit13}.

	\section{Properties of the NS norm}\label{appen:feature-boxnorm}
	
	The NS norm introduced in 
	Eq. (\ref{box-norm}) that has its main application in Proposition \ref{cor:box_norm_exp} strongly relies on the notion of the so-called {\it distinguishing system} 
	\cite{hanggi-2009, hanggi-2009b, Hanggi-phd}. The {\it distinguishing system}, also dubbed as the distinguisher, is an external black box type device having the same interfaces as the original device {(with one extra output)} however, its inputs are interchanged into outputs and vice versa. The structure of the distinguishing system allows it then to be connected to the interfaces of the original device. For each pair of systems to be distinguished, the distinguisher is devised in such a way that it attains maximal guessing advantage to distinguish between two examined devices. {The extra output is used to communicate the guess.} For a far more detailed description of the distinguishing system, we refer the reader to \cite{Hanggi-phd}.
	
	{
		{In this section, we show that in} the heuristic approach, the NS norm is a maximal guessing advantage for a distinguisher to distinguish between two devices and plays a role of a distance $\mathcal{D}$ between two conditional probability distributions \cite{Portmann-Renner, Hanggi-phd}. Devices with unary inputs are isomorphic to probability distributions. For them, the NS norm, is by definition, proportional to the total variational distance. 
		\ben\label{eq:distinguisher-boxnorm}
		\left|\left| P-Q \right|\right|_\mathrm{NS}=\mathrm{\cal D}(P,Q),
		\een
	}
	
	For the sake of cohesion we introduce the NS norm formally:
	\begin{definition}[Of the NS norm]
		Let $P$ and $P'$ be any two non-signaling devices. The following distance measure between $P$ and $P'$ is called the NS norm.
		\be
		\left| \left|P-P'\right| \right|_\mathrm{NS}:= \sup_{ g\in\mathcal{G}} \frac 12  \left| \left|g(P)- g(P')\right| \right|_1,
		\ee 
		where $||.||_1$ is a variational distance between two distributions. Furthermore $\mathcal{G}$, is a set of generating operations that consists of:
		\begin{enumerate}[(i)]
			\item adding an auxiliary device that has single input and single output (a {dice}),
			\item connecting the output of a device/dice to the input of a dice/device respectively, called {wirings},
			\item pre-processing the inputs of device(s),
			\item post-processing inputs and outputs of the devices.
		\end{enumerate}
	\end{definition}

	The results of this section, although seem to be {highly technical}, have a direct implication in distinguishability of the states of devices at the end of the protocol. For an initial tripartite device $P(ABE|XYZ)$, when the honest parties finish the MDLOPC protocol on it, i.e., perform measurements in their respective parts and post-process their data by local operations and public communication, the device is transformed into a {\it classical-classical-device probability distribution} (cc-d state). In fact, it is enough to consider classical-device states (c-d states) $P_{B,A_1|X_1}$, and the result still holds for any {\bf c}-d states, i.e., consisting of many classical subsystems (see Fig. \ref{fig:wiring}). This is because one can always claim that classical variable $B$ is the Cartesian product of many classical variables. 
	
	{
		We identify the operations {$g \in \mathcal{G}$} that the distinguisher can perform to discriminate between the devices. These can  always be decomposed into several basic operations belonging to disjoint sub-classes of different operational meaning, i.e., $g=\mathcal{P} \circ \mathcal{M}^G \circ \mathcal{W}$ considered together with external randomness $D$. This decomposition guarantees adequate causal order of operations.

		\begin{enumerate}[i)]
			\item The distinguisher can make use of {\bf external randomness}, which in general may depend on the output of the classical part of the system $B$. We incorporate this randomness by combining systems to be distinguished with an external system, $D_{A_2|X_2,B}$ called a dice.
			\item A composition of \emph{\textbf{wirings and prior to input classical communication}} (WIPCC), we denote this operation with $\cal W$. In general, wirings can be adaptive to the outcome of classical variable $B$, and can be constructed in different manners.
			\begin{enumerate}[a)]
				\item{$\mathcal{W}^\rightarrow$: deterministic wirings from \textbf{c}-d system to dice.}
				\item{$\mathcal{W}^\leftarrow$:  deterministic wirings from a dice into the input of {the} \textbf{c}-d system.}
				\item{A mixture of the above.}
			\end{enumerate}
			\item{Direct or general measurements
				\begin{enumerate}[a)]
					\item{\textbf{Full direct measurement $\bm{{\cal M}_x^F}$ :} A full direct measurement acting on a device $P (A|X) \equiv P_{A|X}$, is equivalent to choosing an input $x \in X$,  resulting with a conditional probability distribution, 
						\be
						\mathcal{M}_x^F(P (A|X) ) = P(A|X=x).
						\ee
						Different $x$ correspond to different measurements (inputs).}
					\item{ \textbf{General measurement $\bm{{\cal M}_x^G}$ :} A general measurement is a mixture of direct measurements, ${\cal M}_{x'}^G = \sum_x p(x|x') {\cal M}_x^F$, and its action is described as 
						\be
						\mathcal{M}_{x'}^G(P (A|X) ) = \sum_x p(x|x')  {\cal M}_x^F(P (A|X) ) = \sum_x p(x|x')  P(A|X=x),
						\ee
						with a conditional probability distribution $p(x|x')$  satisfying $\sum_x p(x|x') = 1$ $\forall x'$. Here different $x'$ indicate different choices of a general measurement. }
				\end{enumerate}
			}
			\item {\bf Classical data post-processing} we denote with $\mathcal{P}$.
		\end{enumerate}

		\begin{figure}[t]
			\begin{center}
				\includegraphics[width=17cm]{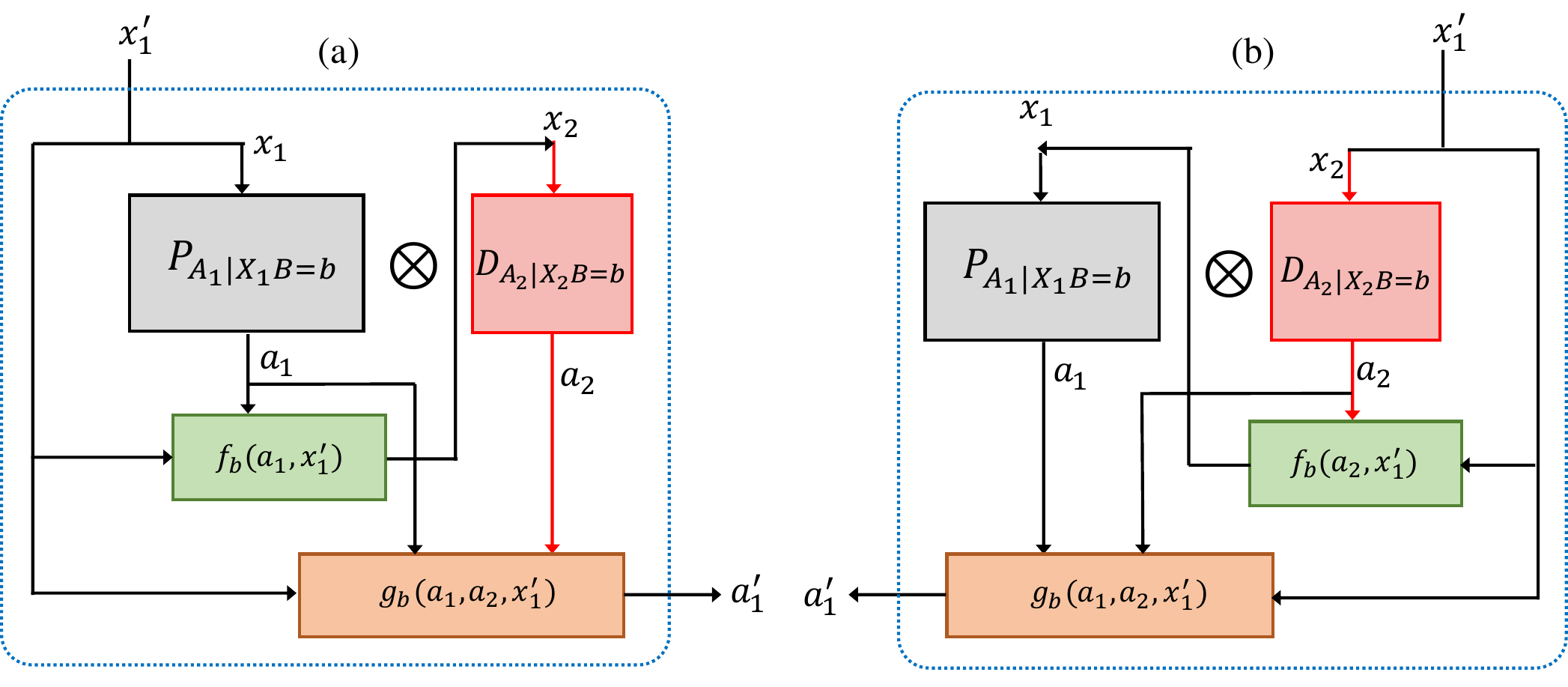}
			\end{center}
			\caption{Schematic diagram of deterministic wiring between the cc-d distribution $P_{A_1|X_1B}$ and an arbitrary  external device (called a dice) $D_{A_2|X_2}$. Fig. (a) represents the wiring from the cc-d distribution to the external device, $\mathcal{W}^\rightarrow$, and Fig. (b) represents the converse one, i.e., wiring from an external device to the cc-d distribution, $\mathcal{W}^\leftarrow$. The diagram is motivated by \cite{Tuziemski}.}
			\label{fig:wiring}
		\end{figure}
		
		{
			In the proof of the following Proposition, we consider supremum over external systems $D_{A_2|X_2 B}$. Hence without loss of generality, we can consider only wirings employing deterministic functions. The notation for wirings is adapted from \cite{Tuziemski}, as depicted in Fig. \ref{fig:wiring} above. The domains and codomains of functions $f_b$ and $g_b$, which determine wirings, are always adapted to the sizes of inputs and outputs. We consider deterministic wiring, so the sets of $\{f_b\}$ and $\{g_b\}$ are always finite. For the sake of simplicity, in the proof, we omit a unary input in the places where it does not lead to any ambiguity.
		}
		
		\begin{prop2}\label{prop:ccb} For the \textbf{c}-d states (alike those shared at the end of the MDLOPC-protocol $\Lambda_N$), i.e., the many parties non-signaling device for which only a single party has not unary input, the NS norm takes the form
			\ben
			\left| \left| P^1_{B,A_1|X_1}-P^2_{B,A_1|X_1}\right| \right|_\mathrm{NS} =
			\frac 12 \sum_{b} \sup_{  \mathcal{M}^F_{x_1}} \sum_{a}
			\left|    \mathcal{M}^F_{x_1} \left (P^1_{B,A_1|X_1}  \right)\left(b,a\right) - \mathcal{M}^F_{x_1} \left (P^2_{B,A_1|X_1}  \right)\left(b,a\right)\right|,
			\een
			where $b \in B$ is a multi-variable corresponding to outputs of \textbf{c} part of the \textbf{c}-d distribution. 
		\end{prop2}
		
		From now on, for the sake of the ease of notation we make the following identification: $\mathcal{M}^F \equiv \mathcal{M}^F_{x}$ and $\mathcal{M}^G \equiv \mathcal{M}^G_{x}$, where $x$ should be understood from the context. Note that wherever fiducial measurements are considered the $\sup$ operator can be used here interchangeably with $\max$ operator, as they act in the set with a finite number of elements.

		\begin{proof}
			{
				To attain the supremum over all operations given in Eq. (\ref{box-norm}), we have to consider all possible actions of the distinguisher. For the proof, it is sufficient to consider the single most general operation instead of a mixture. This is because a norm defined with supremum of some distance is a convex function and attains maximum at the boundaries of the set over which the supremum is evaluated. 
				\begin{align}
				&\sup_{g \in \mathcal{G}}  \norm{g(P)-g(Q)}_1 =\sup_{ \{\lambda_i\}} \sup_{ \{\tilde{g}_i\} \subseteq \tilde{\mathcal{G}}}  \norm{\sum_i \lambda_i \tilde{g}_i(P)-\sum_i \lambda_i \tilde{g}_i(Q)}_1\le \sup_{ \{\lambda_i\}} \sup_{ \{\tilde{g}_i\} \subseteq \tilde{\mathcal{G}}} \sum_i \lambda_i  \norm{\tilde{g}_i(P)- \tilde{g}_i(Q)}_1
				\\
				&\le \sup_{ \{\lambda_i\}} \sum_i \lambda_i   \sup_{ \tilde{g} \in \tilde{\mathcal{G}}} \norm{\tilde{g}(P)- \tilde{g}(Q)}_1 = \sup_{ \tilde{g} \in \tilde{\mathcal{G}}} \norm{\tilde{g}(P)- \tilde{g}(Q)}_1,
				\end{align}
				where ${ \tilde{g} \in \tilde{\mathcal{G}}}$ are pure operations, i.e., they are not a mixture of others.\\
			}
			%
			%
			{Following the arguments of the previous paragraphs the} NS norm can be phrased as
			\ben
			&&\left| \left|P^1_{B,A_1|X_1}-P^2_{B,A_1|X_1}\right| \right|_\mathrm{NS} =   \sup_{g\in \mathcal{G}} \frac 12 \left| \left|g(P^1_{B,A_1|X_1}) - g(P^2_{B,A_1|X_1})\right| \right|_1 
			\\ &&=
			\sup_D  \sup_\mathcal{W} \sup_{\mathcal{M}^G}  \sup_\mathcal{P} \frac12 \left| \left| \left(\mathcal{P} \circ \mathcal{M}^G \circ \mathcal{W}\right) (P^1_{B,A_1|X_1}\otimes D_{A_2|X_2,B})-\left(\mathcal{P} \circ \mathcal{M}^G \circ \mathcal{W}\right) (P^2_{B,A_1|X_1}\otimes D_{A_2|X_2,B})\right| \right|_1,
			\een
			where the suprema are taken over operations being adaptive with respect to the output $B$. When acting on the systems with a fixed value of classical output $B$, with a little abuse of notation, this can be rephrased using the same symbols for non-adaptive operations.
			\begin{align}
			&\left| \left|P^1_{B,A_1|X_1}-P^2_{B,A_1|X_1}\right| \right|_\mathrm{NS} \nonumber \\ =& \frac 12 \sum_b \sup_D  \sup_\mathcal{W} \sup_{\mathcal{M}^G}  \sup_\mathcal{P} \left| \left| \left(\mathcal{P} \circ \mathcal{M}^G \circ \mathcal{W}\right) (P^1_{B=b,A_1|X_1}\otimes D_{A_2|X_2,B=b})-\left(\mathcal{P} \circ \mathcal{M}^G \circ \mathcal{W}\right) (P^2_{B=b,A_1|X_1}\otimes D_{A_2|X_2,B=b})\right| \right|_1. ~~~~~~
			\end{align}
			The first step to simplify the expression above is to notice that $\left| \left|\cdot - \cdot \right|\right|_1$ is contractive under classical post-processing on probability distributions. Since the trivial post-processing is always accessible, we obtain 
			\begin{align}
			&\left| \left|P^1_{B,A_1|X_1}-P^2_{B,A_1|X_1}\right| \right|_\mathrm{NS} \nonumber \\ =& \frac 12 \sum_b \sup_D  \sup_\mathcal{W} \sup_{\mathcal{M}^G}   \left| \left| \left( \mathcal{M}^G \circ \mathcal{W}\right) (P^1_{B=b,A_1|X_1}\otimes D_{A_2|X_2,B=b})-\left( \mathcal{M}^G \circ \mathcal{W}\right) (P^2_{B=b,A_1|X_1}\otimes D_{A_2|X_2,B=b})\right| \right|_1.
			\end{align}
			As it was stated informally above, the general wiring, $\mathcal{W}$, can be constructed adaptively upon the knowledge of the values of the output $B$, as a probabilistic combination of two types of wirings $\mathrm{conv}\{\mathcal{W}^\rightarrow, \mathcal{W}^\leftarrow\}$ (see Fig. \ref{fig:wiring}). In the following lines, we show that the strategy of mixing is not optimal. However, in general, the cardinalities of inputs and outputs in different (types) of wiring can be different. In order to overcome this obstacle, we consider a common supremum over a convex set of wirings composed with measurements. From an operational point of view, this procedure means that the knowledge about the preparation was discarded after the optimal measurement for each type of wiring had already been chosen.  
			\begin{align}
			&\left| \left|P^1_{B,A_1|X_1}-P^2_{B,A_1|X_1}\right| \right|_\mathrm{NS} \nonumber \\ & = \frac 12 \sum_b \sup_D  \sup_{\{p_b^\leftarrow,p_b^\rightarrow\}}\sup_{\mathcal{M}^G\circ \mathcal{W^\leftarrow}} \sup_{\mathcal{M}^G \circ \mathcal{W^\rightarrow}}   \left| \left| \left(p_b^\leftarrow\left( \mathcal{M}^G \circ \mathcal{W^\leftarrow}\right) (P^1_{B=b,A_1|X_1}\otimes D_{A_2|X_2,B=b})\right.\right.\right.
			\nonumber	\\&  \left.\left.\left.+p_b^\rightarrow\left( \mathcal{M}^G \circ \mathcal{W^\rightarrow}\right) (P^1_{B=b,A_1|X_1}\otimes D_{A_2|X_2,B=b})\right)
			\right.\right. \nonumber \\&  - \left.\left.
			\left(p_b^\leftarrow\left( \mathcal{M}^G \circ \mathcal{W^\leftarrow}\right) (P^2_{B=b,A_1|X_1}\otimes D_{A_2|X_2,B=b})+p_b^\rightarrow\left( \mathcal{M}^G \circ \mathcal{W^\rightarrow}\right) (P^2_{B=b,A_1|X_1}\otimes D_{A_2|X_2,B=b})\right)\right| \right|_1 
			\\&\le
			\frac 12 \sum_b \sup_D  \sup_{\{p_b^\leftarrow,p_b^\rightarrow\}} \left( \sup_{ \mathcal{M}^G \circ \mathcal{W^\leftarrow}}  p_b^\leftarrow \left| \left| \left( \mathcal{M}^G \circ \mathcal{W^\leftarrow}\right) (P^1_{B=b,A_1|X_1}\otimes D_{A_2|X_2,B=b}) \right.\right. \right.
			\nonumber \\&   \left.\left.\left. -\left( \mathcal{M}^G \circ \mathcal{W^\leftarrow}\right) (P^2_{B=b,A_1|X_1}\otimes D_{A_2|X_2,B=b}) 
			\right|\right|_1\right. \nonumber \\&  \left.+
			\sup_{  \mathcal{M}^G \circ \mathcal{W^\rightarrow} }  p_b^\rightarrow \left| \left| \left( \mathcal{M}^G \circ \mathcal{W^\rightarrow}\right) (P^1_{B=b,A_1|X_1}\otimes D_{A_2|X_2,B=b})
			-\left( \mathcal{M}^G \circ \mathcal{W^\rightarrow}\right) (P^2_{B=b,A_1|X_1}\otimes D_{A_2|X_2,B=b})	\right|\right|_1\right)
			\\&\le
			\frac 12 \sum_b \sup_D  \max \left\{ \sup_{\mathcal{M}^G \circ \mathcal{W^\leftarrow}}     \left| \left| \left( \mathcal{M}^G \circ \mathcal{W^\leftarrow}\right) (P^1_{B=b,A_1|X_1}\otimes D_{A_2|X_2,B=b})-\left( \mathcal{M}^G \circ \mathcal{W^\leftarrow}\right) (P^2_{B=b,A_1|X_1}\otimes D_{A_2|X_2,B=b})
			\right|\right|_1\right. \nonumber\\&  \left.,
			\sup_{\mathcal{M}^G \circ \mathcal{W^\rightarrow}}   \left| \left| \left( \mathcal{M}^G \circ \mathcal{W^\rightarrow}\right) (P^1_{B=b,A_1|X_1}\otimes D_{A_2|X_2,B=b})-\left( \mathcal{M}^G \circ \mathcal{W^\rightarrow}\right) (P^2_{B=b,A_1|X_1}\otimes D_{A_2|X_2,B=b})
			\right|\right|_1\right\}. \label{eq:wiring_decomposition}
			\end{align}

			In the two following paragraphs, we {investigate} probability distributions, obtained after the wirings $\mathcal{W}^{\rightarrow}$ and $\mathcal{W}^{\leftarrow}$.\\
			
			$\left. \mathcal{W}^{\rightarrow}\right)$ The first thing to do now is to identify a probability distribution we obtain after wiring. The state of the system after distinguisher obtains a classical output $B=b$, which is prior to input in the considered scenario, is given by $P_{A_1|X_1,B=b} \otimes D_{A_2|X_2B=b}$, see Fig. \ref{fig:wiring}(a). The distinguisher can apply wirings from $P$ to $D$, controlled by $f_b$, $g_b$, which can depend on outcome $b$. The probability distribution after the wiring $\mathcal{W}^{\rightarrow}$ (for a fixed value of outcome $B$) is given by
			\ben
			&& \mathcal{W}^{\rightarrow} \left(P_{A_1|X_1,B} \otimes D_{A_2|X_2,B}\right)_{A_1^\prime|X_1,B}(a_1^\prime|x_1^\prime, b) \nonumber \\
			&&=
			\sum_{a_1,a_2:~ \mathrm{g}_b(a_1,a_2,x_1^\prime)=a_1^\prime} P_{A_1|X_1,B}(a_1|x_1^\prime, b) D_{A_2|X_2,B}(a_2|\mathrm{f}_b(a_1,x_1^\prime),b)
			\een
			
			Hence the probability distribution for the device after a wiring is given by
			\ben
			\overbar{P_{\mathrm{f}_b,\mathrm{g}_b}}_{B,A_1^\prime|X_1^\prime}(b,a_1^\prime|x_1^\prime): &=&
			P_{B|X_1}(b|x_1^\prime) \sum_{a_1,a_2:~\mathrm{g}_b(a_1,a_2,x_1^\prime)=a_1^\prime} P_{A_1|X_1,B}(a_1|x_1^\prime, b) D_{A_2|X_2,B}(a_2|\mathrm{f}_b(a_1,x_1^\prime),b)\\ &=&
			\sum_{a_1,a_2:~\mathrm{g}_b(a_1,a_2,x_1^\prime)=a_1^\prime} P_{B,A_1|X_1}(b,a_1|x_1^\prime) D_{A_2|X_2,B}(a_2|\mathrm{f}_b(a_1,x_1^\prime),b)\
			\een

			$\left. \mathcal{W}^{\leftarrow}\right)$. The first thing to do is again to identify a probability distribution after wiring. However, we are now in a comfortable situation, as it is enough to interchange inputs of $P_{A_1|X_1,B}$ and $D_{A_2|X_2,B}$ systems, see Fig. \ref{fig:wiring}(b).
			\ben
			\overbar{\overbar{P_{\mathrm{f}_b,\mathrm{g}_b}}}_{B,A_1^\prime|X_1^\prime}(b,a_1^\prime|x_1^\prime):=
			\sum_{a_1,a_2:~\mathrm{g}_b(a_1,a_2,x_1^\prime)=a_1^\prime} P_{B,A_1|X_1}(b,a_1|\mathrm{f}_b(a_2,x_1^\prime)) D_{A_2|X_2,B}(a_2|x_1^\prime,b).
			\een

			At this point we are ready to calculate both terms in Eq. (\ref{eq:wiring_decomposition}) separately.\\
			~~\\
			a) In the first term $\forall_{b\in B} \forall_D$ we have:
			\ben
			&&	 \sup_{\mathcal{M}^G \circ \mathcal{W^\rightarrow}}  \left| \left| \left( \mathcal{M}^G \circ \mathcal{W^\rightarrow}\right) (P^1_{B=b,A_1|X_1}\otimes D_{A_2|X_2,B=b})-\left( \mathcal{M}^G \circ \mathcal{W^\rightarrow}\right) (P^2_{B=b,A_1|X_1}\otimes D_{A_2|X_2,B=b})
			\right|\right|_1 \\
			&=&
			\sup_{\mathrm{f}_b,\mathrm{g}_b} \sup_{\mathcal{M}^G}  \sum_{a_1^\prime} \left| \mathcal{M}^G\left(\overbar{P^1_{\mathrm{f}_b,\mathrm{g}_b}}_{B,A_1^\prime|X_1^\prime}\right)(b,a_1^\prime)- \mathcal{M}^G\left(\overbar{P^2_{\mathrm{f}_b,\mathrm{g}_b}}_{B,A_1^\prime|X_1^\prime}\right)(b,a_1^\prime)\right|\\
			&= & \sup_{\mathrm{f}_b,\mathrm{g}_b}\sup_{\left\{\omega_i\right\}}  \sum_{a_1^\prime} \left|\sum_i \omega_i \mathcal{M}_i^F\left(\overbar{P^1_{\mathrm{f}_b,\mathrm{g}_b}}_{B,A_1^\prime|X_1^\prime}\right)(b,a_1^\prime)-\sum_i \omega_i \mathcal{M}_i^F\left(\overbar{P^2_{\mathrm{f}_b,\mathrm{g}_b}}_{B,A_1^\prime|X_1^\prime}\right)(b,a_1^\prime)\right|\\
			& \le &  \sup_{\mathrm{f}_b,\mathrm{g}_b}  \sup_{\left\{\omega_i\right\}} \sum_{a_1^\prime} \sum_i \omega_i \left| \mathcal{M}_i^F\left(\overbar{P^1_{\mathrm{f}_b,\mathrm{g}_b}}_{B,A_1^\prime|X_1^\prime}\right)(b,a_1^\prime)- \mathcal{M}_i^F\left(\overbar{P^2_{\mathrm{f}_b,\mathrm{g}_b}}_{B,A_1^\prime|X_1^\prime}\right)(b,a_1^\prime)\right|\\ 
			& \le &  \sup_{\mathrm{f}_b,\mathrm{g}_b} \max_{x_1^\prime}  \sum_{a_1^\prime} \left| \overbar{P^1_{\mathrm{f}_b,\mathrm{g}_b}}_{B,A_1^\prime|X_1^\prime}(b,a_1^\prime|x_1^\prime)- \overbar{P^2_{\mathrm{f}_b,\mathrm{g}_b}}_{B,A_1^\prime|X_1^\prime}(b,a_1^\prime|x_1^\prime)\right| \\
			&=&  \sup_{\mathrm{f}_b,\mathrm{g}_b} \max_{x_1^\prime}  \sum_{a_1^\prime} \left| \sum_{a_1,a_2:~\mathrm{g}_b(a_1,a_2,x_1^\prime)=a_1^\prime} P^1_{B,A_1|X_1}(b,a_1|x_1^\prime) D_{A_2|X_2,B}(a_2|\mathrm{f}_b(a_1,x_1^\prime),b) \right. \nonumber \\
			&& \hspace{5cm} -\left. \sum_{a_1,a_2:~\mathrm{g}_b(a_1,a_2,x_1^\prime)=a_1^\prime} P^2_{B,A_1|X_1}(b,a_1|x_1^\prime) D_{A_2|X_2,B}(a_2|\mathrm{f}_b(a_1,x_1^\prime),b)\right|\\ 
			&=&  \sup_{\mathrm{f}_b,\mathrm{g}_b} \max_{x_1^\prime}  \sum_{a_1^\prime} \left| \sum_{a_1,a_2:~\mathrm{g}_b(a_1,a_2,x_1^\prime)=a_1^\prime}D_{A_2|X_2,B}(a_2|\mathrm{f}_b(a_1,x_1^\prime),b) \left( P^1_{B,A_1|X_1}(b,a_1|x_1^\prime)-P^2_{B,A_1|X_1}(b,a_1|x_1^\prime) \right) \right| ~~~~~~~\\ 
			&\le &  \sup_{\mathrm{f}_b,\mathrm{g}_b} \max_{x_1^\prime} \sum_{a_1^\prime} \sum_{a_1,a_2:~\mathrm{g}_b(a_1,a_2,x_1^\prime)=a_1^\prime}D_{A_2|X_2,B}(a_2|\mathrm{f}_b(a_1,x_1^\prime),b) \left|  P^1_{B,A_1|X_1}(b,a_1|x_1^\prime)-P^2_{B,A_1|X_1}(b,a_1|x_1^\prime) \right| \\
			&= &\sup_{\mathrm{f}_b,\mathrm{g}_b} \max_{x_1^\prime} \sum_{a_1,a_2} D_{A_2|X_2,B}(a_2|\mathrm{f}_b(a_1,x_1^\prime),b) \left|  P^1_{B,A_1|X_1}(b,a_1|x_1^\prime)-P^2_{B,A_1|X_1}(b,a_1|x_1^\prime) \right|
			\\
			&= &
			\max_{x_1^\prime}  \sum_{a_1} \left|  P^1_{B,A_1|X_1}(b,a_1|x_1^\prime)-P^2_{B,A_1|X_1}(b,a_1|x_1^\prime) \right| \hspace{3.3in}  \\ 
			& = &
			\sup_{\mathcal{M}^F}  \sum_{a_1} \left|  \mathcal{M}^F \left( P^1_{B,A_1|X_1} \right)(b,a_1) - \mathcal{M}^F \left( P^2_{B,A_1|X_1} \right)(b,a_1) \right|.
			\een
			The important point is to notice that $\sum_{a_1^\prime} \sum_{a_1,a_2:~\mathrm{g}_b(a_1,a_2,x_1^\prime)=a_1^\prime} \mathrm{h}(a_1,a_2) = \sum_{a_1,a_2}\mathrm{h}(a_1,a_2)$.

			%
			b) Now in the second term $\forall_{b\in B} \forall_D$ we have:
			\ben
			&&	 \sup_{\mathcal{M}^G \circ \mathcal{W^\leftarrow}}   \left| \left| \left( \mathcal{M}^G \circ \mathcal{W^\leftarrow}\right) (P^1_{B=b,A_1|X_1}\otimes D_{A_2|X_2,B=b})-\left( \mathcal{M}^G \circ \mathcal{W^\leftarrow}\right) (P^2_{B=b,A_1|X_1}\otimes D_{A_2|X_2,B=b})
			\right|\right|_1
			\\ &=& 
			\sup_{\mathrm{f}_b,\mathrm{g}_b} \sup_{\mathcal{M}^G} \sum_{a_1^\prime} \left| \mathcal{M}^G\left(\overbar{\overbar{P^1_{\mathrm{f}_b,\mathrm{g}_b}}}_{B,A_1^\prime|X_1^\prime}\right)(b,a_1^\prime)- \mathcal{M}^G\left(\overbar{\overbar{P^2_{\mathrm{f}_b,\mathrm{g}_b}}}_{B,A_1^\prime|X_1^\prime}\right)(b,a_1^\prime)\right|\\ 
			&=& 
			\sup_{\mathrm{f}_b,\mathrm{g}_b} \sup_{\left\{\omega_i\right\}} \sum_{a_1^\prime} \left|\sum_i \omega_i \mathcal{M}_i^F\left(\overbar{\overbar{P^1_{\mathrm{f}_b,\mathrm{g}_b}}}_{B,A_1^\prime|X_1^\prime}\right)(b,a_1^\prime)-\sum_i \omega_i \mathcal{M}_i^F\left(\overbar{\overbar{P^2_{\mathrm{f}_b,\mathrm{g}_b}}}_{B,A_1^\prime|X_1^\prime}\right)(b,a_1^\prime)\right|\\ 
			&\le & \frac 12
			\sup_{\mathrm{f}_b,\mathrm{g}_b} \sup_{\left\{\omega_i\right\}} \sum_{a_1^\prime} \sum_i \omega_i \left| \mathcal{M}_i^F\left(\overbar{\overbar{P^1_{\mathrm{f}_b,\mathrm{g}_b}}}_{B,A_1^\prime|X_1^\prime}\right)(b,a_1^\prime)- \mathcal{M}_i^F\left(\overbar{\overbar{P^2_{\mathrm{f}_b,\mathrm{g}_b}}}_{B,A_1^\prime|X_1^\prime}\right)(b,a_1^\prime)\right|\\ 
			&\le & \frac 12
			\sup_{\mathrm{f}_b,\mathrm{g}_b} \max_{x_1^\prime} \sum_{a_1^\prime} \left| \overbar{\overbar{P^1_{\mathrm{f}_b,\mathrm{g}_b}}}_{B,A_1^\prime|X_1^\prime}(b,a_1^\prime|x_1^\prime)- \overbar{\overbar{P^2_{\mathrm{f}_b,\mathrm{g}_b}}}_{B,A_1^\prime|X_1^\prime}(b,a_1^\prime|x_1^\prime)\right|\\ 
			&= &
			\sup_{\mathrm{f}_b,\mathrm{g}_b} \max_{x_1^\prime} \sum_{a_1^\prime} \left|  \sum_{a_1,a_2:~\mathrm{g}_b(a_1,a_2,x_1^\prime)=a_1^\prime} P^1_{B,A_1|X_1}(b,a_1|\mathrm{f}_b(a_2,x_1^\prime)) D_{A_2|X_2,B}(a_2|x_1^\prime,b) \right.\\
			& &  \hspace{5cm} - \left. \sum_{a_1,a_2:~\mathrm{g}_b(a_1,a_2,x_1^\prime)=a_1^\prime} P^2_{B,A_1|X_1}(b,a_1|\mathrm{f}_b(a_2,x_1^\prime)) D_{A_2|X_2,B}(a_2|x_1^\prime,b) \right|\\ 
			&=& 
			\sup_{\mathrm{f}_b,\mathrm{g}_b} \max_{x_1^\prime} \sum_{a_1^\prime} \left|  \sum_{a_1,a_2:~\mathrm{g}_b(a_1,a_2,x_1^\prime)=a_1^\prime} D_{A_2|X_2,B}(a_2|x_1^\prime,b) \left( P^1_{B,A_1|X_1}(b,a_1|\mathrm{f}_b(a_2,x_1^\prime)) \right. \right.\\
			&&  \hspace{9.7cm} -\left.\left.  P^2_{B,A_1|X_1}(b,a_1|\mathrm{f}_b(a_2,x_1^\prime)) \right) \right| \\
			&\le & 
			\sup_{\mathrm{f}_b,\mathrm{g}_b} \max_{x_1^\prime} \sum_{a_1^\prime}   \sum_{a_1,a_2:~\mathrm{g}_b(a_1,a_2,x_1^\prime)=a_1^\prime} D_{A_2|X_2,B}(a_2|x_1^\prime,b) \left| P^1_{B,A_1|X_1}(b,a_1|\mathrm{f}_b(a_2,x_1^\prime)) \right. \\
			&&  \hspace{10cm} - \left. P^2_{B,A_1|X_1}(b,a_1|\mathrm{f}_b(a_2,x_1^\prime)) \right| \\ 
			&= & 
			\sup_{\mathrm{f}_b} \max_{x_1^\prime}  \sum_{a_1,a_2} D_{A_2|X_2,B}(a_2|x_1^\prime,b) \left| P^1_{B,A_1|X_1}(b,a_1|\mathrm{f}_b(a_2,x_1^\prime)) - P^2_{B,A_1|X_1}(b,a_1|\mathrm{f}_b(a_2,x_1^\prime)) \right|
			\\ &=& 
			\sup_{\mathrm{f}_b} \max_{x_1^\prime}  \sum_{a_2} D_{A_2|X_2,B}(a_2|x_1^\prime,b) \sum_{a_1}\left| P^1_{B,A_1|X_1}(b,a_1|\mathrm{f}_b(a_2,x_1^\prime)) - P^2_{B,A_1|X_1}(b,a_1|\mathrm{f}_b(a_2,x_1^\prime)) \right| \\ 
			&\le & 
			\sup_{\mathrm{f}_b} \max_{x_1^\prime}  \sum_{a_2} D_{A_2|X_2,B}(a_2|x_1^\prime,b) \max_{a_2^\prime}\sum_{a_1} \left| P^1_{B,A_1|X_1}(b,a_1|\mathrm{f}_b(a_2^\prime,x_1^\prime)) - P^2_{B,A_1|X_1}(b,a_1|\mathrm{f}_b(a_2^\prime,x_1^\prime)) \right| \\ 
			&=&
			\sup_{\mathrm{f}_b} \max_{x_1^\prime} \max_{a_2^\prime}  \sum_{a_1}  \left| P^1_{B,A_1|X_1}(b,a_1|\mathrm{f}_b(a_2^\prime,x_1^\prime)) - P^2_{B,A_1|X_1}(b,a_1|\mathrm{f}_b(a_2^\prime,x_1^\prime)) \right|
			\\ &=&
			\max_{x_1}  \sum_{a_1}  \left| P^1_{B,A_1|X_1}(b,a_1|x_1) - P^2_{B,A_1|X_1}(b,a_1|x_1) \right| 
			\\ &= &
			\sup_{\mathcal{M}^F}  \sum_{a_1}  \left| \mathcal{M}^F\left(P^1_{B,A_1|X_1}\right)(b,a_1) - \mathcal{M}^F\left(P^2_{B,A_1|X_1}\right)(b,a_1) \right|.
			\een
			
			\noindent	From a), b) and Eq. (\ref{eq:wiring_decomposition}) we conclude that:
			\begin{align}
			&\left|\left|P^1_{B,A_1|X_1}  -  P^2_{B,A_1|X_1}  \right|\right|_\mathrm{NS}\\
			& \le
			\frac 12 \sum_b \sup_D  \max \left\{ \sup_{\mathcal{M}^F}  \sum_{a_1}  \left| \mathcal{M}^F\left(P^1_{B,A_1|X_1}\right)(b,a_1) - \mathcal{M}^F\left(P^2_{B,A_1|X_1}\right)(b,a_1) \right|\right. \nonumber\\&  \left.,
			\sup_{\mathcal{M}^F}  \sum_{a_1}  \left| \mathcal{M}^F\left(P^1_{B,A_1|X_1}\right)(b,a_1) - \mathcal{M}^F\left(P^2_{B,A_1|X_1}\right)(b,a_1) \right|\right\}\\
			&\le \frac 12 \sum_b \sup_{\mathcal{M}^F}  \sum_{a_1}  \left| \mathcal{M}^F\left(P^1_{B,A_1|X_1}\right)(b,a_1) - \mathcal{M}^F\left(P^2_{B,A_1|X_1}\right)(b,a_1) \right|.
			\end{align}
			As the r.h.s of the expression above realizes a particular strategy of the distinguisher within considered NS norm, the above inequality can be always saturated, what yields:
			\ben
			\left|\left|P^1_{B,A_1|X_1}  -  P^2_{B,A_1|X_1}  \right|\right|_\mathrm{NS} = \frac 12 \sum_b \sup_{\mathcal{M}^F}  \sum_{a_1}  \left| \mathcal{M}^F\left(P^1_{B,A_1|X_1}\right)(b,a_1) - \mathcal{M}^F\left(P^2_{B,A_1|X_1}\right)(b,a_1) \right|.
			\een
		\end{proof}
		
		\begin{corone} For the cc-d states shared at the end of the MDLOPC protocol $\Lambda$, the NS norm can be rephrased with a simplified expression:
			\begin{align}
			\left| \left| P_{S_A,S_B,Q,E|Z}-Q_{S_A,S_B,Q,E|Z}\right| \right|_\mathrm{NS}= \frac{1}{2}
			\sum_{s_A,s_B,q} \max_z \sum_{e}
			\left| P_{S_A,S_B,Q,E|Z}(s_A,s_B,q,e|z)   -Q_{S_A,S_B,Q,E|Z}(s_A,s_B,q,e|z) \right|, 
			\end{align}
			\noindent where $\max_z$, stands for the maximization over all possible direct measurements performed by the eavesdropper. 
		\end{corone}
		
		\begin{proof}
			{The proof follows directly from substituting} $B\equiv (S_A,S_B,Q)$, $A_1\equiv E$ and $X_1 \equiv Z$ {in the result of Proposition \ref{prop:ccb}}.  {In this way we obtain} cc-d states {that are} shared at the end of the MDLOPC protocol $\Lambda$, and hence we arrive at the claim:
			
			\ben
			\left| \left| P_{S_A,S_B,Q,E|Z}-Q_{S_A,S_B,Q,E|Z}\right| \right|_\mathrm{NS}= \frac{1}{2}
			\sum_{s_A,s_B,q} \max_z \sum_{e}
			\left| P_{S_A,S_B,Q,E|Z}(s_A,s_B,q,e|z) -Q_{S_A,S_B,Q,E|Z}(s_A,s_B,q,e|z) \right|, \nonumber \\
			\een
			where the $\max_z$ is the maximization over direct measurements in the part of Eve.
		\end{proof}
		
		\begin{rem}
			{The norm on the space of no-signaling conditional probability distributions based on trace distance introduced by M. Christandl and B. Toner \cite{ChristandlToner}} is based on a supremum over all possible linear operations. According to our best knowledge, these operations have not been characterized yet in the literature. In this section, we do not target to describe this class of operations. Instead, via the set $\mathcal{G}$, we constructed a particular action of the distinguishing system on {\bf c}-d states, which is sufficient for cryptographic purpose as it yields equivalent security criterion to \cite{masanes-2009-102}. 
		\end{rem}

		\section{Equivalence between security criteria for NSDI protocols}\label{sec:equivalence-security}

		
		The iid NSDI key rate in Definition \ref{def:key_rate} is implicitly dependent on proximity in the NS norm security criterion in Eq. (\ref{eq:pdit}). 
		In the quantum case, it was shown  that the proximity in the norm (of a state to the ideal one) is equivalent to {\it the correctness and secrecy} of a protocol \cite{Portmann-Renner, Beaudry}. These two notions are employed in a protocol independent definition of security \cite{Ben-OrHLMO05}. {In this section, we show that security criterion based on NS norm is equivalent to the one based on secrecy and correctness of MDLOPC protocol}. 
		
		In what follows, we employ the notions of {\it real, ideal}, and {\it intermediate systems}. A real system is a device shared by the parties at the end of a protocol. An ideal device is the one which possesses the same distribution on Eve's side as a real device, however, possesses perfect (uniform) correlations between Alice and Bob, that are completely uncorrelated with Eve. {An intermediate device is {another} kind of device in which Alice and Bob always share fully correlated keys. However, the distribution of the keys is not uniform (Eve's part stays unchanged).} The usual part of any protocol employing non-local correlations is an acceptance phase in which honest parties decide (upon some test) whether to abort or to proceed with the protocol. 
		
		
		Composability concept in security is an area of research concerned with composing cryptographic primitives into more complex ones while keeping high security level. In the universal composability approach, a cryptographic primitive is said to be \emph{universally composable} if any functionality using this primitive is as secure as an ideal one \cite{Can01,Ben-OrHLMO05}. The composable security is considered as the strongest notion of security \cite{Can01,Ben-OrHLMO05}. However, in the device independent scenario, so far, it was not rigorously proven that this scheme is ultimately secure. Furthermore, the results of \cite{memory-attack} strongly suggest that it is not the case, so the problem arises when one wants to reuse the device. In particular, if the device used for composition has some memory, then it can  leak the key of the previous use. This implies that, in general, the protocol is composably secure as long as the same device is not reused in the protocol. {We refer to this notion of security to be restricted composable.}
		
		Theorem \ref{thm:equivalence} is essential to compare the secret key of our scenario to these of other cryptographic schemes or even certain protocols, in particular to the results of  H{\"a}nggi, Renner and Wolf \cite{hanggi-2009}, with the upper bounds that will be presented in this paper. We start with a few definitions.
		
		\begin{definition}[State of the device at the end of protocol] The state of the device after the MDLOPC protocol is a conditional probability distribution (\textbf{c}-d state) denoted  by $P_{S_A,S_A,Q,E|Z}^\mathrm{real}$: 
			\ben
			P_{S_A,S_B,Q,E|Z}^\mathrm{real}=\mathrm{p}_\mathrm{abort}{P}_{S_A,S_B,Q,E|Z}^\mathrm{real|abort}
			+\left(1-\mathrm{p}_\mathrm{abort}\right){P}_{S_A,S_B,Q,E|Z}^{\mathrm{real|pass}}.
			\een
		\end{definition}
		\noindent The random variables $S_A, S_B, E$ are respectively outputs of Alice, Bob, and Eve conditioned upon input Z of Eve. $S_A, S_B$  are the key strings hold by Alice and Bob after the protocol, respectively. Q is the random denoting public communication. During the protocol, Q is shared by the three parties, although Alice and Bob use it only to distill the final key and discard it after the protocol is finished. For this reason, we treat Q to be the random variable of Eve that she can use for the choice of her input. Despite the fact that in the notation adopted by as variables of outputs are conditioned upon variables of inputs, Eve's choice of input $Z$ can still depend on the value of $Q$. The superscripts abort and pass indicate whether protocol passed the acceptance phase.   
		
		
		\begin{definition}[Ideal output state]\label{def:idealOutput} The ideal output state of the device is the one that possesses perfect correlations between honest parties that are completely uncorrelated with the eavesdropper. Local outcomes of the eavesdropper and communication simulate the real system.
			\ben
			P_{S_A,S_B,Q,E|Z}^\mathrm{ideal|pass} (s_A,s_B,q,e|z) &=& 
			\frac{\delta_{s_A,s_B}}{|S_A|} \sum_{s_A^\prime,s_B^\prime} P_{S_A,S_B,Q,E|Z}^\mathrm{real|pass} (s_A^\prime,s_B^\prime,q,e|z).
			\\
			\een
		\end{definition}
		Since the honest parties are uncorrelated with Eve, the ideal system can be decomposed according to tensor rule formula for independent systems in the following way: 
		\ben
		P_{S_A,S_B,Q,E|Z}^\mathrm{ideal|pass}
		&=&
		P_{S_A,S_B}^\mathrm{ideal|pass} \otimes P_{Q,E|Z}^\mathrm{ideal|pass}.
		\een
		

		\begin{definition}[State of the intermediate system] An intermediate system is the one that bears {fully correlated key strings between the honest parties}, but the distribution they possess is not uniform; hence correlations are not perfect in a cryptographic sense. Eavesdropper is not completely uncorrelated with the honest parties.
			\ben
			P_{S_A,S_B,Q,E|Z}^\mathrm{int|pass} (s_A,s_B,q,e|z) &=& 
			\delta_{s_A,s_B} \sum_{s_B}P_{S_A,S_B,Q,E|Z}^\mathrm{real|pass} (s_A,s_B,q,e|z).
			\een
		\end{definition}
		\noindent Since the states of the intermediate and the ideal systems are constructed with respect to the state of the real system, the $\mathrm{p}_\mathrm{abort}$ is the same in all cases (later, we consider the protocol after the acceptation phase, for which $\mathrm{p}_\mathrm{abort}=0$). {The same is true for all states conditioned on aborting, i.e., they are trivially the same.}
		
		For the sake of cohesion, we provide definitions of secrecy, correctness, and security of a cryptographic protocol in case of non-signaling devices.	
		\begin{definition}[$\varepsilon$-secrecy of a protocol]\label{def:Secrecy}
			An MDLOPC key distribution protocol is $\varepsilon$-secret if it outputs a device for which conditional probability distribution shared between Alice (Bob) and Eve at the end of the protocol (and the protocol does not abort) satisfies
			\ben
			\left(1-\mathrm{p}_\mathrm{abort}\right) \norm{P_{S_A,Q,E|Z}^\mathrm{real|pass}-P_{S_A,Q,E|Z}^\mathrm{ideal|pass}}_\mathrm{NS} \le \varepsilon,
			\een
			where
			\ben
			P^\mathrm{real(ideal)|pass}_{S_A,Q,E|Z}(s_A,q,e|z):=\sum_{s_B} P^\mathrm{real(ideal)|pass}_{S_A,S_B,Q,E|Z}(s_A,s_B,q,e|z).
			\een
		\end{definition}
		
		\begin{definition}[$\varepsilon$-correctness]\label{def:Correctness}
			An MDLOPC key distribution protocol is $\varepsilon$-correct if the probability (and the protocol does not abort) for Alice and Bob not to share the same output keys satisfies 
			\be
			\left(1-\mathrm{p}_\mathrm{abort}\right)\mathrm{P}\left[S_A \neq S_B|\mathrm{pass}\right] \le \varepsilon.
			\ee
		\end{definition}
		
		
		\begin{definition}[$\varepsilon$-security of a protocol]
			\label{abc}
			Let $P_{S_A,S_B,Q,E|Z}^\mathrm{real|pass}$ be the state of the system shared between Alice, Bob, and Eve after the protocol (and the protocol does not abort). Then the protocol is $\varepsilon$-secure if
			\beq 	\label{eq:epsilon_security}
			\left(1-\mathrm{p}_\mathrm{abort}\right) \norm{P_{S_A,S_B,Q,E|Z}^\mathrm{real|pass}-P_{S_A,S_B,Q,E|Z}^\mathrm{ideal|pass}}_\mathrm{NS}  \le \varepsilon, 
			\eeq	
			where $\mathrm{p}_\mathrm{abort}$ is the probability of aborting (which is the same for the real and
			ideal protocols).
		\end{definition}
		
		{To prove the equivalence between security criterion based on NS norm and the one based on security and correctness}, we provide technical Lemmas, showing that proximity in NS norm implies {secrecy and correctness, and vice versa}.
		
		
		\begin{observation}\label{obs:ort-supp} The following equality holds.
			\ben
			&&\left| \left| P_{S_A,S_B,Q,E|Z}^\mathrm{real}-P_{S_A,S_B,Q,E|Z}^\mathrm{ideal} \right| \right|_\mathrm{NS} =
			\left(1-\mathrm{p}_\mathrm{abort} \right)\left| \left|   P_{S_A,S_B,Q,E|Z}^\mathrm{real|pass}-  P_{S_A,S_B,Q,E|Z}^\mathrm{ideal|pass} \right|\right|_\mathrm{NS},
			\een
		\end{observation}
		
		\begin{proof}
			\ben
			&&\left| \left| P_{S_A,S_B,Q,E|Z}^\mathrm{real}-P_{S_A,S_B,Q,E|Z}^\mathrm{ideal} \right| \right|_\mathrm{NS} \nonumber \\ &=&
			\left| \left| \mathrm{p}_\mathrm{abort} P_{S_A,S_B,Q,E|Z}^\mathrm{real|abort}+\left(1-\mathrm{p}_\mathrm{abort} \right) P_{S_A,S_B,Q,E|Z}^\mathrm{real|pass}-\mathrm{p}_\mathrm{abort} P_{S_A,S_B,Q,E|Z}^\mathrm{ideal|abort}-\left(1-\mathrm{p}_\mathrm{abort} \right) P_{S_A,S_B,Q,E|Z}^\mathrm{ideal|pass} \right|\right|_\mathrm{NS}\\
			&=&
			\left| \left| \mathrm{p}_\mathrm{abort} \left( P_{S_A,S_B,Q,E|Z}^\mathrm{real|abort}-P_{S_A,S_B,Q,E|Z}^\mathrm{ideal|abort} \right) +\left(1-\mathrm{p}_\mathrm{abort} \right) \left( P_{S_A,S_B,Q,E|Z}^\mathrm{real|pass}-  P_{S_A,S_B,Q,E|Z}^\mathrm{ideal|pass}\right) \right|\right|_\mathrm{NS}\\
			&\stackrel{(I)}{=}&
			\left(1-\mathrm{p}_\mathrm{abort} \right)\left| \left|   P_{S_A,S_B,Q,E|Z}^\mathrm{real|pass}-  P_{S_A,S_B,Q,E|Z}^\mathrm{ideal|pass} \right|\right|_\mathrm{NS},
			\een
			$(I)$ - we {use the fact that} $P_{S_A,S_B,Q,E|Z}^\mathrm{real|abort}$ and $P_{S_A,S_B,Q,E|Z}^\mathrm{ideal|abort}$ are the same when the protocol is aborted \cite{Beaudry}. 
		\end{proof}

		\begin{lemma} The NS norm evaluated for real and intermediate states quantifies the probability of Alice and Bob to share different key strings at the end of the protocol. 
			\ben
			\norm{P_{S_A,S_B,Q,E|Z}^\mathrm{real}-P_{S_A,S_B,Q,E|Z}^\mathrm{int}} &=&  
			\left(1-\mathrm{p}_\mathrm{abort}\right) \mathrm{P}\left[S_A \neq S_B|\mathrm{pass}\right] 
			\een
			\label{lemma:sasb}
		\end{lemma}
		
		\begin{proof} From the Observation \ref{obs:ort-supp} we have:
			\ben
			\norm{P_{S_A,S_B,Q,E|Z}^\mathrm{real}-P_{S_A,S_B,Q,E|Z}^\mathrm{int}}_\mathrm{NS} &\stackrel{(I)}{=}&  \left(1-\mathrm{p}_\mathrm{abort}\right)\norm{(P_{S_A,S_B,Q,E|Z}^\mathrm{real|pass}-P_{S_A,S_B,Q,E|Z}^\mathrm{int|pass})}_\mathrm{NS}
			\een
			
			Now, using Proposition \ref{cor:box_norm_exp}:
			\ben
			&& \norm{P_{S_A,S_B,Q,E|Z}^\mathrm{real|pass}-P_{S_A,S_B,Q,E|Z}^\mathrm{int|pass}}_\mathrm{NS} \nonumber\\ &=&
			\frac{1}{2} \sum_{s_A,s_B,q} \max_z \sum_e \left|P_{S_A,S_B,Q,E|Z}^\mathrm{real|pass}(s_A,s_B,q,e|z)- P_{S_A,S_B,Q,E|Z}^\mathrm{int|pass} (s_A,s_B,q,e|z)\right|  \\
			&=& \frac{1}{2} \sum_{s_A,s_B,q} \max_z \sum_e \left|P_{S_A,S_B,Q,E|Z}^\mathrm{real|pass}(s_A,s_B,q,e|z)- \delta_{s_A,s_B} \sum_{s_B}P_{S_A,S_B,Q,E|Z}^\mathrm{real|pass} (s_A,s_B,q,e|z) \right| \\
			&=&  \frac{1}{2}
			\sum_{s_A,s_B,q} \max_z \sum_e \left|\mathrm{P}^\mathrm{real|pass}_{S_A,S_B,Q,E|Z}(s_A,s_B,q,e|z)-\delta_{s_A,s_B} \sum_{s_B}P_{S_A,S_B,Q,E|Z}^\mathrm{real|pass} (s_A,s_B,q,e|z)\right|\delta_{s_A,s_B} \nonumber \\ &&+ \frac{1}{2}
			\sum_{s_A,q}\sum_{s_B\neq s_A} \max_z \sum_e \left|\mathrm{P}^\mathrm{real|pass}_{S_A,S_B,Q,E|Z}(s_A,s_B,q,e|z)-\delta_{s_A,s_B} \sum_{s_B}P_{S_A,S_B,Q,E|Z}^\mathrm{real|pass} (s_A,s_B,q,e|z)\right|  \\&=&
			\frac{1}{2} \sum_{s_A,q} \max_z \sum_e \left|P_{S_A,S_B,Q,E|Z}^\mathrm{real|pass}(s_A,s_A,q,e|z)-\sum_{s_B}P_{S_A,S_B,Q,E|Z}^\mathrm{real|pass} (s_A,s_B,q,e|z)\right| \nonumber \\ && +
			\frac{1}{2} \sum_{s_A,q}\sum_{s_B\neq s_A} \max_z \sum_e \mathrm{P}^\mathrm{real|pass}_{S_A,S_B,Q,E|Z}(s_A,s_B,q,e|z)  \\
			&\stackrel{(I)}{=}& \frac{1}{2}
			\sum_{s_A,q} \max_z \sum_e \left(\sum_{s_B}P_{S_A,S_B,Q,E|Z}^\mathrm{real|pass} (s_A,s_B,q,e|z)-P_{S_A,S_B,Q,E|Z}^\mathrm{real|pass}(s_A,s_A,q,e|z)\right) \nonumber \\ &&+
			\frac{1}{2} \sum_{s_A,q}\sum_{s_B\neq s_A}  \sum_e \mathrm{P}^\mathrm{real|pass}_{S_A,S_B,Q,E|Z}(s_A,s_B,q,e|z) \\
			&=& \frac{1}{2}
			\sum_{s_A,q} \max_z \sum_e \sum_{s_B \neq s_A}P_{S_A,S_B,Q,E|Z}^\mathrm{real|pass} (s_A,s_B,q,e|z) +
			\frac{1}{2} \sum_{s_A,q}\sum_{s_B\neq s_A}  \sum_e \mathrm{P}^\mathrm{real|pass}_{S_A,S_B,Q,E|Z}(s_A,s_B,q,e|z)  \\
			&\stackrel{(II)}{=}&
			\sum_{s_A,q} \sum_{s_B \neq s_A} \sum_e P_{S_A,S_B,Q,E|Z}^\mathrm{real|pass}(s_A,s_A,q,e|z) \\
			&=& \mathrm{P}\left[S_A \neq S_B |\mathrm{pass}\right] \label{eq:correct}
			\een
			where (I) and (II) are due to non-signaling condition on  Eves's input $z$.
			Finally we obtain:
			\ben
			\norm{P_{S_A,S_B,Q,E|Z}^\mathrm{real}-P_{S_A,S_B,Q,E|Z}^\mathrm{int}}_\mathrm{NS} &=&  \left(1-\mathrm{p}_\mathrm{abort}\right) \mathrm{P}\left[S_A \neq S_B |\mathrm{pass}\right]
			\een
		\end{proof}
		
		\begin{lemma}[Secrecy and correctness imply security]
			\label{lemma:implication1}
			If a protocol is $\varepsilon_\mathrm{sec}$-secret and $\varepsilon_\mathrm{cor}$-correct then
			the protocol is $\varepsilon$-secure, where $\varepsilon = \varepsilon_\mathrm{sec} + \varepsilon_\mathrm{cor}$.
			\ben
			&&\left\{ \left(1-\mathrm{p}_\mathrm{abort}\right) \norm{P_{S_A,Q,E|Z}^\mathrm{real|pass}-P_{S_A,Q,E|Z}^\mathrm{ideal|pass}}_\mathrm{NS} \le \varepsilon_\mathrm{sec} ~~\mathrm{and}~~  \left(1-\mathrm{p}_\mathrm{abort}\right) \mathrm{P}\left[S_A \neq S_B|\mathrm{pass}\right] \le \varepsilon_\mathrm{cor} \right\} \nonumber
			\\
			&& \implies \left(1-\mathrm{p}_\mathrm{abort}\right) \norm{P_{S_A,S_B,Q,E|Z}^\mathrm{real|pass},P_{S_A,S_B,Q,E|Z}^\mathrm{ideal|pass} }_\mathrm{NS} \le \varepsilon_\mathrm{sec}+\varepsilon_\mathrm{cor} = \varepsilon.
			\een
		\end{lemma}

		\begin{proof} 
			To prove the security of the protocol, we can decompose the l.h.s of  Eq. (\ref{eq:epsilon_security}) in the following way:
			\begin{align}\label{eqn:decomposition}
			&&\norm{P_{S_A,S_B,Q,E|Z}^\mathrm{real|pass}-P_{S_A,S_B,Q,E|Z}^\mathrm{ideal|pass}  }_\mathrm{NS}
			\le \norm{P_{S_A,S_B,Q,E|Z}^\mathrm{real|pass}-P_{S_A,S_B,Q,E|Z}^\mathrm{int|pass}}_\mathrm{NS}+
			\norm{P_{S_A,S_B,Q,E|Z}^\mathrm{int|pass}-P_{S_A,S_B,Q,E|Z}^\mathrm{ideal|pass}}_\mathrm{NS},
			\end{align}
			where we used the triangle inequality for the NS norm. From Proposition \ref{cor:box_norm_exp} we have:
			\ben
			&& \norm{P_{S_A,S_B,Q,E|Z}^\mathrm{int|pass}-P_{S_A,S_B,Q,E|Z}^\mathrm{ideal|pass}}_\mathrm{NS} \\
			&=& \frac{1}{2} \sum_{s_A,s_B,q} \max_z \sum_e \left|\delta_{s_A,s_B}\sum_{s_B}P_{S_A,S_B,Q,E|Z}^\mathrm{real|pass} (s_A,s_B,q,e|z) -  \frac{\delta_{s_A,s_B}}{|S_A|} \sum_{s_A^\prime,s_B^\prime} P_{S_A,S_B,Q,E|Z}^\mathrm{real| pass} (s_A^\prime,s_B^\prime,q,e|z)\right|  \\ 
			&=& \frac{1}{2} \sum_{s_A,s_B,q} \max_z \sum_e\delta_{s_A,s_B} \left|\sum_{s_B}P_{S_A,S_B,Q,E|Z}^\mathrm{real|pass} (s_A,s_B,q,e|z)  -  \frac{1}{|S_A|} \sum_{s_A^\prime,s_B^\prime} P_{S_A,S_B,Q,E|Z}^\mathrm{real| pass} (s_A^\prime,s_B^\prime,q,e|z)\right| \\ 
			&=& \frac{1}{2} \sum_{s_A} \max_z \sum_e \left| \sum_{s_B} P_{S_A,S_B,Q,E|Z}^\mathrm{real|pass} (s_A,s_B,q,e|z) - \sum_{s_B} \frac{\delta_{s_A,s_B}}{|S_A|} \sum_{s_A^\prime,s_B^\prime} P_{S_A,S_B,Q,E|Z}^\mathrm{real| pass} (s_A^\prime,s_B^\prime,q,e|z)\right|  \\ 
			&=&		\left| \left|P_{S_A,Q,E|Z}^\mathrm{real|pass}-P_{S_A,Q,E|Z}^\mathrm{ideal|pass}\right| \right|_\mathrm{NS} \label{eq:secret}
			\een

			Using now Lemma \ref{lemma:sasb} and Eq. (\ref{eqn:decomposition}) we have:
			\begin{align}
			&&\norm{P_{S_A,S_B,Q,E|Z}^\mathrm{real|pass}-P_{S_A,S_B,Q,E|Z}^\mathrm{ideal|pass}  }_\mathrm{NS}
			\le \mathrm{P}\left[S_A\neq S_B|\mathrm{pass}\right]+
			\norm{P_{S_A,Q,E|Z}^\mathrm{real|pass}-P_{S_A,Q,E|Z}^\mathrm{ideal|pass}}_\mathrm{NS},
			\end{align}
			Hence,
			\ben
			&&(1-\mathrm{p_{abort}})\norm{P_{S_A,S_B,Q,E|Z}^\mathrm{real|pass}-P_{S_A,S_B,Q,E|Z}^\mathrm{ideal|pass}  }_\mathrm{NS} \nonumber \\
			&& \le (1-\mathrm{p_{abort}})\mathrm{P}\left[S_A\neq S_B|\mathrm{pass}\right]+(1-\mathrm{p_{abort}})
			\norm{P_{S_A,Q,E|Z}^\mathrm{real|pass}-P_{S_A,Q,E|Z}^\mathrm{ideal|pass}}_\mathrm{NS},
			\een
			Using the above inequality if a protocol is $\varepsilon_\mathrm{sec}$-secret and $\varepsilon_\mathrm{cor}$-correct it is also at least $(\varepsilon_\mathrm{sec}+\varepsilon_\mathrm{cor})$-secure.
			\ben
			&&\left\{ \left(1-\mathrm{p}_\mathrm{abort}\right) \norm{P_{S_A,Q,E|Z}^\mathrm{real|pass}-P_{S_A,Q,E|Z}^\mathrm{ideal|pass}}_\mathrm{NS} \le \varepsilon_\mathrm{sec} ~~\mathrm{and}~~  \left(1-\mathrm{p}_\mathrm{abort}\right) \mathrm{P}\left[S_A \neq S_B|\mathrm{pass}\right] \le \varepsilon_\mathrm{cor} \right\}
			\\
			&& \implies \left(1-\mathrm{p}_\mathrm{abort}\right) \norm{P_{S_A,S_B,Q,E|Z}^\mathrm{real|pass},P_{S_A,S_B,Q,E|Z}^\mathrm{ideal|pass} }_\mathrm{NS} \le \varepsilon_\mathrm{sec}+\varepsilon_\mathrm{cor} = \varepsilon.
			\een
		\end{proof}
		
		We proved that if the protocol is $\varepsilon_\mathrm{sec}$-secret and $\varepsilon_\mathrm{cor}$-correct then its output is $\varepsilon_\mathrm{sec}+\varepsilon_\mathrm{cor}$ close to ideal device in NS norm, and by definition is $\varepsilon_\mathrm{sec}+\varepsilon_\mathrm{cor}$ secure. To prove equivalence of security criteria, we now show the proof in the opposite direction, i.e., we show that if an output device of the protocol is $\varepsilon$ close in NS norm to the ideal one, then the protocol is at least $\varepsilon$-secret and $\varepsilon$-correct.
		
		\begin{lemma}[Security implies secrecy and correctness]
			If a protocol is $\varepsilon$-secure, then it is at least $\varepsilon$-secret and $\varepsilon$-correct.
			\ben
			&&
			\left(1-\mathrm{p}_\mathrm{abort}\right) \norm{ P_{S_A,S_B,Q,E|Z}^\mathrm{real|pass}-P_{S_A,S_B,Q,E|Z}^\mathrm{ideal|pass}}_\mathrm{NS}
			\le \varepsilon \nonumber \\
			\implies 	&& \left\{ \left(1-\mathrm{p}_\mathrm{abort}\right) \mathrm{P}\left[S_A \neq S_B|pass\right] \le \varepsilon ~~ \mathrm{and}~~ \left(1-\mathrm{p}_\mathrm{abort}\right)\norm{P_{S_A,Q,E|Z}^\mathrm{real|pass}-P_{S_A,Q,E|Z}^\mathrm{ideal|pass} }_\mathrm{NS} \le \varepsilon \right\}
			\een
			\label{lemma:implication2}
		\end{lemma}
		
		\begin{proof}[Proof of Lemma \ref{lemma:implication2}]
			Let us prove the following first.
			\ben
			\norm{ P_{S_A,S_B,Q,E|Z}^\mathrm{real|pass}-P_{S_A,S_B,Q,E|Z}^\mathrm{ideal|pass}}_\mathrm{NS} \ge \mathrm{P}\left[S_A \neq S_B \right|\mathrm{pass}].
			\label{proof:impli2:eqn1}
			\een
			To proceed with this task we employ Definition \ref{def:idealOutput} of the ideal system and Proposition \ref{cor:box_norm_exp}.
			\ben
			&&\norm{ P_{S_A,S_B,Q,E|Z}^\mathrm{real|pass}-P_{S_A,S_B,Q,E|Z}^\mathrm{ideal|pass}}_\mathrm{NS} \\
			&=&\frac{1}{2} \sum_{s_A,s_B,q} \max_z \sum_e \left|P_{S_A,S_B,Q,E|Z}^\mathrm{real|pass} (s_A,s_B,q,e|z)-\frac{\delta_{s_A,s_B}}{|S_A|}\sum_{s_A^\prime,s_B^\prime}P_{S_A,S_B,Q,E|Z}^\mathrm{real|pass} (s_A^\prime,s_B^\prime,q,e|z) \right| \\
			&=&		\frac{1}{2} \sum_{s_A,q} \max_z \sum_e  \left|P_{S_A,S_B,Q,E|Z}^\mathrm{real|pass} (s_A,s_A,q,e|z)-\frac{1}{|S_A|}\sum_{s_A^\prime,s_B^\prime}P_{S_A,S_B,Q,E|Z}^\mathrm{real|pass} (s_A^\prime,s_B^\prime,q,e|z) \right| \nonumber \\
			&&\hspace{2.5in} + ~\frac{1}{2} \sum_{s_A} \sum_{s_B \neq s_A} \max_z \sum_e P_{S_A,S_B,Q,E|Z}^\mathrm{real|pass} (s_A,s_B,q,e|z) \\
			&\stackrel{(I)}{\geq}& \frac{1}{2} \sum_{s_A,q} \max_z   \left| \sum_e  P_{S_A,S_B,Q,E|Z}^\mathrm{real|pass} (s_A,s_A,q,e|z) -\frac{1}{|S_A|}\sum_{s_A^\prime, s_B^\prime, e}P_{S_A,S_B,Q,E|Z}^\mathrm{real|pass} (s_A^\prime,s_B^\prime,q,e|z) \right|
			\nonumber \\
			&& \hspace{2.5in} + ~\frac{1}{2} \sum_{s_A} \sum_{s_B \neq s_A} \max_z \sum_e P_{S_A,S_B,Q,E|Z}^\mathrm{real|pass} (s_A,s_B,q,e|z) \\
			&\stackrel{(II)}{\geq}& \frac{1}{2} \sum_{s_A,q}    \left| \sum_e  P_{S_A,S_B,Q,E|Z}^\mathrm{real|pass} (s_A,s_A,q,e|z) -\frac{1}{|S_A|}\sum_{s_A^\prime, s_B^\prime, e}P_{S_A,S_B,Q,E|Z}^\mathrm{real|pass} (s_A^\prime,s_B^\prime,q,e|z) \right|
			\nonumber \\
			&& \hspace{2.8in} +~ \frac{1}{2} \sum_{s_A} \sum_{s_B \neq s_A}  \sum_e P_{S_A,S_B,Q,E|Z}^\mathrm{real|pass} (s_A,s_B,q,e|z) \\
			&\stackrel{(III)}{\geq}& \frac{1}{2}    \left| \sum_{s_A,q} \left( \sum_e P_{S_A,S_B,Q,E|Z}^\mathrm{real|pass} (s_A,s_A,q,e|z) -\frac{1}{|S_A|}\sum_{s_A^\prime, s_B^\prime, e}P_{S_A,S_B,Q,E|Z}^\mathrm{real|pass} (s_A^\prime,s_B^\prime,q,e|z)\right) \right|
			\nonumber \\
			&& \hspace{3.8in} + ~\frac 12\mathrm{P}\left[S_A\neq S_B|\mathrm{pass}\right] \\
			&=& \frac 12\left| \sum_{s_A,q,e} P_{S_A,S_B,Q,E|Z}^\mathrm{real|pass} (s_A,s_A,q,e|z) -  \sum_{s_A} \frac{1}{|S_A|} \sum_{s_A^\prime, s_B^\prime, q, e}P_{S_A,S_B,Q,E|Z}^\mathrm{real|pass} (s_A^\prime,s_B^\prime,q,e|z) \right| \nonumber \\
			&&\hspace{3.8in}  + ~\frac 12\mathrm{P}\left[S_A\neq S_B|\mathrm{pass}\right] \\
			&=& \frac 12 \sum_{s_A^\prime} \sum_{s_B^\prime \neq s_A^\prime}  \sum_{e, q} P_{S_A,S_B,Q,E|Z}^\mathrm{real|pass} (s_A^\prime, s_B^\prime, q,e|z) + ~\frac 12\mathrm{P}\left[S_A\neq S_B|\mathrm{pass}\right] \\
			&=& \mathrm{P}\left[S_A\neq S_B|\mathrm{pass}\right],
			\een
			
			where we used the triangle inequality used in (I) and (III), and the non-signaling condition in the Eve's subsystems used in (II). Hence:
			\ben
			\left(1-\mathrm{p}_\mathrm{abort}\right) \norm{ P_{S_A,S_B,Q,E|Z}^\mathrm{real|pass}-P_{S_A,S_B,Q,E|Z}^\mathrm{ideal|pass}}_\mathrm{NS}\ge \left(1-\mathrm{p}_\mathrm{abort} \right) \mathrm{P}\left[S_A \neq S_B \right| \mathrm{pass}].
			\een 
			\noindent The above inequality verifies that $\varepsilon$-security implies $\varepsilon$-correctness.
			
			In the next step we prove:
			\ben
			\left(1-\mathrm{p}_\mathrm{abort}\right)
			\norm{ P_{S_A,S_B,Q,E|Z}^\mathrm{real|pass}-P_{S_A,S_B,Q,E|Z}^\mathrm{ideal|pass}}_\mathrm{NS} \ge 
			\left(1-\mathrm{p}_\mathrm{abort}\right)
			\norm{ P_{S_A,Q,E|Z}^\mathrm{real|pass}-P_{S_A,Q,E|Z}^\mathrm{ideal|pass}}_\mathrm{NS}.
			\label{proof:impli2:eqn2}
			\een
			
			Let us use Proposition \ref{cor:box_norm_exp} again.
			\ben
			&&	\norm{ P_{S_A,S_B,Q,E|Z}^\mathrm{real|pass}-P_{S_A,S_B,Q,E|Z}^\mathrm{ideal|pass}}_\mathrm{NS} \\ 
			&=& \frac{1}{2} \sum_{s_A,s_B,q} \max_z \sum_e \left|P_{S_A,S_B,Q,E|Z}^\mathrm{real|pass} (s_A,s_B,q,e|z)-\frac{\delta_{s_A,s_B}}{|S_A|}\sum_{s_A^\prime,s_B^\prime}P_{S_A,S_B,Q,E|Z}^\mathrm{real|pass} (s_A^\prime,s_B^\prime,q,e|z) \right| \\
			&=&	\frac{1}{2} \sum_{s_A,q} \max_z \sum_e  \left|P_{S_A,S_B,Q,E|Z}^\mathrm{real|pass} (s_A,s_A,q,e|z)-\frac{1}{|S_A|}\sum_{s_A^\prime,s_B^\prime}P_{S_A,S_B,Q,E|Z}^\mathrm{real|pass} (s_A^\prime,s_B^\prime,q,e|z) \right| \nonumber \\
			&& \hspace{2.8in} +\frac{1}{2} \sum_{s_A} \sum_{s_B \neq s_A} \max_z \sum_e P_{S_A,S_B,Q,E|Z}^\mathrm{real|pass} (s_A,s_B,q,e|z) \\
			&\stackrel{(I)}{=}&
			\frac{1}{2} \sum_{s_A,q} \max_z \sum_e  \left|P_{S_A,S_B,Q,E|Z}^\mathrm{real|pass} (s_A,s_A,q,e|z)-\frac{1}{|S_A|}\sum_{s_A^\prime,s_B^\prime}P_{S_A,S_B,Q,E|Z}^\mathrm{real|pass} (s_A^\prime,s_B^\prime,q,e|z) \right| \nonumber \\
			&& \hspace{4.5in} +\frac{1}{2}\mathrm{P} \left[S_A \neq S_B | \mathrm{pass}\right] \\
			&=&		\frac{1}{2} \sum_{s_A,q} \max_z \sum_e  \left| \bigg(P_{S_A,S_B,Q,E|Z}^\mathrm{real|pass} (s_A,s_A,q,e|z) - \sum_{s_B}P_{S_A,S_B,Q,E|Z}^\mathrm{real|pass} (s_A,s_B,q,e|z)  \bigg) \right. \nonumber \\ 
			&& \left.	\hspace{0.2in}+ \bigg(\sum_{s_B}P_{S_A,S_B,Q,E|Z}^\mathrm{real|pass} (s_A,s_B,q,e|z)  -\frac{1}{|S_A|}\sum_{s_A^\prime,s_B^\prime}P_{S_A,S_B,Q,E|Z}^\mathrm{real|pass} (s_A^\prime,s_B^\prime,q,e|z)\bigg) \right| +\frac{1}{2}\mathrm{P} \left[S_A \neq S_B | \mathrm{pass}\right] \\
			&\stackrel{(II)}{\ge}&
			\frac{1}{2} \sum_{s_A,q} \max_z \sum_e  \left| \bigg|  P_{S_A,S_B,Q,E|Z}^\mathrm{real|pass} (s_A,s_A,q,e|z) - \sum_{s_B}P_{S_A,S_B,Q,E|Z}^\mathrm{real|pass} (s_A,s_B,q,e|z)\bigg|  \right. \nonumber \\ 
			&& \hspace{0.3in} \left. - \bigg| \sum_{s_B}P_{S_A,S_B,Q,E|Z}^\mathrm{real|pass} (s_A,s_B,q,e|z)  -\frac{1}{|S_A|}\sum_{s_A^\prime,s_B^\prime}P_{S_A,S_B,Q,E|Z}^\mathrm{real|pass} (s_A^\prime,s_B^\prime,q,e|z) \bigg| \right| 
			+\frac{1}{2}\mathrm{P} \left[S_A \neq S_B | \mathrm{pass}\right] \\
			&\stackrel{(III)}{\ge}&
			\frac{1}{2}   \left| \sum_{s_A,q} \max_z \sum_e \bigg|  P_{S_A,S_B,Q,E|Z}^\mathrm{real|pass} (s_A,s_A,q,e|z) - \sum_{s_B}P_{S_A,S_B,Q,E|Z}^\mathrm{real|pass} (s_A,s_B,q,e|z)\bigg|  \right. \nonumber \\ && \left. 
			\hspace{-0.25in} -  \sum_{s_A,q} \max_z \sum_e  \bigg| \sum_{s_B}P_{S_A,S_B,Q,E|Z}^\mathrm{real|pass} (s_A,s_B,q,e|z)  -\frac{1}{|S_A|}\sum_{s_A^\prime,s_B^\prime}P_{S_A,S_B,Q,E|Z}^\mathrm{real|pass} (s_A^\prime,s_B^\prime,q,e|z) \bigg| \right| 
			+\frac{1}{2}\mathrm{P} \left[S_A \neq S_B | \mathrm{pass}\right]   \nonumber \\ \\
			&\stackrel{(IV)}{=}	&
			\left| \frac{1}{2}  \mathrm{P} \left[S_A \neq S_B | \mathrm{pass}\right]	 
			- \norm{ P_{S_A,Q,E|Z}^\mathrm{real|pass}-P_{S_A,Q,E|Z}^\mathrm{ideal|pass}}_\mathrm{NS} \right|
			+\frac{1}{2}\mathrm{P} \left[S_A \neq S_B | \mathrm{pass}\right],
			\een
			where in (I) the second component is treated like in the previous step, reverse triangle inequality has been used in (II), triangle inequality in (III) and in (IV) we  use the results given in Eqs. (\ref{eq:correct}) and (\ref{eq:secret}). We have:
			\ben
			&&\left(1-\mathrm{p}_\mathrm{abort}\right) \norm{ P_{S_A,S_B,Q,E|Z}^\mathrm{real|pass}-P_{S_A,S_B,Q,E|Z}^\mathrm{ideal|pass}}_\mathrm{NS}\\
			&&\ge
			\left| \frac{1}{2}  \left(1-\mathrm{p}_\mathrm{abort}\right) \mathrm{P} \left[S_A \neq S_B | \mathrm{pass}\right]	 
			- \left(1-\mathrm{p}_\mathrm{abort}\right) \norm{ P_{S_A,Q,E|Z}^\mathrm{real|pass}-P_{S_A,Q,E|Z}^\mathrm{ideal|pass}}_\mathrm{NS} \right| \nonumber \\
			&&+\frac{1}{2} \left(1-\mathrm{p}_\mathrm{abort}\right) \mathrm{P} \left[S_A \neq S_B |\mathrm{pass}\right] ~~~~~
			\een
			One should now go through two separate cases:\\
			
			Case 1.  
			$\left(\frac{1}{2}  \mathrm{P} \left[S_A \neq S_B |\mathrm{pass}\right]	 
			\ge  \norm{ P_{S_A,Q,E|Z}^\mathrm{real|pass}-P_{S_A,Q,E|Z}^\mathrm{ideal|pass}}_\mathrm{NS}\right)$: \\
			\ben
			&&\left(1-\mathrm{p}_\mathrm{abort}\right) \norm{ P_{S_A,S_B,Q,E|Z}^\mathrm{real|pass}-P_{S_A,S_B,Q,E|Z}^\mathrm{ideal|pass}}_\mathrm{NS} \ge  
			\frac{1}{2}  \left(1-\mathrm{p}_\mathrm{abort}\right) \mathrm{P} \left[S_A \neq S_B |\mathrm{pass}\right] \nonumber \\	
			&&- \left(1-\mathrm{p}_\mathrm{abort}\right) \norm{ P_{S_A,Q,E|Z}^\mathrm{real|pass}-P_{S_A,Q,E|Z}^\mathrm{ideal|pass}}_\mathrm{NS} +\frac{1}{2} \left(1-\mathrm{p}_\mathrm{abort}\right) \mathrm{P} \left[S_A \neq S_B |\mathrm{pass}\right] \\
			&&=
			\left(1-\mathrm{p}_\mathrm{abort}\right) \mathrm{P} \left[S_A \neq S_B |\mathrm{pass}\right]	 
			- \left(1-\mathrm{p}_\mathrm{abort}\right) \norm{P_{S_A,Q,E|Z}^\mathrm{real|pass}-P_{S_A,Q,E|Z}^\mathrm{ideal|pass}}_\mathrm{NS} \\
			&& \ge 
			2 \left(1-\mathrm{p}_\mathrm{abort}\right) \norm{ P_{S_A,Q,E|Z}^\mathrm{real|pass}-P_{S_A,Q,E|Z}^\mathrm{ideal|pass}}_\mathrm{NS}-
			\left(1-\mathrm{p}_\mathrm{abort}\right) \norm{ P_{S_A,Q,E|Z}^\mathrm{real|pass}-P_{S_A,Q,E|Z}^\mathrm{ideal|pass}}_\mathrm{NS} \\
			&&\ge
			\left(1-\mathrm{p}_\mathrm{abort}\right) \norm{ P_{S_A,Q,E|Z}^\mathrm{real|pass}-P_{S_A,Q,E|Z}^\mathrm{ideal|pass}}_\mathrm{NS}
			\een
			\\

			Case 2. $\left(\frac{1}{2}  \mathrm{P} \left[S_A \neq S_B |\mathrm{pass}\right]	 
			<  \norm{ P_{S_A,Q,E|Z}^\mathrm{real|pass}-P_{S_A,Q,E|Z}^\mathrm{ideal|pass}}_\mathrm{NS}\right)$:
			\ben
			&& \left(1-\mathrm{p}_\mathrm{abort}\right) \norm{ P_{S_A,S_B,Q,E|Z}^\mathrm{real|pass}-P_{S_A,S_B,Q,E|Z}^\mathrm{ideal|pass}}_\mathrm{NS} 
			\ge 
			\left(1-\mathrm{p}_\mathrm{abort}\right) \norm{ P_{S_A,Q,E|Z}^\mathrm{real|pass}-P_{S_A,Q,E|Z}^\mathrm{ideal|pass}}_\mathrm{NS}\\ \nonumber
			&&- \frac{1}{2} \left(1-\mathrm{p}_\mathrm{abort}\right) \mathrm{P} \left[S_A \neq S_B |\mathrm{pass}\right]	 
			+\frac{1}{2}\left(1-\mathrm{p}_\mathrm{abort}\right) \mathrm{P} \left[S_A \neq S_B |\mathrm{pass}\right]\\
			&&=
			\left(1-\mathrm{p}_\mathrm{abort}\right) \norm{ P_{S_A,Q,E|Z}^\mathrm{real|pass}-P_{S_A,Q,E|Z}^\mathrm{ideal|pass}}_\mathrm{NS}
			\een
			
			Finally:
			\ben
			\left(1-\mathrm{p}_\mathrm{abort}\right) \norm{ P_{S_A,S_B,Q,E|Z}^\mathrm{real|pass}-P_{S_A,S_B,Q,E|Z}^\mathrm{ideal|pass}}_\mathrm{NS} \ge 
			\left(1-\mathrm{p}_\mathrm{abort}\right) \norm{ P_{S_A,Q,E|Z}^\mathrm{real|pass}-P_{S_A,Q,E|Z}^\mathrm{ideal|pass}}_\mathrm{NS}
			\een
			
			If protocol is $\varepsilon$-secure we see from (\ref{proof:impli2:eqn1}) and (\ref{proof:impli2:eqn2}) that:
			\ben
			&& \left(1-\mathrm{p}_\mathrm{abort}\right) \norm{ P_{S_A,S_B,Q,E|Z}^\mathrm{real|pass}-P_{S_A,S_B,Q,E|Z}^\mathrm{ideal|pass}}_\mathrm{NS}
			\le \varepsilon  \\ \nonumber
			\implies &&	\left\{\left(1-\mathrm{p}_\mathrm{abort}\right)\mathrm{P}\left[S_A \neq S_B|pass\right] \le \varepsilon ~~ \mathrm{and} ~~ \left(1-\mathrm{p}_\mathrm{abort}\right)\norm{ P_{S_A,Q,E|Z}^\mathrm{real|pass}-P_{S_A,Q,E|Z}^\mathrm{ideal|pass}}_\mathrm{NS} \le \varepsilon \right\}.
			\een
		\end{proof}

		Once we proved the above Lemmas, we can state the Theorem regarding the equivalence between the secrecy and correctness and proximity in NS norm criteria of security for a protocol we have considered. 
		\begin{thm3}[Equivalence of security criteria]
			For an MDLOPC protocol $\Lambda$, the proximity in the NS norm security criterion is equivalent to the criterion based on security and correctness. That is for any $\varepsilon_\mathrm{sec}+\varepsilon_\mathrm{cor} \equiv \varepsilon  \ge \varepsilon_\mathrm{sec},\varepsilon_\mathrm{cor} \ge 0$ the following equivalence relation holds:
			\ben \label{eqn:security}
			&&\left(1-\mathrm{p}_\mathrm{abort}\right) \norm{ P_{S_A,S_B,Q,E|Z}^\mathrm{real|pass}-P_{S_A,S_B,Q,E|Z}^\mathrm{ideal|pass}}_\mathrm{NS} \le O(\varepsilon) ~\Longleftrightarrow~ 
			\left\{
			\left(1-\mathrm{p}_\mathrm{abort}\right)\mathrm{P}\left[S_A \neq S_B|\mathrm{pass} \right] \le O(\varepsilon_{\mathrm{cor}}) \right. \nonumber
			\\&&\left.  ~\wedge~ 
			\left(1-\mathrm{p}_\mathrm{abort}\right) \norm{ P_{S_A,Q,E|Z}^\mathrm{real|pass}-P_{S_A,Q,E|Z}^\mathrm{ideal|pass}}_\mathrm{NS}
			\le O(\varepsilon_{\mathrm{sec}}) \right\},
			\een
			where $\mathrm{p}_\mathrm{abort}$ is the probability for the protocol to abort and the constant in $O(\varepsilon)$ does not depend on any parameter of the protocol. 
		\end{thm3}
		
		\begin{proof}
			From Lemma \ref{lemma:implication1} we have:
			\ben
			&&\left\{ \left(1-\mathrm{p}_\mathrm{abort}\right) \norm{P_{S_A,Q,E|Z}^\mathrm{real|pass}-P_{S_A,Q,E|Z}^\mathrm{ideal|pass}}_\mathrm{NS} \le \varepsilon_\mathrm{sec} ~~\mathrm{and}~~  \left(1-\mathrm{p}_\mathrm{abort}\right) \mathrm{P}\left[S_A \neq S_B|\mathrm{pass}\right] \le \varepsilon_\mathrm{cor} \right\} \nonumber
			\\
			&& \implies \left(1-\mathrm{p}_\mathrm{abort}\right) \norm{P_{S_A,S_B,Q,E|Z}^\mathrm{real|pass},P_{S_A,S_B,Q,E|Z}^\mathrm{ideal|pass} }_\mathrm{NS} \le \varepsilon_\mathrm{sec}+\varepsilon_\mathrm{cor} = \varepsilon.
			\een
			and from Lemma \ref{lemma:implication2}:
			\ben
			&&
			\left(1-\mathrm{p}_\mathrm{abort}\right) \norm{ P_{S_A,S_B,Q,E|Z}^\mathrm{real|pass}-P_{S_A,S_B,Q,E|Z}^\mathrm{ideal|pass}}_\mathrm{NS}
			\le \varepsilon \nonumber \\
			\implies 	&& \left\{ \left(1-\mathrm{p}_\mathrm{abort}\right) \mathrm{P}\left[S_A \neq S_B|pass\right] \le \varepsilon ~~ \mathrm{and}~~ \left(1-\mathrm{p}_\mathrm{abort}\right)\norm{P_{S_A,Q,E|Z}^\mathrm{real|pass}-P_{S_A,Q,E|Z}^\mathrm{ideal|pass} }_\mathrm{NS} \le \varepsilon \right\}
			\een
			By combining the above implications under $\varepsilon_\mathrm{sec}+\varepsilon_\mathrm{cor} \equiv \varepsilon  \ge \varepsilon_\mathrm{sec},\varepsilon_\mathrm{cor} \ge 0$ constraints, we obtain: 
			\ben 
			&&\left(1-\mathrm{p}_\mathrm{abort}\right) \norm{ P_{S_A,S_B,Q,E|Z}^\mathrm{real|pass}-P_{S_A,S_B,Q,E|Z}^\mathrm{ideal|pass}}_\mathrm{NS} \le O(\varepsilon) ~\Longleftrightarrow~ 
			\left\{
			\left(1-\mathrm{p}_\mathrm{abort}\right)\mathrm{P}\left[S_A \neq S_B|\mathrm{pass} \right] \le O(\varepsilon_{\mathrm{cor}}) \right.
			\\&&\left.  ~\wedge~ 
			\left(1-\mathrm{p}_\mathrm{abort}\right)\left(1-\mathrm{p}_\mathrm{abort}\right) \norm{ P_{S_A,Q,E|Z}^\mathrm{real|pass}-P_{S_A,Q,E|Z}^\mathrm{ideal|pass}}_\mathrm{NS}
			\le O(\varepsilon_{\mathrm{sec}}) \right\},
			\een		
			hence the corresponding notion's are cryptographically equivalent.
		\end{proof}
		
		
		\begin{rem}
			In the rest of this article, we assume that the protocol is after the acceptance phase. However, for the full generality in this section, we took a step back and also considered the possibility of aborting. From now, we set the probability of aborting to zero. 
		\end{rem}

		\section{Rephrasing the key rate in the secret key agreement scenario}\label{sec:Maurer-rate}
		
		The secret key agreement (SKA) scenario is a cryptographic scheme in which the honest parties and the eavesdropper share many copies of a classical joint probability distribution $P(ABE)$ \cite{CsisarKorner_key_agreement,Maurer93}. The honest parties task is to agree on the secret key, by employing local operations and public communication (LOPC), in such a manner that the eavesdropper's knowledge about the key is negligible. In the following lines, we propose an alternative definition of the secret key rate $\mathrm{S}(A:B||E)$ in the aforementioned scenario and prove that the definition we propose is equivalent to those present in the literature \cite{Maurer93,MaurerWolf00CK,reduced-intrinsic}. This technical result intends to show and utilize a connection between the definition of secret key rate in SKA and NSDI scenarios, as it was done in the case of quantum cryptography \cite{Christandl12}.
		
		Before we begin with the proof of Theorem \ref{thm:SK_key0}, let us {recall} two definitions of secret key rate in SKA scenario \cite{Maurer93,MaurerWolf00CK}.   
		\begin{definition}[The weak secret key rate \cite{Maurer93,MaurerWolf00CK}]\label{def:Maurer93}
			The {\it(weak)} secret key rate of A and B with respect to E, denoted ${\overbar{S}}(A:B||Z)$, is the maximal $R \ge 0$ such that for every $\varepsilon >0$ and for all $N \ge N_0 (\varepsilon)$ there exists a protocol, using public communication over an insecure but an authenticated channel, such that Alice and Bob , who receive $A^N=[A_1,...,A_N]$ and $B^N=[B_1,...,B_N]$, can compute keys $S_A$ and $S_B$, respectively, with the following properties. First, $S_A=S_B$ hold with probability at least $1-\varepsilon$, and second,
			\begin{align}\label{eqn:M1}
			\frac 1N I(S_A:CE^N) \le \varepsilon~~~~\mathrm{and}~~~~ \frac 1N H(S_A) \ge R-\varepsilon
			\end{align} 
			hold. Here, C denotes the collection of messages sent over the insecure channel by Alice and Bob.
		\end{definition}
		\begin{definition}[The strong secret key rate \cite{MaurerWolf00CK}]\label{def:Maurer2000}
			The {\it strong secret key rate} of A and B with respect to E, denoted by $\overbar{\overbar{S}}(A:B||Z)$, is defined in the same way as ${\overbar{S}}(A:B||Z)$ with the modifications that Alice and Bob compute strings $S_A$ and $S_B$ which are with probability at least $1-\varepsilon$ both equal to a string S with the properties  
			\begin{align}\label{eqn:M2}
			I(S:CE^N) \le \varepsilon~~~~~\mathrm{and}~~~~ H(S)=\log \abs{\mathcal{S}} \ge N\cdot (R - \varepsilon).
			\end{align}
		\end{definition}
		The above definitions of secret key rate were proven to be equivalent \cite{MaurerWolf00CK}, i.e., $\overbar{\overbar{S}}(A:B||Z)={\overbar{S}}(A:B||Z)$, for every distribution $P(ABE)$ shared between the parties before the protocol. We propose an alternative definition of the secret key rate based on proximity in the trace distance (total variational distance).
		\begin{definition}[The secret key rate] \label{def:SKAour}
			Let $P(ABE)$ be the joint distribution of three discrete random variables A, B, and E. The secret key rate ${S} ( A : B||E)$ is given by
			\begin{align}\label{eqn:statement}
			{S} ( A : B||E)_{P(ABE)} :=
			\sup_{\cal P} \limsup_{N\rightarrow \infty} \frac{\log \mathrm{dim}_\mathrm{S_A} \left( \mathcal{P}_N\left({P}^{\ot N}(ABE)\right)\right)}{N},
			\end{align}
			where $\mathcal{P}=\cup_{N=1}^{\infty} \left\{\mathcal{P}_N\right\}$ is a LOPC protocol that satisfies
			\begin{align}
			\norm{ P_N^\mathrm{real} - P_N^\mathrm{ideal}}_1 \le \delta_N \stackrel{N\to \infty}{\longrightarrow}0, \label{eqn:Appoach0}
			\end{align}
			for
			\begin{align}
			&P^\mathrm{real}_N  \equiv P^\mathrm{real}_N (S_AS_B CE^N) :={\cal P}_N\left(P^{\ot N}(ABE)\right),\\
			&P^\mathrm{ideal}_N \equiv P^\mathrm{ideal}_N (S_AS_B CE^N) := \left(\frac{\delta_{s_A,s_B}}{|S_A|}\right) \otimes\sum_{s_A,s_B} P^\mathrm{real}_N (S_A=s_A,S_B=s_B, CE^N).
			\end{align}
		\end{definition}

		\begin{thm2}\label{thm:SK_key} 
			The secret key rate ${S} ( A : B||E)$ introduced in Definition \ref{def:SKAour} is equal to secret key rates $\overline{S} ( A : B||E)$ and $\overbar{\overbar{S}}(A:B||Z)$ provided in Definitions \ref{def:Maurer93} and \ref{def:Maurer2000}, respectively.
		\end{thm2}

		Before we show the proof of Theorem \ref{thm:SK_key0}, we present the basic tools that will be used. For two joint probability distributions $P\equiv P(XY)$ and $Q \equiv Q(XY)$, that are close by according to the trace distance, their Shannon entropies, and the mutual information functions satisfy the asymptotic continuity relations \cite{shirokov,Alicki-Fannes}, which is 
		\ben
		\left|\mathrm{H}(X)_{P}-\mathrm{H}(X)_{Q}\right| \le \epsilon \log  \left(\mathrm{dim_X} \left( P\right) -1 \right) + {h}_2 (\epsilon) , \label{ac:entropy}\\
		\left|\mathrm{I}(X:Y)_{P}-\mathrm{I}(X:Y)_{Q}\right| \le 2 \epsilon \log  d + 2g(\epsilon),\label{ac:mutual}
		\een
		where  $\epsilon=\frac{1}{2}\left|\left| {P(XY)}-{Q(XY)}\right|\right|_1 \in [0,1]$, ${h}_2\left( \epsilon \right):=-\epsilon \log \epsilon-\left(1-\epsilon\right)\log \left(1-\epsilon\right)$ is the binary Shannon entropy, ${g}  \left( \epsilon \right) := -\epsilon \log \epsilon+\left(1+\epsilon\right)\log \left(1+\epsilon\right)$, and $d = \text{min}\{ \mathrm{dim_X} \left( P\right), \mathrm{dim_Y} \left( P \right)\}$. Functions $h_2$ and $g$ are equal at $\epsilon=0$ and for $\epsilon > 0$ $h_2(\epsilon) < g(\epsilon)$. It is also useful to observe that $\left|\left| {P(X)}-{Q(X)}\right|\right|_1 \leq \left|\left| {P(XY)}-{Q(XY)}\right|\right|_1$ for $P(X)$ and $Q(X)$ being marginal probability distributions of $P(XY)$ and $Q(XY)$ respectively.
		
		Another relation that we need is the so-called Pinsker's inequality. It states that if $P$ and $Q$ are two probability distributions, then
		\begin{align}
		\frac 12 \norm{P-Q}_1 \le \sqrt{\frac 12 D_\mathrm{KL} (P||Q)},
		\end{align}
		where $D_\mathrm{KL} (P||Q)$ is the Kullback–Leibler divergence. One of the properties of this function is its relation to mutual information, i.e., for a joint probability distribution $P(XY)$ and $P(X)$, $P(Y)$ being its marginal distributions we have: $D_\mathrm{KL} (P(XY)||P(X)P(Y)) = I(X:Y)_{P(XY)}$.
		
		The last mathematical property we describe before the proof is the Fano's inequality stating that:
		\begin{align}
		&H(X|Y) \le h(e) + P(e) \log \left(\abs{X}-1\right), \nonumber\\
		&P(e)=\mathrm{Prob}\left[X \neq \tilde{X}\right],
		\end{align}
		where $h(x)$ is the binary entropy and $\tilde{X} =f(Y)$ is an approximate version of $X$.
		
		In the proof, we also use the notions of real and ideal systems. The real system $P^\mathrm{real}_N$ is a tripartite probability distribution shared by the honest parties after $N$-th round of an LOPC protocol $\mathcal{P}$. The ideal system $P^\mathrm{ideal}_N$ is the one in which the honest parties are perfectly correlated (with uniform distribution), and Eve's marginal distribution is the same as for the real system, however completely uncorrelated with the honest parties.
		\begin{align}
		&P^\mathrm{real}_N  \equiv P^\mathrm{real}_N (S_AS_B CE^N) :={\cal P}_N\left(P^{\ot N}(ABE)\right),\\
		&P^\mathrm{ideal}_N \equiv P^\mathrm{ideal}_N (S_AS_B CE^N) := \left(\frac{\delta_{s_A,s_B}}{|S_A|}\right) \otimes\sum_{s_A,s_B} P^\mathrm{real}_N (S_A=s_A,S_B=s_B, CE^N),
		\label{eq:defReal}
		\end{align}
		where $P(ABE)$ is tripartite probability distribution shared by all parties at the beginning of SKA protocol, i.e., input state of the protocol, $\abs{S}=\dim_S \left(P^\mathrm{real}_N \right)$ and dimensions of $P^\mathrm{real}_N$ and $P^\mathrm{deal}_N$ are equal. By $\left(\frac{\delta_{s_A,s_B}}{|S_A|}\right)$ we denote a distribution of perfectly and uniformly correlated random variables $S_A$ and $S_B$.

		\begin{proof}[Proof of Theorem \ref{thm:SK_key0}]
			We begin the proof by showing that the weak secret key rate ${\overbar{S}}(A:B||Z)$, constitutes an upper bound on $S(A:B||Z)$. We do this by showing that every protocol that satisfies the condition in Eq. (\ref{eqn:Appoach0}) also satisfies conditions in Definition \ref{def:Maurer93}. 
			
			We denote protocol that satisfy security condition in Eq. (\ref{eqn:Appoach0}) with $\mathcal{P}$. From asymptotic continuity of the mutual information and the fact that $I(S_A:CE^N)_{P^\mathrm{ideal}_N}=0$ by the construction of $P^\mathrm{ideal}_N$, we read
			\begin{align}
			\forall_\mathcal{P}\forall_N~~
			&I(S_A:CE^N)_{{P^\mathrm{real}_N}}=I(S_A:CE^N)_{P^\mathrm{real}_N}-I(S_A:CE^N)_{P^\mathrm{ideal}_N} \nonumber \\
			&\le \abs{I(S_A:CE^N)_{P^\mathrm{real}_N}-I(S_A:CE^N)_{P^\mathrm{ideal}_N} } \le 2 \delta_N \log  d_{S_A} + 2g(\delta_N).
			\end{align}
			Where $d_{S_A}:=\dim_{S_A} \left(P^\mathrm{real}_N\right) \ge \min\left\{\dim_{S_A} \left(P^\mathrm{real}_N\right),\dim_{CE^N} \left(P^\mathrm{real}_N\right)\right\}$ and $\delta_N \ge \frac 12 \norm{P^\mathrm{real}_N-P^\mathrm{ideal}_N}_1$. Because in any reasonable LOPC protocol dimension of the output is smaller than the dimension of the input, and we observe that\footnote{This follows from: $\mathrm{S}(A:B||E)_P \le I(A:B\downarrow E)_P \le \log \dim_A \left(P\right)$.}
			\begin{align}\label{eqn:realOT}
			\forall_\mathcal{P}\forall_N~~ d_S=\dim_{S_A} \left(P^\mathrm{real}_N\right) = \dim_{S_A} \left({\cal P}_N \left(\left(P(ABE)\right)^{\ot N}\right) \right) \le  \dim_A \left( \left(P(ABE)\right)^{\ot N} \right) = \left(\dim_A  \left(P(ABE)\right) \right)^N .
			\end{align}
			Hence,
			\begin{align}
			\forall_\mathcal{P}\forall_N~~\frac{1}{N}I(S_A:CE^N)_{P^\mathrm{real}_N} \le  \frac{2 \delta_N \log  \left(\dim_A  \left(P(ABE)\right) \right)^N + 2g(\delta_N)}{N}=2 \delta_N \log  \left(\dim_A  \left(P(ABE)\right) \right)+ \frac{2g(\delta_N)}{N} 
			\label{eqn:con1}
			\end{align}
			Hence if a protocol satisfies the trace norm security condition $\norm{\mathcal{P}_N\left({P}^{\ot N}\left(ABE\right)\right) - P_N^\mathrm{ideal}}_1 \le \delta_N \stackrel{N\to \infty}{\longrightarrow}0$ then 
			\begin{equation}
			\forall_{\mathcal{P}}\forall_{\varepsilon >0} \exists_{N_1(\varepsilon)} \forall_{N \ge N_1 (\varepsilon)} ~~ \frac{1}{N}I(S_A:CE^N)_{P^\mathrm{real}_N}   <\varepsilon,
			\end{equation}
			as r.h.s. of Eq. (\ref{eqn:con1}) approaches $0$ when $N$ goes to infinity. 
			
			Another condition in Definition \ref{def:Maurer93} we call correctness of a protocol, requiring that $S_A=S_B$ with probability at least $1-\varepsilon$ (equivalently $\mathrm{Prob}[S_A \neq S_B]\le \varepsilon$) is satisfied\footnote{Devices with unary input are isomorphic with unconditional probability distributions.} by virtue of Theorem \ref{thm:equivalence}, with $|Z|=1$ and $\mathrm{p}_\mathrm{abort}=0$.
			
			This is because the NS norm computed for classical probability distributions is equal to the trace distance. Therefore from the condition in Eq. (\ref{eqn:Appoach0}) and Theorem \ref{thm:equivalence} we have
			\begin{align}
			\forall_{\mathcal{P}} \forall_\varepsilon \exists_{N_2 (\varepsilon)} \forall_{N\ge N_2 (\varepsilon)} ~~\mathrm{Prob}[S_A \neq S_B] \le \delta_N.
			\end{align}
			
			Let us show now the upper bound. We first observe that for all protocols the following is true. 
			\begin{align}
			\forall_{\mathcal{P}} \forall_{N}~~ H(S_A)_{P^\mathrm{real}_N} \le H(S_A)_{P^\mathrm{ideal}_N} = \log \dim_{S_A} \left(P^\mathrm{ideal}_N \right) = \log \dim_{S_A} \left(P^\mathrm{real}_N \right),
			\label{eq:hahahaha}
			\end{align}
			where the inequality is due to the definition of ideal system in which $S_A$ is uniformly distributed and of the same dimension as in real system. From asymptotic continuity of the Shannon entropy we have:	
			\begin{align}
			&\forall_{\mathcal{P}} \forall_{N}~~ \frac 1N H(S_A)_{P^\mathrm{real}_N} \ge \frac 1NH(S_A)_{P^\mathrm{ideal}_N}- \frac 1N\left(\delta_N \log  \left(\mathrm{dim_{S_A}} \left( P^\mathrm{real}_N \right) -1 \right) + {h}_2 (\delta_N)\right)\\
			& \ge \frac 1NH(S_A)_{P^\mathrm{ideal}_N}-\left(2 \delta_N \log  \left(\dim_A  \left(P(ABE)\right) \right)+\frac{2g(\delta_N)}{N}\right),
			\end{align}
			where the second inequality is a consequence of the similar arguments as in Eq. (\ref{eqn:realOT}) and the fact that $\forall_{x >0}~~h_2 (s) < g(x)$.
			
			
			{Let us define $L(N):=\frac{ \log \dim_{S_A} \left(P^\mathrm{real}_N \right)}{N}$. In particular there exists $0<\eta(N)\stackrel{N \to \infty}{\longrightarrow}0$ such that $L(N)= \limsup_{N \to \infty} \frac{ \log \dim_{S_A} \left(P^\mathrm{real}_N \right)}{N} - \eta(N)$. Hence, we have the following inequality.}
			\begin{align}
			\forall_{\mathcal{P}} \forall_{\varepsilon >0} \exists_{N_3 (\varepsilon)}\forall_{N>N_3(\varepsilon)}~~ \frac{1}{N}H(S_A)_{P^\mathrm{real}_N} \ge \limsup_{N \to \infty} \frac{ \log \dim_{S_A} \left(P^\mathrm{real}_N \right)}{N}- \varepsilon
			\label{eq:hahahaha2}
			\end{align}
			Let us define now $N_0(\varepsilon):=\max \left\{N_1(\varepsilon),N_2(\varepsilon),N_3(\varepsilon)\right\}$. All conditions in Definition \ref{def:Maurer93}, are now satisfied as for all $\varepsilon>0$ and for all $N \ge N_0(\varepsilon)$, with $R= \limsup_{N \to \infty} \frac{ \log \dim_{S_A} \left(P^\mathrm{real}_N \right)}{N}$. The weak secret key rate is by Definition \ref{def:Maurer93} maximal $R$, for which second inequality in Eq. (\ref{eqn:M1}) is satisfied, hence to achieve $\overbar{S}(A:B||E)$ one has to take a supremum over rates of all protocols. 
			\begin{align}
			\overbar{S}(A:B||E)_{P(ABE)} = \sup_{\overbar{\mathcal{P}}} R,
			\end{align}
			where $\overbar{\mathcal{P}}$ are the protocols that satisfy conditions in Definition \ref{def:Maurer93}. As we have shown that condition (\ref{eqn:Appoach0}) in Definition \ref{def:SKAour} implies conditions in Definition \ref{def:Maurer93}, it is clear that $\left\{\mathcal{P}\right\} \subseteq \left\{\overbar{\mathcal{P}}\right\}$ , and hence:
			\begin{align}\label{eqn:Rub}
			\overbar{S}(A:B||E)_{P(ABE)} = \sup_{\overbar{\mathcal{P}}} R \ge \sup_{{\mathcal{P}}} R = \sup_{{\mathcal{P}}}\limsup_{N \to \infty} \frac{ \log \dim_{S_A} \left(P^\mathrm{real}_N \right)}{N}.
			\end{align}

			Let us now show that the secret key rate $S(A:B||E)$ is lower bounded with the strong secret key rate $\overbar{\overbar{S}}(A:B||E)$. In this part, we refer again to results in Section \ref{sec:equivalence-security}. It is enough to show that conditions in Definition \ref{def:Maurer2000} imply secrecy and correctness of a protocol, as by virtue of Theorem \ref{thm:equivalence} and the same arguments regarding the connection between the NS norm and the trace distance, these conditions imply proximity in the trace distance.
			
			We start with the condition of secrecy (see Definition \ref{def:Secrecy}). Let $\overbar{P^\mathrm{real}_N}(S_AS_BSCE^N)$ be an extension of $P^\mathrm{real}_N (S_AS_BCE^N)$, such it satisfies conditions in equations (\ref{eqn:M2}).
			\begin{align}\label{eqn:complicated0}
			\forall_N~~&\varepsilon^\mathrm{sec}_N := \frac 12 \norm{P^\mathrm{real}_N (S_ACE^N) - P^\mathrm{ideal}_N (S_ACE^N)}_1 =\frac 12 \norm{{P^\mathrm{real}_N} (S_ACE^N)-\left(\frac{1}{|S_A|}\right) \otimes {P^\mathrm{real}_N} ( CE^N)}_1\nonumber \\
			&=\frac 12 \norm{{P^\mathrm{real}_N} (S_ACE^N)-{P^\mathrm{real}_N}(S_A)\otimes {P^\mathrm{real}_N}(CE^N)+{P^\mathrm{real}_N}(S_A)\otimes {P^\mathrm{real}_N}(CE^N) -\left(\frac{1}{|S_A|}\right) \otimes {P^\mathrm{real}_N} ( CE^N)}_1\nonumber \\
			&\le \frac 12 \norm{{P^\mathrm{real}_N} (S_ACE^N)-{P^\mathrm{real}_N}(S_A)\otimes {P^\mathrm{real}_N}(CE^N)}_1
			+\frac 12 \norm{{P^\mathrm{real}_N}(S_A) -\left(\frac{1}{|S_A|}\right)}_1 \nonumber \\
			&\le \frac 12 \norm{\overbar{P^\mathrm{real}_N} (S_ASCE^N)-\overbar{P^\mathrm{real}_N}(S_AS)\otimes {P^\mathrm{real}_N}(CE^N)}_1
			+\frac 12 \norm{{P^\mathrm{real}_N}(S_A) -\left(\frac{1}{|S_A|}\right)}_1,
			\end{align}
			where $\left(\frac{1}{|S_A|}\right)$ denotes uniform distribution, and we identify $\abs{S_A}$ with $\abs{\mathcal{S}}$. The first term in the Eq. above can be upper bounded via Pinsker's inequality, and the first inequality in (\ref{eqn:M2}):
			\begin{align}\label{eqn:complicated1}
			& \forall_{\varepsilon >0}\exists_{N_0(\varepsilon)}\forall_{N>N_0(\varepsilon)} \nonumber\\
			&\frac 12 \norm{\overbar{P^\mathrm{real}_N} (S_ASCE^N)-\overbar{P^\mathrm{real}_N}(S_AS)\otimes \overbar{P^\mathrm{real}_N}(CE^N)}_1 \le \sqrt{\frac 12 D_\mathrm{KL}\left(\overbar{P^\mathrm{real}_N} (S_ASCE^N)||\overbar{P^\mathrm{real}_N}(S_AS)\otimes \overbar{P^\mathrm{real}_N}(CE^N)\right)} \nonumber \\
			&\stackrel{(I)}{=}\sqrt{\frac 12 I(S_AS:CE^N)_{\overbar{P^\mathrm{real}_N}}}=\frac{1}{\sqrt{2}} \sqrt{I(S:CE^N)_{\overbar{P^\mathrm{real}_N}} + I(S_A:CE^N|S)_{\overbar{P^\mathrm{real}_N}}} \le
			\frac{1}{\sqrt{2}} \sqrt{\varepsilon + I(S_A:CE^N|S)_{\overbar{P^\mathrm{real}_N}}},
			\end{align}
			where $(I)$ follows from the properties of the Kullback–Leibler divergence. Let us upper bound $I(S_A:CE^N|S)_{\overbar{P^\mathrm{real}_N}}$ in the next step.
			\begin{align}\label{eqn:complicated11}
			&I(S_A:CE^N|S)_{\overbar{P^\mathrm{real}_N}} = H(S_A|S)_{\overbar{P^\mathrm{real}_N}}-H(S_A|S,CE^N)_{\overbar{P^\mathrm{real}_N}}\nonumber \\
			&\stackrel{(II)}{\le} H(S_A|S)_{\overbar{P^\mathrm{real}_N}} \stackrel{(III)}{\le} H(e)_{P(e)} + P(e)  \log\left(\abs{S}-1\right)  
			\stackrel{(IV)}{\le} h(\varepsilon) + \varepsilon\log\left(\abs{S}-1\right),
			\end{align}
			where $h(x)$ is the binary entropy and in $(II)$ we used non-negativity of the conditional entropy, $(III)$ follows from Fano's inequality for $P(e)= \mathrm{Prob}[S \neq S_A]$, and the last step $(IV)$ is a consequence of $\mathrm{Prob}\left[S_A=S_B=S\right] \ge 1-\epsilon$ and an assumption that $\varepsilon \le \frac 12$. This assumption is well justified in cryptography. From inequalities (\ref{eqn:complicated1}) and (\ref{eqn:complicated11}), we have:
			\begin{align}
			\forall_{\frac 12 \ge \varepsilon >0}\exists_{N_0(\varepsilon)}\forall_{N>N_0(\varepsilon)} ~~
			\frac 12 \norm{\overbar{P^\mathrm{real}_N} (S_ASCE^N)-\overbar{P^\mathrm{real}_N}(S_AS)\otimes \overbar{P^\mathrm{real}_N}(CE^N)}_1 \le 
			\frac{1}{\sqrt{2}}\sqrt{\varepsilon + h(\varepsilon) + \varepsilon\log\left(\abs{S}-1\right)}.
			\end{align}
			
			In order to upper bound the second term we make the following observations:
			\begin{align}
			\forall_{\varepsilon >0}\exists_{N_0(\varepsilon)}&\forall_{N>N_0(\varepsilon)} \nonumber\\
			&a)~~ \mathrm{Prob}[S_A=S_B=S] > 1-\varepsilon ~\implies~ \mathrm{Prob}[S_{A}=S] > 1-\varepsilon ~\Leftrightarrow~ \mathrm{Prob}[S_{A} \neq S] < \varepsilon \nonumber \\
			&~~~~\Leftrightarrow~ \sum_{s_A} \sum_{s \neq s_A} P^\mathrm{real}_N (s_As) < \varepsilon,\\
			&b)~~ H(S) = \log \abs{S} ~~\implies~~{P^\mathrm{real}_N}(s) = \frac{1}{\abs{S}}, \\
			&c)~~\forall_s~~ \frac{1}{\abs{S_A}} = \sum_{s_A} P^\mathrm{real}_N (s_As) = \sum_{s_A \neq s} P^\mathrm{real}_N (s_As)  + P^\mathrm{real}_N (ss) \ge P^\mathrm{real}_N (ss),
			\end{align}
			Therefore we have:
			\begin{align}\label{eqn:complicated2}
			& \forall_{\varepsilon >0}\exists_{N_0(\varepsilon)}\forall_{N>N_0(\varepsilon)} \nonumber\\
			&\frac 12 \norm{{P^\mathrm{real}_N}(S_A) -\left(\frac{1}{|S_A|}\right)}_1 = \frac 12 \sum_{s_A}\abs{{P^\mathrm{real}_N}(s_A) -\frac{1}{|S_A|}} = \frac 12 \sum_{s_A}\abs{ \sum_s {P^\mathrm{real}_N}(s_As) -\frac{1}{|S_A|}} \nonumber \\
			&=\frac 12 \sum_{s_A}\abs{ \sum_{s\neq s_A} {P^\mathrm{real}_N}(s_As) +{P^\mathrm{real}_N}(s_As_A)-\frac{1}{|S_A|}} \stackrel{(I)}{\le} \frac 12 \sum_{s_A} \sum_{s\neq s_A} {P^\mathrm{real}_N}(s_As) +\frac 12 \sum_{s_A}\abs{ {\frac{1}{|S_A|}-P^\mathrm{real}_N}(s_As_A)} \nonumber \\
			&=\frac 12 \sum_{s_A} \sum_{s\neq s_A} {P^\mathrm{real}_N}(s_As) +\frac 12 \sum_{s}\abs{ {\frac{1}{|S_A|}-P^\mathrm{real}_N}(ss)} 
			\stackrel{(II)}{=} \frac 12 \sum_{s_A} \sum_{s\neq s_A} {P^\mathrm{real}_N}(s_As) +\frac 12 \sum_{s_A} \left( {\frac{1}{|S_A|}-P^\mathrm{real}_N}(s_As_A) \right)\nonumber \\
			&=\frac 12 \sum_{s_A} \sum_{s\neq s_A} {P^\mathrm{real}_N}(s_As) +\frac 12  \left( {1-\sum_{s_A}P^\mathrm{real}_N}(s_As_A) \right) 
			= \sum_{s_A} \sum_{s\neq s_A} {P^\mathrm{real}_N}(s_As) \le \varepsilon,
			\end{align}
			where $(I)$ follows from triangle inequality, $(II)$ is due to Observation c), and in the last step we used a). From equations (\ref{eqn:complicated0}), (\ref{eqn:complicated1}) and (\ref{eqn:complicated2}) we conclude that $\varepsilon^\mathrm{sec}_N \le \frac{1}{\sqrt{2}}\sqrt{\varepsilon + h(\varepsilon) + \varepsilon\log\left(\abs{S}-1\right)}+\varepsilon$.
			
			The correctness of a protocol is explicitly stated in Definition \ref{def:Maurer2000}, i.e., $\mathrm{Prob}[S_A=S_B=S]>1-\varepsilon$ (see Definition \ref{def:Correctness} for reference). Hence we have $\varepsilon_N^\mathrm{cor}:=\varepsilon$. From Theorem \ref{thm:equivalence} we obtain:
			\begin{align}
			& \forall_{\frac 12 \ge \varepsilon >0}\exists_{N_0(\varepsilon)}\forall_{N>N_0(\varepsilon)} \nonumber\\
			&\frac 12 \norm{P^\mathrm{real}_N (S_AS_BCE^N) - P^\mathrm{ideal}_N (S_AS_BCE^N)}_1 \le \varepsilon_N^\mathrm{cor} + \varepsilon_N^\mathrm{sec} \le  \frac{1}{\sqrt{2}}\sqrt{\varepsilon + h(\varepsilon) + \varepsilon\log\left(\abs{S}-1\right)}+2\varepsilon, 
			\end{align}
			or equivalently 
			\begin{align}
			\norm{P_N^\mathrm{real}- P_N^\mathrm{ideal}}_1 \le \delta_N \stackrel{N\to \infty}{\longrightarrow}0.
			\end{align}
			From the second inequality (\ref{eqn:M2}) and Eq. (\ref{eq:hahahaha}) we have that:
			\begin{align}
			\forall_{\varepsilon >0}\exists_{N_0(\varepsilon)}\forall_{N>N_0(\varepsilon)} ~~ L(N) \ge R -\epsilon,
			\end{align}
			for $L(N)=\frac{ \log \dim_{S_A} \left(P^\mathrm{real}_N \right)}{N}$ and hence by performing a limit $N \to \infty$, and condition of R being maximal number so that the above is satisfied we have  $R= \limsup_{N \to \infty} \frac{ \log \dim_{S_A} \left(P^\mathrm{real}_N \right)}{N}$. The strong secret key rate is defined as
			\begin{align}
			\overbar{\overbar{S}} (A:B||E)_{P(ABE)} = \sup_{\overbar{\overbar{\mathcal{P}}}}R,
			\end{align}
			where $\overbar{\overbar{\mathcal{P}}}$ are protocols that satisfy conditions in Definition \ref{def:Maurer2000}. Because conditions in Definition \ref{def:Maurer2000} imply condition (\ref{eqn:Appoach0}), we have $\left\{\overbar{\overbar{\mathcal{P}}}\right\} \subseteq \left\{\mathcal{P}\right\}$, and therefore:
			\begin{align}\label{eqn:Rlb}
			\overbar{\overbar{S}} (A:B||E)_{P(ABE)} = \sup_{\overbar{\overbar{\mathcal{P}}}}R = \sup_{\overbar{\overbar{\mathcal{P}}}} \limsup_{N \to \infty} \frac{ \log \dim_{S_A} \left(P^\mathrm{real}_N \right)}{N} \le 
			\sup_{\mathcal{P}} \limsup_{N \to \infty} \frac{ \log \dim_{S_A} \left(P^\mathrm{real}_N \right)}{N} .
			\end{align}
			By combining equations (\ref{eqn:Rub}) and (\ref{eqn:Rlb}) we have:
			\begin{align}
			\overbar{\overbar{S}} (A:B||E)_{P(ABE)} \le \sup_{\mathcal{P}} \limsup_{N \to \infty} \frac{ \log \dim_{S_A} \left(P^\mathrm{real}_N \right)}{N} \le {\overbar{S}} (A:B||E)_{P(ABE)}.
			\end{align}
			However, in the Reference \cite{MaurerWolf00CK} it was shown that $\forall_{P(ABE)} ~~ \overbar{\overbar{S}} (A:B||E)_{P(ABE)}={\overbar{S}} (A:B||E)_{P(ABE)}$, hence we conclude that:
			\begin{align}
			\overbar{\overbar{S}} (A:B||E)_{P(ABE)} = {S} (A:B||E)_{P(ABE)} = {\overbar{S}} (A:B||E)_{P(ABE)},
			\end{align}
			with ${{S}} (A:B||E)_{P(ABE)}=\sup_{\mathcal{P}} \limsup_{N \to \infty} \frac{ \log \dim_{S_A} \left(\mathcal{P}_N\left(P^{\otimes N}(ABE)\right) \right)}{N}$,
			and therefore all three definitions are equivalent.
		\end{proof}

		\section{Upper bound on device independent key}\label{sec:upper-iid}
		In this section, we prove our main result. Namely, we show that the secrecy quantifiers, that provide upper bounds on the key rate in the SKA model \cite{Maurer93,MaurerWolf00CK}, can serve {us to construct} upper bounds in device-independent key agreement scenario via operation of squashing. The secret key agreement scenario (SKA) is a well established area of cryptography, where upper bounds on the key rate are well known and given by entropic functions. The connection between upper bounds in SKA and NSDI cryptographic paradigms that we show in this section may simplify further studies on the latter.
		\begin{thm1} 
			The secret key rate, in the non-signaling device-independent {\it iid} scenario achieved with MDLOPC operations, $K_{DI}^{(iid)}$, from a device $P$ is upper bounded by any non-signaling squashed secrecy quantifier evaluated for the complete extension of $P$:
			\be
			\forall_{P}~~
			\widehat{\mathrm{M}} \left(A:B||E\right)_{\mathcal{E}\left({P}\right)} \ge K_{DI}^{(iid)}(P) ,
			\ee
			where $P\equiv P(AB|XY)$ is a single copy of a bipartite non-signaling device shared by the honest parties, and $\mathcal{E}(P) \equiv \mathcal{E}(P)(ABE|XYZ)$ is its complete extension to the eavesdropper's system. 
		\end{thm1}
		\begin{proof}[Proof of Theorem \ref{thm:main}]
			We start the proof by modifying the equality in Eq. (\ref{eqn:statement}), in Definition \ref{def:SKAour} in the following way:
			\begin{align}\label{eqn:MkeyonBoxes}
			&\max_{x,y}\min_z {S} ( A : B||E)_{(\mathcal{M}^F_{x,y} \otimes \mathcal{M}^G_{z})\mathcal{E}(P)(ABE|XYZ)} \nonumber \\
			&=\max_{x,y}\min_z \sup_{{\cal P}^{x,y,z}} \limsup_{N\rightarrow \infty} \frac{\log \mathrm{dim}_\mathrm{S_A} \left( \mathcal{P}_N^{x,y,z} \left( \left({(\mathcal{M}^F_{x,y} \otimes \mathcal{M}^G_{z})\mathcal{E}(P)}(ABE|XYZ)\right)^{\ot N}\right) \right)}{N},
			\end{align}
			where $\mathcal{P}^{x,y,z}$ is a LOPC protocol secure with respect to probability distribution that arises after $x$, $y$, $z$ choice of inputs (see Section \ref{sec:Maurer-rate} of the Appendix for reference), and $\mathcal{M}^F_{x,y}$, $\mathcal{M}^G_{z}$ are fiducial and general measurements of Alice, Bob and Eve respectively, described before in Section \ref{sec:Evesdrpping-action} of the Appendix.
			
			Let us notice that for each choice of $x$ and $y$ there exists $z=\bar{z}_{x,y}$ such that:
			\begin{align}\label{eqn:choosingZbar}
			&\max_{x,y} \sup_{{\cal P}^{x,y,\bar{z}_{x,y}}} \limsup_{N\rightarrow \infty} \frac{\log \mathrm{dim}_\mathrm{S_A} \left( \mathcal{P}_N^{x,y,\bar{z}_{x,y}} \left( \left({(\mathcal{M}^F_{x,y} \otimes \mathcal{M}^G_{\bar{z}_{x,y}})\mathcal{E}(P)}(ABE|XYZ)\right)^{\ot N}\right) \right)}{N} \\
			&:= \max_{x,y}\min_z \sup_{{\cal P}_{x,y,z}} \limsup_{N\rightarrow \infty} \frac{\log \mathrm{dim}_\mathrm{S_A} \left( \mathcal{P}_N^{x,y,z} \left( \left({(\mathcal{M}^F_{x,y} \otimes \mathcal{M}^G_{z})\mathcal{E}(P)}(ABE|XYZ)\right)^{\ot N}\right) \right)}{N}.
			\end{align}
			Now, when the optimization domains are explicitly stated, we can make use of max-min inequality to obtain:
			\begin{align}\label{eqn:Zvoid}
			&\max_{x,y}\min_z {S} ( A : B||E)_{(\mathcal{M}^F_{x,y} \otimes \mathcal{M}^G_{z})\mathcal{E}(P)(ABE|XYZ)} \\
			&\ge \max_{x,y} \sup_{{\cal P}^{x,y,\bar{z}_{x,y}}} \min_z \limsup_{N\rightarrow \infty} \frac{\log \mathrm{dim}_\mathrm{S_A} \left( \mathcal{P}_N^{x,y,\bar{z}_{x,y}} \left( \left({(\mathcal{M}^F_{x,y} \otimes \mathcal{M}^G_{z})\mathcal{E}(P)}(ABE|XYZ)\right)^{\ot N}\right) \right)}{N}.
			\end{align}
			We notice that the minimization of Eve's choice of input ($\min_z$) is void in the r.h.s. of the Eq. (\ref{eqn:Zvoid}) above. This is because the value of r.h.s. depends only on the value of $\mathrm{dim}_\mathrm{S_A} (\cdot)$ that is determined by choice of $x$, $y$, and hence by the protocol. Therefore we can write the following sequence of equalities where we swap from classical probability distributions to cc-d states. 
			\begin{align}
			\forall_{x,y} \forall_{{\cal P}^{x,y,\bar{z}_{x,y}}}~~&\min_z  \limsup_{N\rightarrow \infty} \frac{\log \mathrm{dim}_\mathrm{S_A} \left( \mathcal{P}_N^{x,y,\bar{z}_{x,y}} \left( \left({(\mathcal{M}^F_{x,y} \otimes \mathcal{M}^G_{z})\mathcal{E}(P)}(ABE|XYZ)\right)^{\ot N}\right) \right)}{N}   \\
			&= \limsup_{N\rightarrow \infty} \frac{\log \mathrm{dim}_\mathrm{S_A} \left( \mathcal{P}_N^{x,y,\bar{z}_{x,y}} \left( \left({(\mathcal{M}^F_{x,y} \otimes \mathds{1} )\mathcal{E}(P)}(ABE|XYZ)\right)^{\ot N}\right) \right)}{N} \\
			&= \limsup_{N\rightarrow \infty} \frac{\log \mathrm{dim}_\mathrm{S_A} \left( \mathcal{P}_N^{x,y,\bar{z}_{x,y}} \left( {(\mathcal{M}^F_{x,y} \otimes \mathds{1} )^{\ot N}\mathcal{E}^{\ot N}(P)}(ABE|XYZ)\right) \right)}{N} \\
			&= \limsup_{N\rightarrow \infty} \frac{\log \mathrm{dim}_\mathrm{S_A} \left( \mathcal{P}_N^{x,y,\bar{z}_{x,y}} \left( {(\mathcal{M}^F_{x,y} \otimes \mathds{1} )^{\ot N}\mathcal{E}(P^{\ot N})}(ABE|XYZ)\right) \right)}{N}\\
			&= \limsup_{N\rightarrow \infty} \frac{\log \mathrm{dim}_\mathrm{S_A} \left(\left( \mathcal{P}_N^{x,y,\bar{z}_{x,y}} \circ (\mathcal{M}^F_{x,y} )^{\ot N} \right)\left(\mathcal{E}(P^{\ot N})(ABE|XYZ)\right) \right)}{N}.
			\label{eqn:someIntermediate}
			\end{align}
			In the third equality above, we again used the fact that the dimension of Alice's subsystem (when the protocol is already fixed) is independent of Eve's action and her marginal distribution. This is the reason why we can substitute $\mathcal{E} \left(P^{\otimes N}\right)$ in the place of $\mathcal{E}^{\otimes N} (P)$. Moreover, in the last equality we use a notation that explicitly shows the composition between a measurement and a LOPC protocol. With a little abuse of notation $\mathds{1}$ in Eve's part is abandoned. 
			
			We notice now that each composition of measurement $x$, $y$ and protocol $\mathcal{P}^{x,y,\bar{z}_{x,y}}$ is a candidate for MDLOPC protocol $\Lambda:=\left\{\Lambda_N\right\}=\left\{\mathcal{P}_N^{x,y,\bar{z}_{x,y}} \circ (\mathcal{M}^F_{x,y} )^{\ot N}\right\}$. However we require that the distribution after the protocol is secure in NS-norm i.e.:
			\begin{align}\label{eqn:OptSec}
			\norm{\Lambda_N \left(\mathcal{E} \left(P^{\otimes N}\right)\right) - P_\mathrm{ideal}^{(d_N)}}_\mathrm{NS} \le \varepsilon_N \stackrel{N \to \infty}{\longrightarrow} 0,
			\end{align}
			what implies security not only with respect to Eve choosing $\bar{z}_{x,y}$, but against eavesdropper that has access to all inputs of $\mathcal{E} \left(P^{\otimes N}\right)$, hence possibly more powerful attacks. This is also a reason why we stay general even if there is any other good choice of $\bar{z}_{x,y}$ in Eq. (\ref{eqn:choosingZbar}). Having this in mind, we can write the inequalities below:
			\begin{align}\label{eqn:UBonK}
			&\max_{x,y}\min_z {S} ( A : B||E)_{(\mathcal{M}^F_{x,y} \otimes \mathcal{M}^G_{z})\mathcal{E}(P)(ABE|XYZ)} \\
			&\ge \max_{x,y} \sup_{{\cal P}^{x,y,\bar{z}_{x,y}}} \limsup_{N\rightarrow \infty} \frac{\log \mathrm{dim}_\mathrm{S_A} \left(\left( \mathcal{P}_N^{x,y,\bar{z}_{x,y}} \circ (\mathcal{M}^F_{x,y} )^{\ot N} \right)\left(\mathcal{E}(P^{\ot N})(ABE|XYZ)\right) \right)}{N} \label{eqn:mainR2}\\
			&\ge \sup_\Lambda \limsup_{N\rightarrow \infty} \frac{\log \mathrm{dim}_\mathrm{S_A} \left( \Lambda_N \left(\mathcal{E}(P^{\ot N})(ABE|XYZ)\right) \right)}{N} = K_{DI}^{(iid)}(P),\label{eqn:mainR3}
			\end{align}
			where the second inequality is due to the fact that now optimization is over a smaller set (not larger), i.e., only these combinations of measurements and LOPC operations that satisfy security condition in Eq. (\ref{eqn:OptSec}). Moreover, in the equality we identified MDLOPC (iid) secret key rate from Definition \ref{def:key_rate}.
			
			For the second part of the proof, we need to recall some properties of a family of  secrecy {quantifiers} $\left\{\mathrm{M}\left(A:B||E\right)\right\}$ of SKA model \cite{lit3}. Each function that upper bounds secret key rate in the SKA paradigm can be squashed according to the following procedure. For any {function} among them:  
			\begin{align}
			\forall_{Q(ABE)} ~~
			\mathrm{M}\left(A:B||E\right)_{{Q}(ABE)} \ge
			\mathrm{S}\left(A:B||E\right)_{{Q}(ABE)}.
			\end{align}
			By extending the above inequality to any tripartite non-signaling device $P(ABE|XYZ)$ {and general measurement for input $Z$}, one can write
			\begin{align}
			\forall_{P(ABE|XYZ)} \forall_{x,y,z} ~~
			\mathrm{M}\left(A:B||E\right)_{(\mathcal{M}^F_{x,y} \otimes \mathcal{M}^G_z) P(ABE|XYZ)} \ge
			\mathrm{S}\left(A:B||E\right)_{(\mathcal{M}^F_{x,y} \otimes \mathcal{M}^G_z)P(ABE|XYZ)}.
			\end{align}
			Without loss of generality, we fix the device $P(ABE|XYZ)$ for now. Let us denote $\tilde{z}_{x,y}$ as such an adaptive choice of $z$ that
			\begin{align}
			\forall_{x,y} ~~
			\mathrm{M}\left(A:B||E\right)_{(\mathcal{M}^F_{x,y} \otimes \mathcal{M}^G_{\tilde{z}_{x,y}}) P(ABE|XYZ)}	:= \min_z \mathrm{M}\left(A:B||E\right)_{(\mathcal{M}^F_{x,y} \otimes \mathcal{M}^G_z) P(ABE|XYZ)}. 
			\end{align}
			The immediate consequence is:
			\begin{align}
			\forall_{x,y} \hspace{1em}
			&\min_z \mathrm{M}\left(A:B||E\right)_{(\mathcal{M}^F_{x,y} \otimes \mathcal{M}^G_z)P(ABE|XYZ)}= \mathrm{M}\left(A:B||E\right)_{(\mathcal{M}^F_{x,y} \otimes \mathcal{M}^G_{\tilde{z}_{x,y}}) P(ABE|XYZ)}	\\
			&\ge \mathrm{S}\left(A:B||E\right)_{(\mathcal{M}^F_{x,y} \otimes \mathcal{M}^G_{\tilde{z}_{x,y}})P(ABE|XYZ)}
			\ge \min_z
			\mathrm{S}\left(A:B||E\right)_{(\mathcal{M}^F_{x,y} \otimes \mathcal{M}^G_{z})P(ABE|XYZ)}.
			\end{align}
			Employing a similar technique again, let us choose $\tilde{x}$, $\tilde{y}$ such that:
			\begin{align}
			\min_z
			\mathrm{S}\left(A:B||E\right)_{(\mathcal{M}^F_{\tilde{x},\tilde{y}} \otimes \mathcal{M}^G_{z}) P(ABE|XYZ)} :=
			\max_{x,y} \min_z \mathrm{S}\left(A:B||E\right)_{(\mathcal{M}^F_{x,y} \otimes \mathcal{M}^G_{z})P(ABE|XYZ)}.
			\end{align}
			This yields:
			\begin{align}
			&\max_{x,y} \min_z \mathrm{S}\left(A:B||E\right)_{(\mathcal{M}^F_{x,y} \otimes \mathcal{M}^G_{z})P(ABE|XYZ)}
			=
			\min_z
			\mathrm{S}\left(A:B||E\right)_{(\mathcal{M}^F_{\tilde{x},\tilde{y}} \otimes \mathcal{M}^G_{z}) P(ABE|XYZ)} \\
			&\le 
			\min_z
			\mathrm{M}\left(A:B||E\right)_{(\mathcal{M}^F_{\tilde{x},\tilde{y}} \otimes \mathcal{M}^G_{z}) P(ABE|XYZ)} 
			\le
			\max_{x,y} \min_z
			\mathrm{M}\left(A:B||E\right)_{(\mathcal{M}^F_{x,y} \otimes \mathcal{M}^G_{z}) P(ABE|XYZ)}.
			\end{align}
			{On the r.h.s. we recognize $\widehat{\mathrm{M}}\left(A:B||E\right)_{P(ABE|XYZ)}$ from Definition \ref{def:SQ}. Using the result in equations (\ref{eqn:UBonK}), (\ref{eqn:mainR2}), (\ref{eqn:mainR3}) from the first part of the proof, and substituting the complete extension of $P(AB|XY)$ as a tripartite device, we obtain:}
			\begin{align}
			\forall_{{P}(AB|XY)} ~~
			\widehat{\mathrm{M}}\left(A:B||E\right)_{\mathcal{E}(P)(ABE|XYZ)} \ge \mathrm{K}_\mathrm{DI}^{(iid)} \left(P(AB|XY)\right).
			\end{align}
		\end{proof}

		\section{Proof of the properties of non-signaling squashed nonlocality}	\label{appen:proofproperties}
		In this section, we give the proofs of the properties of the non-signaling squashed nonlocality.  Before we start with the proof, let us recall the definition of intrinsic information $\mathrm{I}\left(A:B\downarrow E\right)$, given in Sec. X.  We will rewrite the definition in two new ways. One of them is in full analogy to the forms of the squashed entanglement \cite{Tucci-squashed,Winter-squashed-ent}. Indeed, one can write the latter measure in terms  of the minimization over all possible extensions: $E_{sq}(\rho_{AB}) := \inf_{\sigma_{ABE}: \tr_E\sigma_{ABE} = \rho_{AB}} I(A:B|E)_{\sigma_{ABE}}$. The second form
		of the squashed nonlocality involves ensembles induced by measurements on the extending system and resembles the definition of the so-called {\it classical squashed entanglement} \cite{Winter-squashed-ent}.
		
		{The intrinsic information} involves an optimization over all possible conditional probability distributions $\Theta_{E'|E}$. 
		Moreover, in the squashing procedure, an optimization over the measurements on the CE of a bipartite device $P(AB|XY)$, has been involved. The non-signaling squashed intrinsic information is 
		\ben
		\widehat{\mathrm{I}}\left(A:B\downarrow E\right)_\mathrm{\mathcal{E}\left(P\right)(ABE|XYZ)}
		&=& \max_{x,y} \min_z \mathrm{I}\left(A:B\downarrow E\right)_{(\mathcal{M}^F_{x,y} \otimes \mathcal{M}^G_{z}) \mathrm{\cal E}(P)(ABE|XYZ)} \label{eq:squashed-int} \\
		&=&  \max_{x,y} \min_z  \inf_{\Theta^z_{E'|E}} \mathrm{I}(A:B|E')_{(\Theta^z_{E'|E})(\mathcal{M}^F_{x,y} \otimes \mathcal{M}^G_{z}) \mathrm{\cal E}(P)(ABE|XYZ)},
		\een
		where $\mathcal{M}^F_{x,y} $ is the direct measurement on the inputs $X$ and $Y$, and  $\mathcal{M}^G_{z}$ is a general measurement on $Z$. 
		According to Theorem $4$ of \cite{CE}, 
		$(\Theta^z_{E'|E})(\mathds{1}\otimes \mathcal{M}^G_{z}) \mathrm{\cal E}(P)(ABE|XYZ)= \sum_e \Theta^z_{E'|E=e} \sum_{z} p(z|z') \mathrm{\cal E}(P)(ABE = e|X Y Z=z) = \tilde{P}(ABE'|XYZ'=z')$, 
		is an arbitrary ensemble ({possibly} mixed) of the device $P(AB|XY)$,
		where $\mathds{1}$ is the identity operator {on the system of the honest parties}.
		Hence, for a fixed input randomizer (dice $p(z|z')$) and  a fixed channel, one can {generate} an arbitrary extension $\tilde{P}(ABE'|XY)$ with unary input. All possible choices of input randomizer and post-processing channel lead to all 
		possible extensions, hence  $\min_z \inf_{\Theta^z_{E'|E}} = \inf_{\tilde{P}(ABE'|XY)}$. And hence, it follows that Definition \ref{def:squashed_nonlocality} of the squashed nonlocality is equivalent to
		\ben\label{eq:squashednonlocality2}
		{\cal N}_{sq}(P(AB|XY)) &=& \max_{x,y}  \inf_{\tilde{P}(ABE|XY)} \mathrm{I} (A:B|E)_{\mathcal{M}^F_{x,y}{\tilde{P}}(ABE|XY)}.
		\een
		This arbitrary extension of a form $\tilde{P}(ABE|XY)$,  gives rise to an arbitrary but fixed ensemble of the bipartite device $P(AB|XY) = \sum_e \tilde{P}(ABE = e|XY) = \sum_e p(e) P^e(AB|XY)$, where  $P^e(AB|XY)$ is an arbitrary device corresponding to each output $E = e$, and belongs to the same polytope (state space) as $P(AB|XY)$. Moreover, all possible choices of  $\tilde{P}(ABE|XY)$ give rise to all possible ensembles of $P(AB|XY)$. The set of all ensembles of a given device $P(AB|XY)$, reads 
		\ben
		S^{all} := \left\{\{p_i, P^i(AB|XY)\}: \sum_i p_i P^i(AB|XY) = P(AB|XY)\right\}.
		\een
		Hence, $\inf_{\tilde{P}(ABE'|XY)}= \inf_{\{p_i, P^i(AB|XY)\} \in S^{all}}$, and 
		by virtue of Eq. (\ref{eq:squashednonlocality2}) we can rewrite Definition \ref{def:squashed_nonlocality} of the squashed nonlocality in the following way
		\ben\label{eq:squashednonlocality}
		{\cal N}_{sq}(P(AB|XY)) &=& \max_{x,y}  \inf_{\{p_i, P^i(AB|XY)\} \in S^{all}} \sum_i p_i \mathrm{I}(A:B)_{\mathcal{M}^F_{x,y} P^i(AB|XY)}.
		\een
		From Eq. (\ref{eq:squashednonlocality}), it is clear that the squashed nonlocality reduces  to the convex roof extension of the mutual information function. This is analogous 
		to the definition of  entanglement  for mixed quantum states \cite{HoroRMP}, the only difference is that here we are not restricting the device to be decomposable 
		in terms of only  pure (extremal) devices { (see in this context \cite{multiparty-squashed})}.

		\subsection{Relation to the bound of Ref. \cite{acin-2006-8}} \label{sec:app:RelationAcin}
		
        To describe the relation between our results and the results in Ref.~\cite{acin-2006-8},
        we prove that $\max_{x,y}\mathrm{I}_{\mathrm{AMP},(x,y)}={\cal N}_{sq}$. This allows us to compare the bounds on equal footing, and by showing that ${\cal N}_{sq}$ is convex, to use the convexification method to achieve tighter bound than given in Ref.~\cite{acin-2006-8}. 
        
        We first show that the $\geq$ inequality. Indeed, let us fix $(x,y)$ arbitrarily. Let $\{p(E=e)^*,P(ABE=e|XY)^*\}$ be an optimal ensemble achieving $\mathrm{I}_{\mathrm{AMP},(x,y)}$ By definition of the complete extension \cite{CE}, there exists a measurement\footnote{Here, we mean the generalized measurement that gives the eavesdropper the access to any ensemble of the device (see Appendix \ref{sec:Evesdrpping-action} for details). \color{black}} $z$ on its Eve's system $E$ that generates this ensemble: $\{P(E=e |Z=z), P(ABE=e|XY,Z=z)\}$ so that $P(E=e|Z=z)=P(E=e)^*$ and $P(ABE=e|XY,Z=z)=P(ABE=e|XY)^*$. Since $(x,y)$ was arbitrary and the $z$ could be suboptimal for the definition of ${\cal N}_{sq}$ we get the inequality $\max_{x,y} \mathrm{I}_{\mathrm{AMP},(x,y)}\geq N_{sq}$. To see that $\max_{x,y}\mathrm{I}_{\mathrm{AMP},(x,y)}\leq {\cal N}_{sq}$, let $x,y$ be fixed arbitrarily and $z(x,y)$ such that the value of $\inf_{z} I(A:B\downarrow E)_{P(ABE|X=x,Y=y,Z=z)}$ is minimal. Then $\{P(E=e|Z=z(xy)),P(ABE=e|X=x,Y=y,Z=z(x,y)$ is a particular ensemble of $P(AB|XY)$, which may be suboptimal, i.e. not attaining infimum in definition of $\mathrm{I}_{\mathrm{AMP},(x,y)}$, we get that $\mathrm{I}_{\mathrm{AMP},(x,y)} \leq \inf_{z} I(A:B\downarrow E)_{P(ABE|X=x,Y=y,Z=z)}$. Since $(x,y)$ was arbitrary, we can take max over $(x,y)$ on both sides, and on the RHS we obtain ${\cal N}_{sq}$ while the bound $\max_{(x,y)} \mathrm{I}_{\mathrm{AMP},(x,y)}$ is on the LHS, which proves the claimed equality.
		
		\subsection{Positivity of the measure}
		\begin{proposition}\label{cor:NN} The squashed nonlocality is a positive semidefinite function of bipartite non-signaling devices $P(AB|XY)$,
			\be
			{\cal N}_{sq}(P(AB|XY))\ge 0,
			\ee
			and	the equality holds {if} the device $P$ admits a local hidden variable model \cite{Bell-nonlocality}. 
		\end{proposition}
		
		\begin{proof} The intrinsic conditional mutual information satisfy $\mathrm{I}(A:B\downarrow E) \geq 0$ for all distributions $P(ABE)$, hence the positive semi-definiteness directly follows from {its definition}:
			
			\ben
			{\cal N}_{sq}(P(AB|XY))=\max_{x,y} \min_z \mathrm{I}\left(A:B\downarrow E\right)_{(\mathcal{M}^F_{x,y} \otimes \mathcal{M}^G_{z}) \mathrm{\cal E}(P)(ABE|XYZ)}  \ge \max_{x,y} \min_{z} 0 = 0.
			\een
			
			Now we have to show that it is zero for {all} local devices.
			Let us assume $P_L(AB|XY)$ is a local device, i.e. there exists a hidden variable model $\lambda$, such that $ P_L(AB|XY)=\sum_\lambda P(A|X,\lambda) \otimes P(B|Y,\lambda) \rho (\lambda)$.  This leads to an ensemble $\{\rho (\lambda), P(A|X,\lambda) \otimes P(B|Y,\lambda) \}$ whose members are tensor products of local devices, hence from Eq. (\ref{eq:squashednonlocality}), we can directly write
			\ben
			{\cal N}_{sq}(P_L(AB|XY)) = \max_{x,y}  \sum_i \rho (\lambda_i)\mathrm{I}(A:B)_{\mathcal{M}^F_{x,y}\big(P(A|X,\lambda_i) \otimes P(B|Y,\lambda_i)\big) }= 0.
			\een
		\end{proof}

		\subsection{Convexity }\label{sec:convexity}
		\begin{proposition}
			\label{IBOX:convexity}
			$ {\cal N}_{sq}(P)$ is a  convex function, i.e., if $P(AB|XY)$ and $Q(AB|XY)$ are two bipartite non-signaling devices in the same polytope, then 
			\be
			{\cal N}_{sq}\left(\lambda P(AB|XY) + (1-\lambda)Q(AB|XY) \right)
			~\le ~\lambda~ {\cal N}_{sq}(P(AB|XY) ) + (1-\lambda)~  {\cal N}_{sq}(Q(AB|XY) )
			\ee
			$\forall \lambda \in [0,1]$.
		\end{proposition}	
		\begin{proof} 
			Consider the convex combination of the devices
			\be
			\bar{P}(AB|XY) = \lambda P(AB|XY) + (1 - \lambda) Q(AB|XY).
			\ee
			In particular there exists an extension $\bar{P}_{ext}(ABE\Lambda|XY) $ of $\bar{P}(AB|XY)$, such that 
			\ben
			\bar{P}_{ext}(ABE\Lambda = 0|XY) = p(\Lambda = 0) \tilde{P}(ABE|XY), \\
			\bar{P}_{ext}(ABE\Lambda = 1|XY) = p(\Lambda = 1) \tilde{Q}(ABE|XY),
			\een
			with $p(\Lambda = 0) = \lambda$ and $p(\Lambda = 1) = 1 - \lambda$. We consider  that the devices $\tilde{P}(ABE|XY)$ and  $\tilde{Q}(ABE|XY)$ are arbitrary extensions of the devices $P(AB|XY) $ and  $Q(AB|XY)$ respectively, as discussed above. \\
			Hence, from Eq. (\ref{eq:squashednonlocality2}), we have
			\ben
			\forall x,y&& \inf_{{\bar P}(ABE|XY)} \mathrm{I}(A:B|E)_{{\cal M}^F_{x,y}{\bar P}(ABE|XY)} \leq  \mathrm{I}(A:B|E\Lambda)_{{\cal M}^F_{x,y}{\bar P}_{ext}(ABE\Lambda|XY)}\\
			&& \hspace{1.77in}= \lambda ~\mathrm{I}(A:B|E)_{{\cal M}^F_{x,y}\tilde{P}(ABE|XY)} + (1-\lambda) ~\mathrm{I}(A:B|E)_{{\cal M}^F_{x,y}\tilde{Q}(ABE|XY)},
			\een 
			where ${\bar P}(ABE|XY)$ are such that $\sum_e {\bar P}(ABE=e|XY)={\bar P}(AB|XY)$.
			The above relation holds for an arbitrary extensions of $P$ and $Q$, the  $\tilde{P}(ABE|XY)$ and $\tilde{Q}(ABE|XY)$ respectively. Hence, it is also true for the optimal extensions 
			\ben
			\forall x,y&& \inf_{{\bar P}(ABE|XY)} \mathrm{I}(A:B|E)_{{\cal M}^F_{x,y}{\bar P}(ABE|XY)} \nonumber \\
			&& \leq \lambda ~\inf_{\hat{P}(ABE|XY)} \mathrm{I}(A:B|E)_{{\cal M}^F_{x,y}\hat{P}(ABE|XY)} + (1-\lambda) ~\inf_{\hat{Q}(ABE|XY)} \mathrm{I}(A:B|E)_{{\cal M}^F_{x,y}\hat{Q}(ABE|XY)}, ~~~~~~ \label{eq:infext}
			\een
			where ${\hat P}(ABE|XY)$ are such that $\sum_e {\hat P}(ABE=e|XY)={ P}(AB|XY)$ and ${\hat Q}(ABE|XY)$ are such that $\sum_e {\hat Q}(ABE=e|XY)={ Q}(AB|XY)$. 	
			{Consider direct measurements $\bar{x}$ and $\bar{y}$ that maximize l.h.s. of inequality (\ref{eq:infext}). Then from Eq. (\ref{eq:squashednonlocality2}) we have:}
			\ben
			&&{\cal N}_{sq}({\bar P}(AB|XY)) = \max_{x,y} \inf_{{\bar P}(ABE|XY)} \mathrm{I}(A:B|E)_{{\cal M}^F_{x,y}\bar P(ABE|XY)} = \inf_{{\bar P}(ABE|XY)} \mathrm{I}(A:B|E)_{{\cal M}^F_{\bar{x},\bar{y}}{\bar P}(ABE|XY)} \\
			&&\stackrel{(I)}{\leq} \lambda ~\inf_{{\hat P}(ABE|XY)} \mathrm{I}(A:B|E)_{{\cal M}^F_{\bar{x},\bar{y}}\hat{P}(ABE|XY)} + (1-\lambda) ~\inf_{\hat{Q}(ABE|XY)} \mathrm{I}(A:B|E)_{{\cal M}^F_{\bar{x},\bar{y}}\hat{Q}(ABE|XY)}. ~~~~~~~~ \\
			&&\stackrel{(II)}{\leq} \lambda ~\max_{x,y}\inf_{{\hat P}(ABE|XY)} \mathrm{I}(A:B|E)_{{\cal M}^F_{\bar{x},\bar{y}}\hat{P}(ABE|XY)} + (1-\lambda) ~\max_{x,y}\inf_{\hat{Q}(ABE|XY)} \mathrm{I}(A:B|E)_{{\cal M}^F_{\bar{x},\bar{y}}\hat{Q}(ABE|XY)}. ~~~~~~~~ \\
			&&= \lambda ~ {\cal N}_{sq}(P(AB|XY)) + (1 - \lambda)~ {\cal N}_{sq}(Q(AB|XY)),
			\een	
			where in (I), we use the inequality (\ref{eq:infext}), with $x=\bar{x}$ and $y=\bar{y}$. In (II), we use the fact that direct measurements $\bar{x}$ and $\bar{y}$, may not maximize terms at r.h.s. of the inequality (\ref{eq:infext}).
		\end{proof}

		\begin{figure}[h]
			\includegraphics[width=1\textwidth]{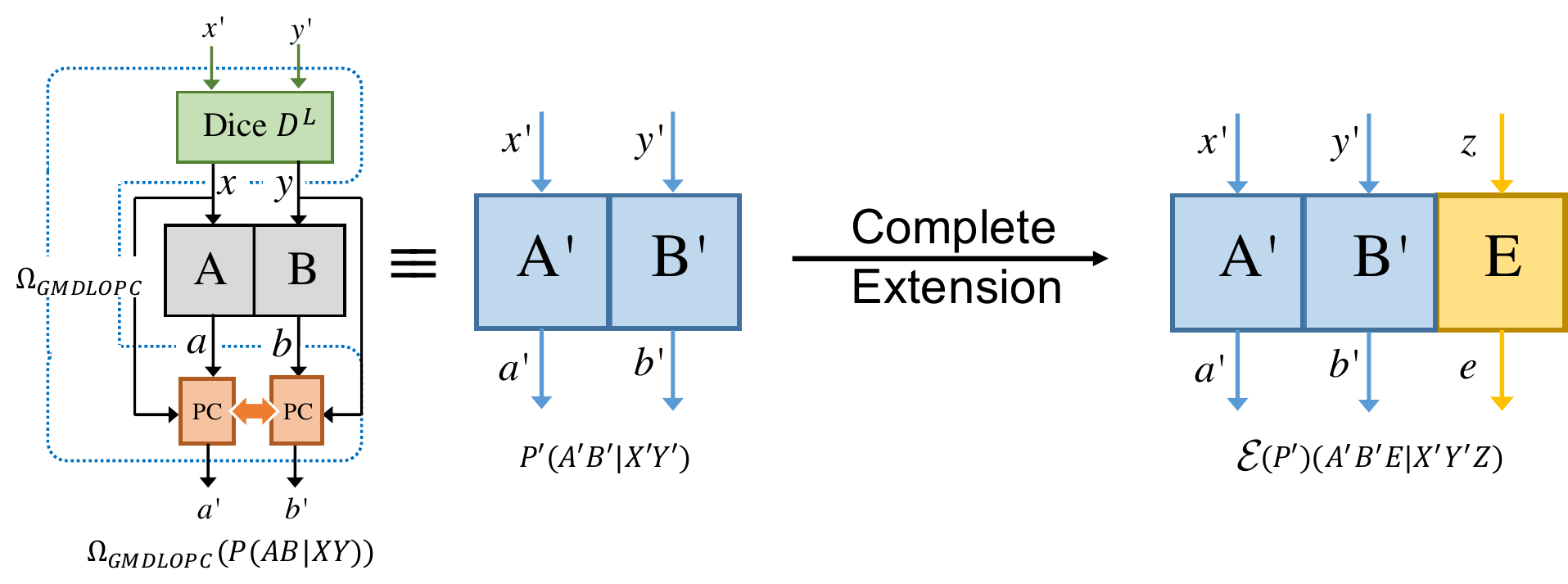}
			\caption{Schematic diagram of the $\Omega_{GMDLOPC}$ operation, where the inputs of the devices shared by the honest parties are chosen by a local randomizer $D^L_{XY|X'Y'}(xy|x'y')$ as given in Eq. (\ref{eq:dicelocal}). Similarly, the outputs are also connected through a post-processing channel $PC^L_{A'B'|ABXY}(a'b'|abxy)$ which also depends on the inputs $x,y$ and has a local hidden variable model  given in Eq. (\ref{eq:pplocal}). }
			\label{fig:LOPC}
		\end{figure}	
		

		\subsection{Inheritance of monotonicity: Monotonicity under MDLOPC class of operation }\label{sec:monotonicity}
		
		In this section, we will show that any secrecy monotone (functional, nonincreasing under LOPC operations), after squashing procedure yields a functional which is monotonic under MDLOPC operations.
		
		\begin{proposition}\label{prop:inheritence} [Inheritance of monotonicity]
			Any secrecy quantifier $\mathrm{M}(A:B|E)$, which is nonincreasing under LOPC operations, after the squashing procedure is nonincreasing under MDLOPC operations.
		\end{proposition}
		\begin{proof}
			Let us consider arbitrary MDLOPC operation $\Lambda_{MDLOPC}$.
			By definition it is a composition of the form $\Lambda_{MDLOPC}=\Lambda_{LOPC}\circ {\cal M}^{\mathbf{F}}_{x_0,y_0}$. Let us also choose arbitrary device $P(ABE|XYZ)$ and let us fix arbitrarily $z=z_0$. As a consequence we can write a sequence of (in)equalities
			which we comment below, where 	for the sake of clarity of the proof we will use a short notation:
			$\mathrm{M}(A:B|E)\equiv \mathrm{M}$ and $\widehat{ \mathrm{M}}(A:B|E)\equiv \widehat{\mathrm {M}}$. 
			\ben
			{\widehat {\mathrm{M}}}(\Lambda_{MDLOPC}(P(ABE|XYZ))) &=& \max_{x,y} \min_z 
			\mathrm{M}(\Lambda_{MDLOPC}(P(ABE|X=x,Y=y,Z=z)))\\
			&=& \min_z\mathrm{M}(\Lambda_{LOPC}(P(ABE|X=x_0,Y=y_0,Z=z)))  \label{eq:mon1}\\
			&\leq&	\mathrm{M}(\Lambda_{LOPC}(P(ABE|X=x_0,Y=y_0,Z=z_0)))  \label{eq:mon2}\\
			&\leq& 	\mathrm{M}(P(ABE|X=x_0,Y=y_0,Z=z_0)   \label{eq:mon3}\\  &\leq& \max_{x,y}\mathrm{M}(P(ABE|X=x,Y=y,Z=z_0))  \label{eq:mon4}\\ &=& \max_{x,y} \min_z \mathrm{M}(P(ABE|X=x,Y=y,Z=z)) \label{eq:mon5}\\ &=& {\widehat {\mathrm M}}(P(ABE|XYZ)). 
			\een
			In the first equality, we use the definition of $\widehat{\mathrm{M}}$. 
			In Eq. (\ref{eq:mon1}) we use the fact that the device $P$ after
			measurement ${\cal M}_{x_0,y_0}^\mathbf{F}$ has unary inputs 	in part of the honest parties (it becomes a distribution in that part), hence there is no
			parameter $x,y$ to maximise over. The inequality (\ref{eq:mon2}) follows from the property of minimum 
			(over $z$). The inequality (\ref{eq:mon3}) is due to
			the monotonicity of $\mathrm M$ under $\Lambda_{LOPC}$. 
			The inequality   (\ref{eq:mon4}) is because $x_0,y_0$ may
			be suboptimal in $\max_{x,y}$ over $\mathrm{M}(P(ABE|X=x,Y=y,Z=z_0))$. The equality (\ref{eq:mon5})
			comes from the fact that the choice of $z_0$ was arbitrary, so it is true for the $z_0$ that attains $\min_z$ in (\ref{eq:mon5}). The last equality comes from definition of $\widehat {\mathrm{M}}\equiv\widehat {\mathrm{M}}(A:B|E)$, which ends the proof.
		\end{proof}
		
		From Proposition \ref{prop:inheritence}, it directly follows that the squashed nonlocality, is monotonic under $\Lambda_{MDLOPC}$, as it is defined based on the secrecy quantifier, intrinsic mutual information $\mathrm{I}(A:B\downarrow E)$. And it is monotonic under LOPC operation \cite{reduced-intrinsic,Christandl12}.
		
		Without using the above Proposition, we can also independently prove that ${\cal N}_{sq}$ is monotonic under MDLOPC operation, or in principle, under a larger class of operations, the GMDLOPC. 
		We have mentioned in the main text, that the MDLOPC class of operations involve (i) direct measurement, changing devices into distributions followed by (ii) Local Operations and Public Communication. If we relax the measurement procedure and include all possible general measurements,  then we will have the GMDLOPC class of operations, as shown in the schematic diagram in Fig. \ref{fig:LOPC}. Clearly MDLOPC $ \subset $ GMDLOPC, and one particular operation of GMDLOPC class will be denoted as  $\Omega_{\mathrm{GMDLOPC}}$. Hence the monotonicity: 
		
		\begin{proposition}\label{cor:monotonicity}\footnote{The 
				result of this section is partially based on Ref. \cite{Kaur-Wilde}}  The non-signaling squashed nonlocality of any non-signaling bipartite device $P$ satisfies
			\ben
			&& \forall_{\Omega_{\mathrm{GMDLOPC}}} ~~~ {\cal N}_{sq}\left(\Omega_{\mathrm{GMDLOPC}}(P)\right) \leq {\cal N}_{sq}(P), 
			\een	
		\end{proposition}

		\begin{proof}
			
			To prove the monotonicity under GMDLOPC, we will use the equivalent definition of ${\cal N}_{sq}$ given in Eq. (\ref{eq:squashednonlocality}).
			Under the GMDLOPC operation $\Omega_{\mathrm{GMDLOPC}}\in$ GMDLOPC, the honest parties can choose general measurements in the input of the shared device $P(AB|XY)$.
			The general measurements can be chosen by using a public shared local randomeness generator  $D^L_{XY|X'Y'}(xy|x'y')$ (as depicted in Fig. \ref{fig:LOPC}), with $x'\in X', y' \in Y'$ the input and $x\in X, y \in Y$ are the output. As the output of $D^L$ will be feeded to the input of $P(AB|XY)$, hence, we will assume without any loss of generality that both cardinality are same. Moreover, as $D^L$ is local randomness generator, hence
			\ben\label{eq:dicelocal}
			D^L_{XY|X'Y'}(xy|x'y') = \sum_{\lambda_1} \mu(\lambda_1) D^1(x|x'\lambda_1)D^2(y|y'\lambda_1),
			\een
			where $\lambda_1 \in \Lambda_1$ is the local hidden variable and $\sum_{\lambda_1} \mu(\lambda_1)  = 1$. 
			Similarly, the outputs are also passed through a local post-processing channel $PC_{A'B'|ABXY}^L(a'b'|abxy)$, which also depends on the inputs of the initial device, as shown in Fig. \ref{fig:LOPC}. Additionally, the locality condition give rise to
			\ben\label{eq:pplocal}
			PC^L_{A'B'|ABXY}(a'b'|abxy) = \sum_{\lambda_2} \nu(\lambda_2) PC^1(a'|ax\lambda_2)PC^2(b'|by\lambda_2),
			\een	 
			with $\lambda_2 \in \Lambda_2$ and $\sum_{\lambda_2} \nu(\lambda_2)  = 1$.
			Hence, under $\Omega_{\mathrm{GMDLOPC}}$, the given device $P(AB|XY)$ transforms into
			\ben\label{eq:localp}
			P'_{A'B'|X'Y'}(a'b'|x'y') &=& \Omega_{\mathrm{GMDLOPC}}(P(AB|XY))  \\  &=& \sum_{xy} D^L_{XY|X'Y'}(xy|x'y') \sum_{ab}P_{AB|XY}(ab|xy) PC^L_{A'B'|ABXY}(a'b'|abxy) \\ 
			&=& \sum_{xy}  \sum_{\lambda_1} \mu(\lambda_1) D^1(x|x'\lambda_1)D^2(y|y'\lambda_1)  \sum_{ab}P(ab|xy) \nonumber \\
			&&	\hspace{2in} \times \sum_{\lambda_2} \nu(\lambda_2) PC^1(a'|ax\lambda_2)PC^2(b'|by\lambda_2). \label{eq:eq:localpexp}
			\een
			Now the ${\cal N}_{sq}$ of $P'$ as in Eq. (\ref{eq:squashednonlocality}) is 
			\ben
			{\cal N}_{sq}(P') = \widehat{\mathrm{I}}\left(A':B'\downarrow E\right)_\mathrm{\mathcal{E}\left(P'\right)(A'B'E|X'Y'Z)}
			= \max_{x'y'}  \inf_{\{p_i, P'^i\} \in S^{all}(P')} \sum_i p_i \mathrm{I}(A':B')_{P'^i},
			\een
			where $\mathcal{E}\left(P'\right)(A'B'E|X'Y'Z)$, is the CE of  $P'(A'B'|X'Y')$, and $S^{all}(P')$ denotes all possible ensembles of $P'$.

			Consider the following tripartite device, resulting upon performing the $\Omega_{\mathrm{GMDLOPC}}$ on the the CE of $P(AB|XY)$,
			\ben
			\Omega_{\mathrm{GMDLOPC}}\otimes \mathds{1}_E\left({\cal E}(P)(ABE|XYZ)\right)
			&=& \sum_{xy} D^L_{XY|X'Y'}(xy|x'y') \sum_{ab}{\cal E}(P)(abe|xyz) PC^L_{A'B'|ABXY}(a'b'|abxy) \\ 
			&=& \sum_{xy}  \sum_{\lambda_1} \mu(\lambda_1) D^1(x|x'\lambda_1)D^2(y|y'\lambda_1)  \sum_{ab}{\cal E}(P)(abe|xyz)  \nonumber \\
			&&	\hspace{0in} \times \sum_{\lambda_2} \nu(\lambda_2) PC^1(a'|ax\lambda_2)PC^2(b'|by\lambda_2).
			\een 
			Here $\mathds{1}_E$ means the identity operator in Eve's subsystem.

			Consider the ensemble $\left \{ p(e|z)\mu(\lambda_1)\nu(\lambda_2), P^{ez\lambda_1\lambda_2}(a'b'|x'y')\right\},$ 
			where 
			\be	 
			P^{ez\lambda_1\lambda_2}(a'b'|x'y') = \sum_{xy}    D^1(x|x'\lambda_1)D^2(y|y'\lambda_1)  \sum_{ab}P^{ez}(ab|xy) PC^1(a'|ax\lambda_2)PC^2(b'|by\lambda_2).
			\ee
			Now we will show that the above ensemble will be an ensemble of $P'(A'B'|X'Y')$, if $\{p(e|z),P^{ez}_{AB|XY}\}$ is an ensemble of $P(AB|XY)$. 
			
			Suppose $\{p(e|z),P^{ez}_{AB|XY}\}$, is an ensemble of $P$, then
			\ben
			&&  \sum_{e\lambda_1\lambda_2} p(e|z)\mu(\lambda_1)\nu(\lambda_2) P^{ez\lambda_1\lambda_2}(a'b'|x'y') \\
			&=& \sum_{e\lambda_1\lambda_2} p(e|z)\mu(\lambda_1)\nu(\lambda_2) \sum_{xy}    D^1(x|x'\lambda_1)D^2(y|y'\lambda_1)  \sum_{ab}P^{ez}(ab|xy) PC^1(a'|ax\lambda_2)PC^2(b'|by\lambda_2) \\
			&=& \sum_{xy}  \sum_{\lambda_1} \mu(\lambda_1) D^1(x|x'\lambda_1)D^2(y|y'\lambda_1)  \sum_{ab} \Big(\sum_e p(e|z) P^{ez}(ab|xy) \Big)\sum_{\lambda_2} \nu(\lambda_2) PC^1(a'|ax\lambda_2)PC^2(b'|by\lambda_2)\\
			&=& \sum_{xy}  \sum_{\lambda_1} \mu(\lambda_1) D^1(x|x'\lambda_1)D^2(y|y'\lambda_1)  \sum_{ab} P(ab|xy)\sum_{\lambda_2} \nu(\lambda_2) PC^1(a'|ax\lambda_2)PC^2(b'|by\lambda_2) \\
			&=& P'_{A'B'|X'Y'}(a'b'|x'y'),
			\een
			by using Eq. (\ref{eq:eq:localpexp}) and the fact that $\sum_e p(e|z) P^{ez}(ab|xy) = P(ab|xy)$. 
			
			Moreover,
			$\{p(e|z),P^{ez}_{AB|XY}\}$	is an arbitrary ensemble, and Eve can easily access it once she has the  CE ${\cal E}(P)(ABE|XYZ)$:
			
			Because $\left \{ p(e|z)\mu(\lambda_1)\nu(\lambda_2), P^{ez\lambda_1\lambda_2}(a'b'|x'y')\right\}$ is an ensemble of $P'$, 
			\ben
			\label{monotoni}
			\forall_{x',y'}   \inf_{\{p_i, P^i\} \in S^{all}(P')} \sum_i p_i \mathrm{I}(A':B')_{P'^i} 
			& \leq &\sum_{e \lambda_1 \lambda_2} p(e|z)\mu(\lambda_1)\nu(\lambda_2) \mathrm{I}(A':B')_{P^{ez\lambda_1\lambda_2}(A'B'|X' =x', Y' = y')} \\
			&\stackrel{(I)}{\leq} & \sum_{e \lambda_1 \lambda_2} p(e|z)\mu(\lambda_1)\nu(\lambda_2) \mathrm{I}(AX:BY)_{P^{ez\lambda_1\lambda_2}(AXBY|X' =x', Y' = y')} \\
			&&\hspace{-5cm} \stackrel{(II)}{=} \sum_{e \lambda_1} p(e|z)\mu(\lambda_1) \left( \mathrm{I}(A:B|XY) + \mathrm{I}(X:B|Y)+\mathrm{I}(A:Y|X) + \mathrm{I}(X:Y)\right)_{P^{ez\lambda_1}(AXBY|X' =x', Y' = y')} \\
			&\stackrel{(III)}{=}& \sum_{e \lambda_1} p(e|z)\mu(\lambda_1)  \mathrm{I}(A:B|XY)_{P^{ez\lambda_1}(AXBY|X' =x', Y' = y')} \label{eq:monoIII}\\
			&&  \hspace{-1in} = \sum_{e x y \lambda_1} p(e|z)\mu(\lambda_1) D^1(x|x'\lambda_1)D^2(y|y'\lambda_1) \mathrm{I}(A:B)_{P^{ez\lambda_1}(AB|X = x, Y=y, X' =x', Y' = y')}\\
			&\stackrel{(IV)}{\leq}&  \max_{xy} \sum_{e } p(e|z) \mathrm{I}(A:B)_{P^{ez}(AB|XY)}, \label{eq:monolast}
			\een
			where in (I) we use the data processing inequality and also use the fact that the distribution $P^{ez\lambda_1\lambda_2}_{AXBY|X'Y'}(axby|x'y') = D^1(x|x'\lambda_1)D^2(y|y'\lambda_1)  P^{ez}_{AB|XY}(ab|xy) \sum_{a'b'} PC^1(a'|ax\lambda_2)PC^2(b'|by\lambda_2)$ is independent of $\lambda_2$. The chain rule of mutual information has been used in (II) whereas in (III), we use the fact that given $x'$, $y'$ and $\lambda_1$, the random variables  $X$ and $Y$ are independent, hence $\mathrm{I}(X:B|Y) = \mathrm{I}(A:Y|X) = 0$,  which follows from the non-signaling condition.
			In (IV) we simply write $P^{ez\lambda_1}(AB|X = x, Y=y, X' =x', Y' = y') = P^{ez}(AB|X=x,Y=y)$.
			
			The  r.h.s. of  (\ref{eq:monolast}) is valid  for an arbitrary ensemble $\{p(e|z), P^{ez}\} \in S^{all}(P)$, so it is still valid when taking infimum over all ensembles. Hence,
			\ben
			\max_{x'y'}  \inf_{\{p_i, P'^i\} \in S^{all}(P')} \sum_i p_i \mathrm{I}(A':B')_{P'^i(A'B'|X'Y')} &\leq & \max_{xy} \inf_{\{p_i, P^i\} \in S^{all}(P)}  \sum_{i } p_i \mathrm{I}(A:B)_{P^i(AB|XY)},\\
			\Rightarrow \hspace{1.8in} {\cal N}_{sq}\left(\Omega_{\mathrm{GMDLOPC}}(P)\right) &\leq & {\cal N}_{sq}(P).
			\een
		\end{proof}
		
		As MDLOPC $ \subset $ GMDLOPC, so we have
		\begin{corthree}
			The non-signaling squashed nonlocality of any non-signaling bipartite device $P$ satisfies
			\ben
			&& \forall_{\Lambda_\mathrm{MDLOPC}} ~~~ {\cal N}_{sq}\left(\Lambda_\mathrm{MDLOPC}(P)\right) \leq {\cal N}_{sq}(P), 
			\een
		\end{corthree}
		The above monotonicity property also holds for the non-signaling squashed conditional mutual information \linebreak $\widehat{\mathrm{I}}(A:B|E)_{{\cal E}(P)(ABE|XYZ)}$.

		\subsection{Superadditivity and additivity}\label{sec:superadditivity}
		
		\begin{proposition}\footnote{ The 
				result of this section is partially based on Ref. \cite{Kaur-Wilde}.}
			If two bipartite non-signaling devices $P(A_1B_1|X_1Y_1)$ and $Q(A_2B_2|X_2Y_2)$ are the marginals of a four partite non-signaling device $\bar{P}(A_1A_2B_1B_2|X_1X_2Y_1Y_2)$, 
			then the non-signaling squashed nonlocality ${\cal N}_{sq}$ is superadditive, 
			\ben
			{\cal N}_{sq}(\bar{P}(A_1A_2B_1B_2|X_1X_2Y_1Y_2)) \geq {\cal N}_{sq}(P(A_1B_1|X_1Y_1)) +  {\cal N}_{sq}(Q(A_2B_2|X_2Y_2)),
			\een 
			and  additive for tensor product of devices $P(A_1B_1|X_1Y_1) \otimes Q(A_2B_2|X_2Y_2)$,  that is 
			\be
			{\cal N}_{sq}(P(A_1B_1|X_1Y_1) \otimes Q(A_2B_2|X_2Y_2)) = {\cal N}_{sq}(P(A_1B_1|X_1Y_1)) +  {\cal N}_{sq}(Q(A_2B_2|X_2Y_2)).
			\ee
		\end{proposition}
		
		\begin{proof}
			\textbf{Superadditivity on joint device:}
			Let us consider two devices $P(A_1B_1|X_1Y_1)$ and $Q(A_2B_2|X_2Y_2)$, which are the marginals of a big four party non-signaling device $\bar{P}(A_1A_2B_1B_2|X_1X_2Y_1Y_2)$, i.e. 
			\ben
			\sum_{a_2b_2 } \bar{P}(a_1,a_2,b_1,b_2|x_1,x_2,y_1,y_2)  &=&  P(a_1,b_1|x_1,y_1) ~~\forall a_1, b_1, x_1, x_2, y_1, y_2, \\
			\sum_{a_1b_1 } \bar{P}(a_1,a_2,b_1,b_2|x_1,x_2,y_1,y_2)  &=& Q(a_2,b_2|x_2,y_2), ~~\forall a_2, b_2, x_1,  x_2, y_1, y_2.
			\een
			where $\bar{P}(A_1 = a_1,A_2 = a_2, B_1=b_1,B_2 = b_2|X_1 = x_1,X_2 = x_2, Y_1 = y_1,Y_2 = y_2) \equiv \bar{P}(a_1,a_2,b_1,b_2|x_1,x_2,y_1,y_2) $, $P(A_1 = a_1,B_1 = b_1|X_1 = x_1,Y_1 = y_1) = P(a_1,b_1|x_1,y_1)$ and $Q(A_2=a_2, B_2 = b_2|X_2=x_2,Y_2=y_2) = Q(a_2,b_2|x_2,y_2)$. Moreover,  $\bar{P}(A_1A_2B_1B_2|X_1X_2Y_1Y_2)$ is also satisfy non-signaling conditions among all of its parties, as defined in Eqs. (\ref{eq:nons:A}) and (\ref{eq:nons:B}).
			
			Consider an arbitrary non-signaling extension of $\bar{P}(A_1A_2B_1B_2|X_1X_2Y_1Y_2) \rightarrow \bar{P}(A_1A_2B_1B_2E|X_1X_2Y_1Y_2Z)$, with unary input $|Z|$ in the extended part. The input is unary, so the non-signaling condition is automatic and we can omit the $Z$. The conditional mutual information of the distribution after performing an arbitrary  pair of direct measurements, i.e.,  ${\cal M}^F_{x_1,y_1} \otimes {\cal M}^F_{x_2,y_2}$ on the inputs $X_1, Y_1$ and $X_2, Y_2$ reads
			\ben
			&&\forall x_1, x_2, y_1,y_2  \nonumber \\
			&&\mathrm{I}(A_1A_2:B_1B_2|E)_{({\cal M}^F_{x_1,y_1} \otimes {\cal M}^F_{x_2,y_2})\bar{P}(A_1A_2B_1B_2E|X_1X_2Y_1Y_2)} \\
			&& \stackrel{(I)}{=} (\mathrm{I}(A_1:B_1|E) +\mathrm{I}(A_2:B_1|EA_1) + \mathrm{I}(A_1:B_2|EB_1) + \mathrm{I}(A_2:B_2|EA_1B_1))_{({\cal M}^F_{x_1,y_1} \otimes {\cal M}^F_{x_2,y_2})\bar{P}(A_1A_2B_1B_2E|X_1X_2Y_1Y_2)} ~~~ \nonumber \\
			&& \stackrel{(II)}{\geq} \mathrm{I}(A_1:B_1|E)_{{\cal M}^F_{x_1,y_1} \bar{P}(A_1B_1E|X_1Y_1)} + \mathrm{I}(A_2:B_2|EA_1B_1)_{({\cal M}^F_{x_1,y_1} \otimes {\cal M}^F_{x_2,y_2})\bar{P}(A_1A_2B_1B_2E|X_1X_2Y_1Y_2)}, \label{eq:mutu-chain}
			\een
			where we use the chain rule of mutual information in (I) and in 
			(II), we use positivity condition of mutual information. ${\cal M}^F_{x_1,y_1}  \bar{P}(A_1B_1E|X_1Y_1)$ is the marginal of the device $({\cal M}^F_{x_1,y_1} \otimes {\cal M}^F_{x_2,y_2})\bar{P}(A_1A_2B_1B_2E|X_1X_2Y_1Y_2)$ after the direct measurements on the inputs. Recall that 
			\be
			({\cal M}^F_{x_1,y_1} \otimes {\cal M}^F_{x_2,y_2})\bar{P}(A_1A_2B_1B_2E|X_1X_2Y_1Y_2) = \bar{P}(A_1A_2B_1B_2E|X_1 = x_1,X_2=x_2, Y_1 = y_1,Y_2=y_2). 
			\ee
			
			Noticing that $ \bar{P}(A_1B_1E|X_1Y_1)$ is an arbitrary extension of $P(A_1B_1|X_1Y_1)$ and similarly $\bar{P}(A_1A_2B_1B_2E|X_1X_2Y_1Y_2)$ is for the device $Q(A_1B_1E|X_1Y_1)$,  we can write
			
			\ben
			&&\forall x_1, x_2, y_1,y_2  \nonumber \\
			&& \mathrm{I}(A_1:B_1|E)_{{\cal M}^F_{x_1,y_1} \bar{P}(A_1B_1E|X_1Y_1)} \geq  \inf_{\bar{P}(A_1B_1E|X_1Y_1)} \mathrm{I}(A_1:B_1|E)_{{\cal M}^F_{x_1,y_1}\bar{P}(A_1B_1E|X_1Y_1)}, \label{eq:A1B1mutu}\\
			&& \mathrm{I}(A_2:B_2|EA_1B_1)_{({\cal M}^F_{x_1,y_1} \otimes {\cal M}^F_{x_2,y_2})\bar{P}(A_1A_2B_1B_2E|X_1X_2Y_1Y_2)} \geq \inf_{\bar{Q}(A_2B_2E|X_2Y_2)} \mathrm{I}(A_2:B_2|E)_{{\cal M}^F_{x_2,y_2}\bar{Q}(A_2B_2E|X_2Y_2)}. ~~~~~~ \label{eq:A2B2mutu}
			\een
			From inequalities (\ref{eq:mutu-chain}), (\ref{eq:A1B1mutu}) and (\ref{eq:A2B2mutu}), we have
			\ben
			&&\forall x_1, x_2, y_1,y_2  \nonumber \\
			&&\mathrm{I}(A_1A_2:B_1B_2|E)_{({\cal M}^F_{x_1,y_1} \otimes {\cal M}^F_{x_2,y_2})\bar{P}(A_1A_2B_1B_2E|X_1X_2Y_1Y_2)} \nonumber \\
			&& \geq  \inf_{\bar{P}(A_1B_1E|X_1Y_1)} \mathrm{I}(A_1:B_1|E)_{{\cal M}^F_{x_1,y_1}\bar{P}(A_1B_1E|X_1Y_1)} +  \inf_{\bar{Q}(A_2B_2E|X_2Y_2)} \mathrm{I}(A_2:B_2|E)_{{\cal M}^F_{x_2,y_2}\bar{Q}(A_2B_2E|X_2Y_2)}.
			\een
			The above inequality holds for all extensions of $\bar{P}(A_1A_2B_1B_2|X_1X_2Y_1Y_2)$, hence also for an optimal extension on the LHS,   so
			\ben
			&&\forall x_1, x_2, y_1,y_2  \nonumber \\
			&&\inf_{\bar{P}(A_1B_1A_2B_2E|X_1X_2Y_1Y_2)} \mathrm{I}(A_1A_2:B_1B_2|E)_{({\cal M}^F_{x_1,y_1} \otimes {\cal M}^F_{x_2,y_2})\bar{P}(A_1A_2B_1B_2E|X_1X_2Y_1Y_2)} \nonumber \\
			&& \geq  \inf_{\bar{P}(A_1B_1E|X_1Y_1)} \mathrm{I}(A_1:B_1|E)_{{\cal M}^F_{x_1,y_1}\tilde{P}(A_1B_1E|X_1Y_1)} +  \inf_{\bar{Q}(A_2B_2E|X_2Y_2)} \mathrm{I}(A_2:B_2|E)_{{\cal M}^F_{x_2,y_2}\bar{Q}(A_2B_2E|X_2Y_2)}. \label{eq:abbas}
			\een
			Suppose that $\bar{x}_1, \bar{y}_1$ are the optimal direct measurement choice for ${\cal N}_{sq}(P)$ and $\bar{x}_2,\bar{y}_2$ are for ${\cal N}_{sq}(Q)$, 
			\ben
			{\cal N}_{sq}(P(A_1B_1|X_1Y_1)) &=&  \max_{x_1,y_1}\inf_{\bar{P}(A_1B_1E|X_1Y_1)} \mathrm{I}(A_1:B_1|E)_{{\cal M}^F_{x_1,y_1}\bar{P}(A_1B_1E|X_1Y_1)} \nonumber \\
			&=& \inf_{\bar{P}(A_1B_1E|X_1Y_1)} \mathrm{I}(A_1:B_1|E)_{{\cal M}^F_{\bar{x}_1,\bar{y}_1}\bar{P}(A_1B_1E|X_1Y_1)}, \label{eq:NsqP}
			\een
			\ben
			{\cal N}_{sq}(Q(A_2B_2|X_2Y_2)) &=&  \max_{x_2,y_2}\inf_{\bar{P}(A_2B_2E|X_2Y_2)} \mathrm{I}(A_2:B_2|E)_{{\cal M}^F_{x_2,y_2}\bar{Q}(A_2B_2E|X_2Y_2)} \nonumber \\
			&=& \inf_{\bar{Q}(A_2B_2E|X_2Y_2)} \mathrm{I}(A_2:B_2|E)_{{\cal M}^F_{\bar{x}_2,\bar{y}_2}\bar{P}(A_1B_1E|X_1Y_1)}. \label{eq:NsqQ}
			\een
			Finally, 
			\ben
			&& \hspace{-1.5em}{\cal N}_{sq}(\bar{P}(A_1B_1A_2B_2|X_1X_2Y_1Y_2)) \nonumber \\
			&=& \max_{x_1,y_1,x_2y_2} \inf_{\bar{P}(A_1B_1A_2B_2E|X_1X_2Y_1Y_2)} \mathrm{I}(A_1A_2:B_1B_2|E)_{({\cal M}^F_{x_1,y_1} \otimes {\cal M}^F_{x_2,y_2})\bar{P}(A_1A_2B_1B_2E|X_1X_2Y_1Y_2)} \\
			&\stackrel{(I)}{\geq}& \inf_{\bar{P}(A_1B_1A_2B_2E|X_1X_2Y_1Y_2)} \mathrm{I}(A_1A_2:B_1B_2|E)_{({\cal M}^F_{\bar{x}_1,\bar{y}_1} \otimes {\cal M}^F_{\bar{x}_2,\bar{y}_2})\bar{P}(A_1A_2B_1B_2E|X_1X_2Y_1Y_2)} \\
			&\stackrel{(II)}{\geq}& \inf_{\bar{P}(A_1B_1E|X_1Y_1)} \mathrm{I}(A_1:B_1|E)_{{\cal M}^F_{\bar{x}_1,\bar{y}_1}\tilde{P}(A_1B_1E|X_1Y_1)} +  \inf_{\bar{Q}(A_2B_2E|X_2Y_2)} \mathrm{I}(A_2:B_2|E)_{{\cal M}^F_{\bar{x}_2,\bar{y}_2}\bar{Q}(A_2B_2E|X_2Y_2)}, \\
			&\stackrel{(III)}{=}&  {\cal N}_{sq}(P(A_1B_1|X_1Y_1)) + {\cal N}_{sq}(Q(A_2B_2|X_2Y_2)). \label{superadditivity}
			\een
			In $(I)$, we use an specific choice of direct measurement, ${\cal M}^F_{\bar{x}_1,\bar{y}_1} \otimes {\cal M}^F_{\bar{x}_2,\bar{y}_2}$, which may not be optimal for device $\bar{P}(A_1B_1A_2B_2|X_1X_2Y_1Y_2)$.
			We use Eq. (\ref{eq:abbas}) for the direct measurements ${\cal M}^F_{\bar{x}_1,\bar{y}_1} \otimes {\cal M}^F_{\bar{x}_2,\bar{y}_2}$ in $(II)$ and finally in 
			$(III)$, Eqs. (\ref{eq:NsqP}) and (\ref{eq:NsqQ}) has been used.
			
			\noindent \textbf{Additivity for tensor product of devices:}
			Let us assume that the joint non-signaling four party device (two random variables for input and output  in the honest parties' part) is the tensor product \cite{TensorProduct} of two bipartite devices,
			\be
			\bar{P}(A_1B_1A_2B_2|X_1X_2Y_1Y_2) = P(A_1B_1|X_1Y_1) \otimes Q(A_2B_2|X_2Y_2)
			\ee
			Consider the (non-signaling) extensions with unary inputs of both the devices, $P(A_1B_1|X_1Y_1) \rightarrow \bar{P}(A_1B_1E_1|X_1Y_1)$ and $ Q(A_2B_2|X_2Y_2) \rightarrow  \bar{Q}(A_2B_2E_2|X_2Y_2)$, which are the optimal  extensions for calculating  ${\cal N}_{sq}$ for both the devices, as given in Eq. (\ref{eq:squashednonlocality2}), for all $x$ and $y$. Hence, their tensor product  $\bar{P}(A_1B_1E_1|X_1Y_1)\otimes \bar{Q}(A_2B_2E_2|X_2Y_2)$ is an extension of $\bar{P}(A_1B_1A_2B_2|X_1X_2Y_1Y_2)$, which may not be optimal one, resulting in\\
			$\forall x_1, x_2, y_1,y_2 $
			\ben
			&&\inf_{\bar{P}(A_1B_1A_2B_2E|X_1X_2Y_1Y_2)}  \mathrm{I}(A_1A_2:B_1B_2|E)_{({\cal M}^F_{x_1,y_1} \otimes {\cal M}^F_{x_2,y_2})\bar{P}(A_1A_2B_1B_2E|X_1X_2Y_1Y_2)} \nonumber \\
			&\leq & \mathrm{I}(A_1A_2:B_1B_2|E_1E_2)_{({\cal M}^F_{x_1,y_1} \otimes {\cal M}^F_{x_2,y_2})\bar{P}(A_1B_1E_1|X_1Y_1)\otimes \bar{Q}(A_2B_2E_2|X_2Y_2)} \\
			&=&  \mathrm{I}(A_1:B_1|E_1)_{{\cal M}^F_{x_1,y_1}\bar{P}(A_1B_1E_1|X_1Y_1)} + \mathrm{I}(A_2:B_2|E_2)_{{\cal M}^F_{x_2,y_2}\bar{Q}(A_2B_2E_2|X_2Y_2)} \\
			&=& \inf_{P(A_1B_1E_1|X_1Y_1)}\mathrm{I}(A_1:B_1|E_1)_{{\cal M}^F_{x_1,y_1}P(A_1B_1E_1|X_1Y_1)} + \inf_{Q(A_2B_2E_2|X_2Y_2)} \mathrm{I}(A_2:B_2|E_2)_{{\cal M}^F_{x_2,y_2}Q(A_2B_2E_2|X_2Y_2)} 
			\een
			
			Considering the optimal direct measurements ${\cal M}^F_{\bar{x}_1, \bar{y}_1}\otimes{\cal M}^F_{\bar{x}_2, \bar{y}_2}$ in the LHS of the above relation, gives
			\ben
			&&{\cal N}_{sq}(P(A_1B_1|X_1Y_1) \otimes Q(A_2B_2|X_2Y_2))  \nonumber \\
			&=& \max_{x_1, x_2,y_1,y_2}\inf_{\bar{P}(A_1B_1A_2B_2E|X_1X_2Y_1Y_2)}  \mathrm{I}(A_1A_2:B_1B_2|E)_{({\cal M}^F_{x_1,y_1} \otimes {\cal M}^F_{x_2,y_2})\bar{P}(A_1A_2B_1B_2E|X_1X_2Y_1Y_2)} \\
			&=& \inf_{\bar{P}(A_1B_1A_2B_2E|X_1X_2Y_1Y_2)}  \mathrm{I}(A_1A_2:B_1B_2|E)_{({\cal M}^F_{\bar{x}_1,\bar{y}_1} \otimes {\cal M}^F_{\bar{x}_2,\bar{y}_2})\bar{P}(A_1A_2B_1B_2E|X_1X_2Y_1Y_2)}\\
			&\leq & \inf_{P(A_1B_1E_1|X_1Y_1)}\mathrm{I}(A_1:B_1|E_1)_{{\cal M}^F_{\bar{x}_1,\bar{y}_1}P(A_1B_1E_1|X_1Y_1)} + \inf_{Q(A_2B_2E_2|X_2Y_2)} \mathrm{I}(A_2:B_2|E_2)_{{\cal M}^F_{\bar{x}_2,\bar{y}_2}Q(A_2B_2E_2|X_2Y_2)} \\
			&\leq & \max_{x_1,y_1} \inf_{P(A_1B_1E_1|X_1Y_1)}\mathrm{I}(A_1:B_1|E_1)_{{\cal M}^F_{x_1,y_1}P(A_1B_1E_1|X_1Y_1)} + \max_{x_2,y_2}\inf_{Q(A_2B_2E_2|X_2Y_2)} \mathrm{I}(A_2:B_2|E_2)_{{\cal M}^F_{x_2,y_2}Q(A_2B_2E_2|X_2Y_2)} \nonumber \\  \\
			&=& {\cal N}_{sq}(P(A_1B_1|X_1Y_1)) + {\cal N}_{sq}(Q(A_2B_2|X_2Y_2)).
			\een
			
			Using relation (\ref{superadditivity}), we finish the proof with equality:
			\be
			{\cal N}_{sq}(P(A_1B_1|X_1Y_1) \otimes Q(A_2B_2|X_2Y_2)) = {\cal N}_{sq}(P(A_1B_1|X_1Y_1)) + {\cal N}_{sq}(Q(A_2B_2|X_2Y_2)).
			\ee
		\end{proof}

		\subsection{Subextensivity}
		
		\begin{proposition}
			Non-signaling squashed nonlocality is bounded by $\log\left(\min\left\{d_A,d_B\right\}\right)$.
		\end{proposition}
		
		\begin{proof}
			From the defnintion of non-signaling squashed nonlocality given in Eq. (\ref{eq:squashednonlocality}) we have
			\ben
			{\cal N}_{sq}(P(AB|XY)) &=& \max_{x,y}  \inf_{\{p_i, P^i(AB|XY)\} \in S^{all}} \sum_i p_i \mathrm{I}(A:B)_{\mathcal{M}^F_{x,y} P^i(AB|XY)} \\
			&\stackrel{(I)}{\leq}& \max_{x,y}  \inf_{\{p_i, P^i(AB|XY)\} \in S^{all}} \sum_i p_i \log \left(\min \left\{d_A^x,d_B^y\right\}\right)\\
			&\leq& \log\left(\min\left\{d_A,d_B\right\}\right). 
			\een
			
			where in $(I)$, we use the fact that $\mathrm{I}(A:B)_{\mathcal{M}_{x,y}\left(P^i(AB|XY)\right)} \le \log \left(\min \left\{d_A^x,d_B^y\right\}\right)$ for all $i$, and $d_A^x =\text{supp} P(A|X=x)$ and $d_B^y = \text{supp} P(B|Y=y)$ and $d_A = \max_{x} \text{supp} P(A|X=x)$ and $d_B = \max_{y} \text{supp} P(B|Y=y)$.
		\end{proof}
		
		\section{Nonlocality cost as an upper bound}\label{sec:nonlocalitycost}
		
		\begin{definition}\label{def:NC} The nonlocality cost of bipartite non-signaling device is 
			\be
			\mathcal{N}_\mathrm{C}(P):= \mathrm{C}(P)\log\left(\min\left\{d_A,d_B\right\}\right),
			\ee
			where $d_A = \max_{x} (\mathrm{supp} {\cal M}^F_x(P(A|X)))$ and $d_B = \max_{y} (\mathrm{supp} {\cal M}^F_y(P(B|Y))$ are dimensions of the outputs, and $\mathrm{C}(P)$ is the nonlocality fraction of P \cite{Elitzur-nonloc, Brunneretal2011}.
		\end{definition}
		
		
		\begin{proposition}
			\label{lemma:NC}
			The secret key rate $K_{DI}^{(iid)}(P)$ of a device is upper bounded by
			\be
			\mathcal{N}_\mathrm{C}(P) \ge K_{DI}^{(iid)}(P),
			\ee
		\end{proposition}

		\begin{proof}
			Suppose Alice and Bob share a non-signaling device $P\equiv P(AB|XY)$, and Eve has access to its complete extension \cite{CE}. The device $P$ can be decomposed into a non-local vertex and a local device,
			\be\label{eq:nonlocality-decom}
			P=\alpha P_\mathrm{NL}^V+(1-\alpha)P_\mathrm{L},
			\ee
			where $P_\mathrm{NL}^V$ is the non-local vertex and $P_\mathrm{L}$ is any local device. Let us denote the nonlocality fraction
			\be
			\mathrm{C}(P):=\min_\mathrm{All~decompsitions ~as ~in ~Eq.~ (\ref{eq:nonlocality-decom})} \alpha.
			\ee
			Eve can always get access to this ensemble, $\left\{\left(\mathrm{C}(P),\bar{P}_\mathrm{NL}^V\right),\left(1-\mathrm{C}(P),\bar{P}_\mathrm{L}\right)\right\}$, in part of the honest parties. \\
			We assume that Eve works in favor of Alice and Bob,  and informs them about her output when she obtains the above ensemble. The key rate $\tilde{K}$, in this scenario, must be greater than in NSDI-iid scenario, since in the latter case Eve does not work on  account of Alice and Bob,
			\be
			K_{DI}^{(iid)}(P) \le \tilde{K}(P).
			\label{help}
			\ee
			With a probability  $\mathrm{C}(P)$ the honest parties share the non-local correlations, useful for secret key agreement and with probability $1-\mathrm{C}(P)$, they share a local device with zero key rates. Since the key satisfying Maurer's security definition is upper bounded by mutual information function, and both of them are non-increasing under the LOPC operations, we obtain
			\be
			\tilde{K}(P) \le \mathrm{C}(P) \left( \max_{x,y} \mathrm{I}(A:B)_{{\cal M}^F_{x,y}P^V_{NL}(AB|XY)}  \right).
			\ee
			Furthermore,
			\be
			\mathrm{I}(A:B)_{{\cal M}^F_{x,y}\left(P(AB|XY)\right)} \le \log \left(\min \left\{d_A^x,d_B^y\right\}\right),
			\ee	
			where $d_A^x =\text{supp} P(A|X=x)$ and $d_B^y = \text{supp} P(B|Y=y)$.
			Employing Eq. (\ref{help}),
			we finally obtain
			\ben
			K_{DI}^{(iid)}(P) &\le& \mathrm{C}(P) \left( \sup_{\mathcal{M}^F_{x,y}} \log  \left(\min \left\{d_A^x,d_B^y\right\}\right) \right)\\
			&=& \mathrm{C}(P)\log\left(\min\left\{d_A,d_B\right\}\right) = {\cal N}_C(P),
			\een
			by Definition \ref{def:NC},	with $d_A = \max_{x} \text{supp} P(A|X=x)$ and $d_B = \max_{y} \text{supp} P(B|Y=y)$.
		\end{proof}

		\section{Examples of secrecy monotones, Convexification of $\widehat{\mathrm{I}} \left(A:B \downarrow E\right)$ and a non-trivial bound}\label{sec:nsq}


		Monotones, based on mutual information functions, are used to upper bound the secret key rate on the SKA scenario. However, the only one amongst them, which is easily computable, is the mutual information itself. All of them can be ``squashed" and used to generate the upper bounds for $K_{DI}^{(iid)}$.

		\begin{fact} The secrecy quantifiers and monotones \cite{Christandl12} (and the mutual information function) are the upper bounds on $\mathrm{S}\left(A:B||E\right)$:
			\label{fct:3}
			\ben
			\mathrm{I} \left(A:B\right)_{P(ABE)} \ge \mathrm{S}\left(A:B||E\right)_{P(ABE)}, \\
			\mathrm{I} \left(A:B|E\right)_{P(ABE)} \ge \mathrm{S}\left(A:B||E\right)_{P(ABE)},\\
			\min \left\{\mathrm{I} \left(A:B\right)_{P(ABE)}, \mathrm{I} \left(A:B|E\right)_{P(ABE)} \right\} \ge \mathrm{S}\left(A:B||E\right)_{P(ABE)}, \\
			\mathrm{I} \left(A:B \downarrow E\right)_{P(ABE)} \ge \mathrm{I} \left(A:B \downarrow \downarrow E\right)_{P(ABE)} \ge \mathrm{S}\left(A:B||E\right)_{P(ABE)}.
			\een
		\end{fact}
		
		We can use all of the functions displayed in  Fact \ref{fct:3} to construct the non-signaling squashed secrecy quantifiers and monotones for the devices. See Section \ref{entropies} for the proper definition of the above functions.
		
		\begin{corfour}\label{crl:UB} The following upper bounds on $K_{DI}^{(iid)}(\mP)$ hold
			\ben
			\widehat{\mathrm{I}} \left(A:B\right)_{\mathcal{E}(P)(ABE|XYZ)} \ge K_{DI}^{(iid)}(P), \\
			\widehat{\mathrm{I}} \left(A:B|E\right)_{\mathcal{E}(P)(ABE|XYZ)} \ge K_{DI}^{(iid)}(P),\\
			\min \left\{\widehat{\mathrm{I}} \left(A:B\right)_{\mathcal{E}(P)(ABE|XYZ)}, \widehat{\mathrm{I}} \left(A:B|E\right)_{\mathcal{E}(P)(ABE|XYZ)} \right\} \ge K_{DI}^{(iid)}(P), \\
			\widehat{\mathrm{I}} \left(A:B \downarrow E\right)_{\mathcal{E}(P)(ABE|XYZ)} \ge \widehat{\mathrm{I}} \left(A:B \downarrow \downarrow E\right)_{\mathcal{E}(P)(ABE|XYZ)} \ge K_{DI}^{(iid)}(P).
			\een
		\end{corfour}
		The proof of the above Corollary is straightforward from Theorem \ref{thm:main}. It is important to note that,  the complete extension of a device, $P(AB|XY)$, has been denoted as $\mathcal{E}(P)(ABE|XYZ)$, where the extended systems are in full control of Eve.

		The intrinsic information $\widehat{\mathrm{I}}\left(A:B\downarrow E\right)$ and the reduced intrinsic information $\widehat{\mathrm{I}}\left(A:B\downarrow \downarrow E\right)$ are functions without closed-form expression, and hence they cannot be computed straightforwardly. We present a technique for finding a nontrivial bound using the properties of one of them. First, we notice that for any fixed bipartite device and its complete extension, the following is true.
		
		\begin{observation}[Hierarchy between different mutual information functions] 
			\label{fct:4}
			\ben
			{\cal N}_{sq}(P) = 	\widehat{\mathrm{I}}\left(A:B\downarrow E\right)_\mathrm{\mathcal{E}\left(P\right)(ABE|XYZ)} &\le& \widehat{\mathrm{I}}\left(A:B\right)_\mathrm{\mathcal{E}\left(P\right)(ABE|XYZ)}, \\
			{\cal N}_{sq}(P)= 	\widehat{\mathrm{I}}\left(A:B\downarrow E\right)_\mathrm{\mathcal{E}\left(P\right)(ABE|XYZ)} &\le & \widehat{\mathrm{I}}\left(A:B|E\right)_\mathrm{\mathcal{E}\left(P\right)(ABE|XYZ)}
			.
			\een 
		\end{observation}

		The  squashed nonlocality is upper bounded by the squashed conditional mutual information $\widehat{\mathrm{I}}\left(A:B|E\right)_\mathrm{\mathcal{E}\left(P\right)(ABE|XYZ)}$, and also with  squashed mutual information $\widehat{\mathrm{I}}\left(A:B\right)_\mathrm{\mathcal{E}\left(P\right)(ABE|XYZ)}$, hence: 
		
		\begin{observation}\label{fct:mutcond}
			Non-signaling squashed nonlocality is upper-bounded by the following expression.
			\ben \label{eqn:min}
			\mathcal{N}_\mathrm{sq}(P) \le \min \left\{ \widehat{\mathrm{I}}\left(A:B\right)_\mathrm{\mathcal{E}\left(P\right)(ABE|XYZ)}, \widehat{\mathrm{I}}\left(A:B|E\right)_\mathrm{\mathcal{E}\left(P\right)(ABE|XYZ)} \right\}.
			\een
		\end{observation}
		
		Unfortunately, the squashed nonlocality lacks a closed-form expression for an arbitrary non-signaling device. It involves optimization over general measurement and post-processing channels in the eavesdropper side. This makes it hard to compute for a generic non-signaling device. Moreover, we obtained the squashed nonlocality to be a convex function over the mixture of devices, see Sec. \ref{sec:convexity}, whereas the intrinsic information  is not a convex function. This might be due to the fact that it was constructed in the same way as the non-signaling squashed entanglement, and the latter is a convex function of quantum states \cite{Winter-squashed-ent}. In this Section, we will show how convexity of squashed nonlocality can be used not only to calculate nontrivial upper bounds on $K_{DI}^{(iid)}$, but also how it can be used to define new non-signaling squashed secrecy quantifiers.

		Observation \ref{fct:mutcond}, brings the idea of how to use the convexity property of  squashed nonlocality. Since the  squashed nonlocality is an upper bound on $K_{DI}^{(iid)}$, hence, the r.h.s. of Eq. (\ref{eqn:min}) must also be an upper bound on secret key rate as well. Together with the convexity property, it implies that a lower convex hull of $\widehat{\mathrm{I}}\left(A:B|E\right)$ and $\widehat{\mathrm{I}}\left(A:B\right)$ also bounds $K_{DI}^{(iid)}$ from above. 
		
		\begin{theorem}\label{thm:family}
			Within a family of functions $\left\{\mathrm{F}_i\right\}$, which are convex with respect to mixtures of devices, and  
			\ben
			\mathrm{F}_i(P) &\le& \widehat{\mathrm{I}}\left(A:B\right)_\mathrm{\mathcal{E}\left(P\right)(ABE|XYZ)}, \label{Eqn:upN1}\\
			\mathrm{F}_i(P) &\le& \widehat{\mathrm{I}}\left(A:B|E\right)_\mathrm{\mathcal{E}\left(P\right)(ABE|XYZ)}, \label{Eqn:upN2}
			\een
			there exists a function $\mathrm{F}$ that upper bounds any function in  $\left\{\mathrm{F}_i\right\}$ and for which the following relation holds 
			\ben
			\mathrm{F}\left(P\right) \ge 
			K_{DI}^{(iid)}(\mP).
			\een
		\end{theorem}
		\begin{proof}
			Since $	\widehat{\mathrm{I}} \left(A:B\downarrow E\right) \in \left\{\mathrm{F}_i\right\}$ because of Proposition \ref{IBOX:convexity} and $\widehat{\mathrm{I}} \left(A:B\downarrow E\right)_{\mathcal{E}\left(\mathrm{P}\right)} \ge K_{DI}^{(iid)}(\mP)$, then, for a function $F$ which lies above the values of the squashed intrinsic mutual information, satisfies $\mathrm{F}\left(\mathrm{P}\right) \ge K_{DI}^{(iid)}(\mP)$.
		\end{proof}

		Theorem \ref{thm:family}, can be easily generalized by imposing different constraints than Equations (\ref{Eqn:upN1}) and (\ref{Eqn:upN2}) for example by using other upper bounds on the  squashed nonlocality and also an arbitrary number of them. 
		\begin{rem}\label{rem:LCH}
			The lower convex hull of plots of an arbitrary number of functions, each being an upper bound on a convex function which upper bounds $K_{DI}^{(iid)}$, is an upper bound on the key rate itself.
		\end{rem}
		This observation automatically yields a recipe on how to construct nontrivial upper bounds on $K_{DI}^{(iid)}$. We come up with the following Corollary, being a direct consequence of Theorem \ref{thm:family} and Remark \ref{rem:LCH}.
		
		\begin{corfive} \label{cor:NT}
			A non-trivial upper bound is given by the lower convex hull $(\mathrm{LCH})$ of plots of non-signaling squashed secrecy quantifiers.
			\ben
			{\cal N}_\mathrm{sq}(P) \leq \mathrm{F}\left(P\right) :=  \mathrm{LCH} 
			\left\{\widehat{\mathrm{I}}\left(A:B\right)_\mathrm{\mathcal{E}\left(P\right)(ABE|XYZ)},\widehat{\mathrm{I}}\left(A:B|E\right)_\mathrm{\mathcal{E}\left(P\right)(ABE|XYZ)} \right\} \nonumber \\
			\een
		\end{corfive}
		
		\begin{proof}
			We prove by contradiction. If there would be a function which at any point is greater than the lower convex hull of 
			$\widehat{\mathrm{I}} \left(A:B \right)$
			and $\widehat{\mathrm{I}} \left(A:B \left| E\right.\right)$, either it would not be convex or it would be greater (at least at a single point) then at least one from the above non-signaling squashed nonlocality quantifiers. Therefore, it is not in the set $\left\{\mathrm{F}_i\right\}$.
		\end{proof}
		The upper bound on $K_{DI}^{(iid)}$ introduced in the above Corollary can be computed much more easily than the non-signaling squashed nonlocality. We will refer to the procedure of calculating upper bounds via this technique as \emph{convexification}. 
		Observation \ref{fct:4} and Proposition \ref{lemma:NC} provide a collection of functions   which are upper bounds for ${\cal N}_{sq}$. Hence, there exists a convex (in the same sense) function, which is an upper bound on the squashed nonlocality, but at the same time, it is a lower bound on any function in this group, which is very clear from the proof of Theorem \ref{thm:family}.


		\section{Numerical upper bounds on squashed nonlocality}\label{sec:numerical}
		
		In this Section, we will provide the upper bound on the ${\cal N}_{sq}$, for some exemplary family of devices, namely two binary input and two binary output devices $(2,2,2,2)$ and for a device which has ternary input for one subsystem and binary input for the other subsystem but all the outputs are binary $(3,2,2,2)$. We have obtained that there exist some devices that are not MDLOPC key distillable, although they are non-local.
		Describing the procedure of convexification, we focused on obtaining upper bounds by employing a lower convex hull of the upper bounds on ${\cal N}_{sq}$.
		The reason behind such an approach is to simplify our calculations.
		In this Section, we present a specific example of upper bounds on ${\cal N}_{sq}$, 
		which we have obtained via this technique  for some bipartite binary input output non-local devices.
		Let us recall here that  ${\cal N}_{sq}$ is defined as 
		\be \label{eqn:SN_full}
		\mathcal{N}_\mathrm{sq}(P) = \max_{x,y}\min_z \inf_{\Theta_{E^\prime|E}}\mathrm{I}(A:B|E^\prime)_{(\mathcal{M}_{x,y}^F \otimes \mathcal{M}_{z}^G)\mathcal{E}(P)}.
		\ee
		
		The core strategy is based on the observation that the definition of non-signaling squashed nonlocality involves two minimizations: one in the measurement process and another one in applying suitable  post-processing channel, in part of the eavesdropper.
		We notice that one can obtain upper bounds also in the case in which used measurement and channels are not optimal, which follows from the property of infimum. Knowing this, we can run a three-step strategy to obtain an upper bound on $K_{DI}^{(iid)}$ for the desired set of devices.
		\begin{enumerate}
			\item Choose an (arbitrary, possibly continuous) set of devices, for which an upper bound is to be calculated.
			\item Calculate the values of upper bounds on non-signaling squashed nonlocality employing different devices, different measurement choices, and different post-processing channels. These can be obtained either via educated guess, some heuristic method or with computer aid, including a random search over the space.
			\item Construct lower convex hull of all previously generated plots, and the result is the convex hull of the chosen set of points.
		\end{enumerate}
		
\subsection{Upper bound for the non-signalling device used by   H{\"a}nggi, Renner and Wolf}\label{sec:HRW_bound}
		
		We will now employ the above technique to bound the $K_{DI}^{(iid)}$.
		As we have argued, the notion of security employed by us is equivalent to that used by   H{\"a}nggi, Renner and Wolf \cite{hanggi-2009}. The protocol proposed by them yields a positive key rate for devices exhibiting quantum correlations, we compare our upper bounds with the lower bound presented by them \cite{hanggi-2009,Hanggi-phd}, in Fig. \ref{fig:numerical-ub-all}. 
		The non-signaling device we consider, as in Ref. \cite{hanggi-2009}, is given by 
		\ben 
		\mathrm{P}_\mathrm{HRW}\left(ab|xy\right)=
		\begin{array}{cc|cc|cc}
			&\multicolumn{1}{c}{x} & \multicolumn{2}{c}{0}	& \multicolumn{2}{c}{1} \\
			y &$\diagbox[width=1.8em, height=1.8em, innerrightsep=0pt]{$b$}{$a$}$  & 0	& 1	&0  & 1 \\
			\hline \\ [-0.9em]
			\multirow{2}{*}{0 }  & 0 & \frac{1}{2}	- \frac{\delta}{2}	& \frac{\delta}{2}		& \frac{3}{8}	-\frac{\epsilon}{2}	& \frac{1 }{8} +  \frac{\epsilon}{2}\\[0.3em]
			& 1	& \frac{\delta}{2}	& \frac{1}{2}	- \frac{\delta}{2}	& \frac{1 }{8} +  \frac{\epsilon}{2}	& \frac{3}{8}	-\frac{\epsilon}{2}  \\ [0.2em]
			\hline  \\ [-0.9em]
			\multirow{2}{*}{1 }   &0 & \frac{3}{8}	-\frac{\epsilon}{2}	   &  \frac{1}{8}	+\frac{\epsilon}{2}	  &   \frac{1}{8}	+\frac{\epsilon}{2}& \frac{3}{8}	-\frac{\epsilon}{2} \\ [0.3em]
			& 1 & \frac{1}{8}	+ \frac{\epsilon}{2}		&       \frac{3}{8}	-\frac{\epsilon}{2}	  &  \frac{3}{8}	-\frac{\epsilon}{2}	 	& \frac{1}{8}	+\frac{\epsilon}{2}
		\end{array}. ~~~~~~
		\label{eqn:RH-box}
		\een
		
		It remains a valid non-signaling probability distribution in the parameter range $0 \leq \delta\leq 1$ and $- \frac 14 \leq \epsilon \leq \frac 34$.
		It exhibits non-local correlation for a very small range of parameters, quantified by the parameter $\varepsilon$, probability of not wining the CHSH game \cite{CHSH}, which is 
		\begin{equation}\label{eq:CHSHerror}
		\varepsilon = \text{Pr}(a \oplus b \neq x\cdot y) =  \frac 14 \left( \frac 34 + \delta + 3 \epsilon\right).
		\end{equation}
		The device is non-local when the error $\varepsilon \in [0, \frac 14)$, and there are multiple choices of
		$\delta $ and $\epsilon$ to attain this. Without loss of generality, we choose $0 \leq \delta\leq 1$ and $- \frac 14 \leq \epsilon \leq \frac{1}{12} - \frac{\delta}{3}$.
		The nonlocality fraction of these devices in the above range of parameters is $C(P) = \frac 14 - \delta - 3\epsilon$.
		

		\begin{figure*}[t]
			\begin{center}
				\includegraphics[width=\textwidth]{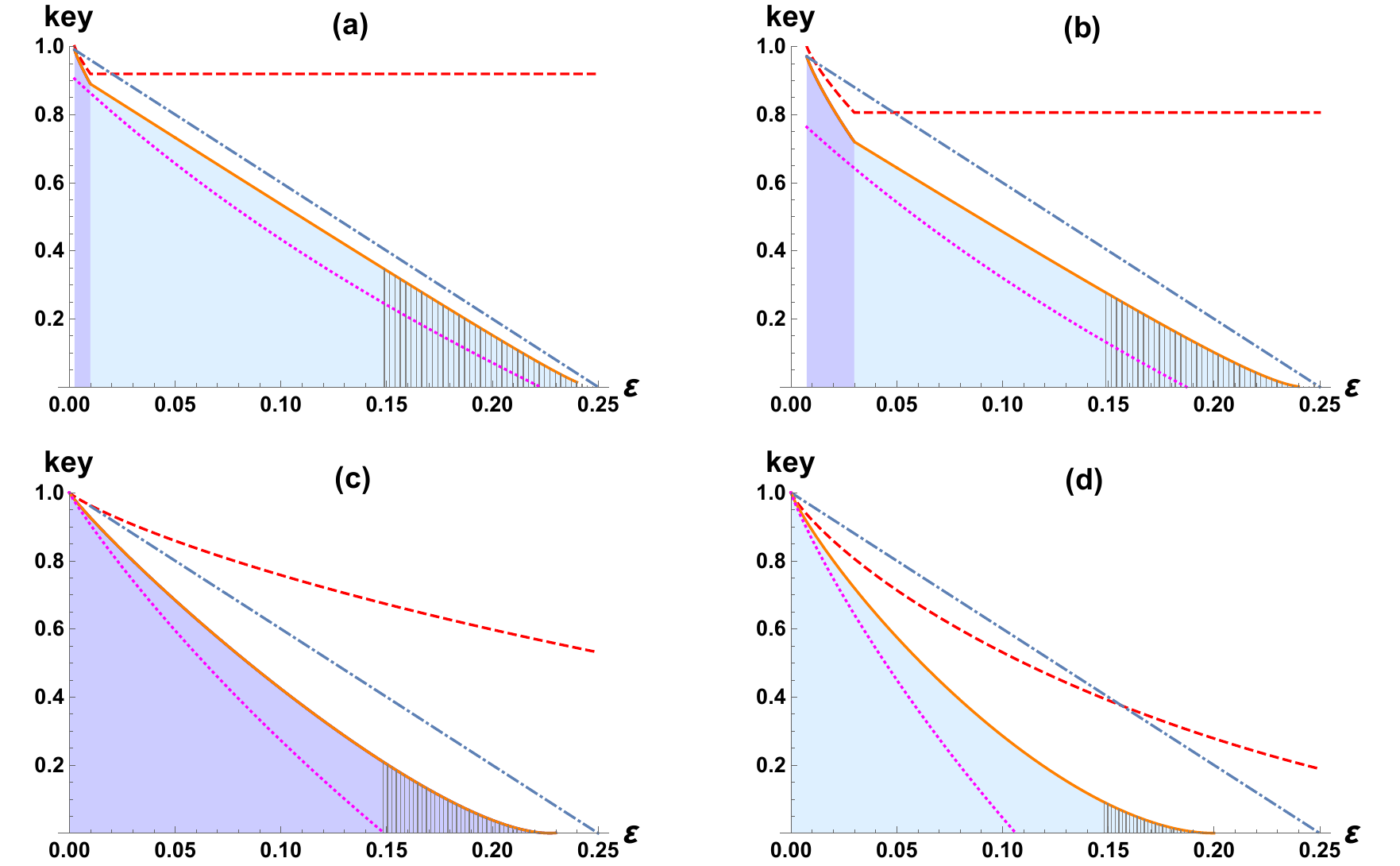}
			\end{center}
			\caption{Plot of several non-signaling secrecy quantifiers $\widehat{\mathrm{M}}(A:B||E)$, as an upper bound on secure key rate $K_{DI}^{(iid)}$,  for the bipartite binary input output device $P_{HRW}$ given in Eq. (\ref{eqn:RH-box}) (also in Ref. \cite{hanggi-2009}). 	The parameters chosen for drawing these figures are provided in Table \ref{table:dia_values}. The dashed red line corresponds to the non-signaling squashed mutual information $\widehat{\mathrm{I}}(A:B)_{P_{HRW}}$. The blue \textcolor{blue}{dashed-dotted} straight line represents the nonlocality cost, as well as the non-signaling squashed conditional mutual information $\widehat{\mathrm{I}}(A:B|E)_{{\cal E}({P_{HRW}})}$ over the complete extension ${\cal E}({P_{HRW}})$ of the given device $P$. The solid orange line represents the upper bound on the non-signaling squashed nonlocality ${\cal N}_{sq}$ which is in fact  the lower convex hull of the several other upper bounds on ${\cal N}_{sq}$. The magenta dotted line is the key rate  $\mathcal{R}\left(\left.\mathcal{P}\right|_{\mathrm{P_{HRW}}}\right)$ of the protocol design by H{\"a}nggi, Renner and Wolf \cite{hanggi-2009}.
			}
			\label{fig:numerical-ub-all}
		\end{figure*}

		The polytope of $\mathrm{P_{HRW}}$, bipartite binary input-output devices,  consists of $24$ extremal devices \cite{Barret-Roberts},  among which $16$  are local or deterministic devices, and the remaining $8$ are non-local.
		The local devices are given by
		\be 
		\mathrm{L}_{\alpha\beta\gamma\sigma}(ab|xy) = \left\{ \begin{array}{ll}
			1 & \mbox{if $a= \alpha x\oplus \beta$, $b = \gamma y \oplus \sigma$} \\
			0 & \mbox{otherwise}.\end{array} \right.
		\ee
		where $\alpha,\beta,\gamma,\sigma \in \{0,1\}$. And the non-local devices are
		\be 
		\mathrm{B}_{rst}(ab|xy) = \left\{ \begin{array}{ll}
			1/2 & \mbox{if $a\oplus b = xy\oplus rx \oplus sy \oplus t$} \\
			0 & \mbox{otherwise},\end{array} \right.
		\ee
		where $r,s,t \in \left\{0,1\right\}$.

		In Fig. \ref{fig:numerical-ub-all}, we plot several non-signaling squashed secrecy quantifiers and monotones $\widehat{\mathrm{M}} \left(A:B||E\right)$ for different choices of the parameters $\delta$ and $\epsilon$, with respect to the $\varepsilon$, which forms the upper bound on  $K^{(iid)}_{DI}$.
		Different plots correspond to different choices of the
		parameters $\epsilon$ and $\delta$, as given in Table \ref{table:dia_values}. The last row of Table \ref{table:dia_values}, give rise to the isotropic device, i.e., $P_{iso} = (1 -  \varepsilon) PR +  \varepsilon \overline{PR} $, described in the main text.
		
		\begin{table}
			\begin{center}
				\begin{tabular}{| C{1.7cm} | C{1.9cm} | C{2.3cm} |N} 
					\hline
					Fig. & $\delta$ & $\epsilon$  &  \\ [1ex]
					\hline
					(a) & $0.01$ & $\frac{1}{16}(3.04 + 12\varepsilon)$ & \\ [1ex]
					\hline
					(b) & $0.03$ & $\frac{1}{16}(3.12 + 12\varepsilon)$ & \\ [1ex]
					\hline 
					(c) & $\frac 25 \varepsilon$ & $\frac 65 \varepsilon -\frac 14$ & \\[1ex]
					\hline
					(d) & $ \varepsilon$ & $\varepsilon -\frac 14$ & \\[1ex]
					\hline
				\end{tabular}
				\caption{Table of the different values of the parameters $\delta$  and $\epsilon$, for the sub-figures as given in Fig. \ref{fig:numerical-ub-all}.
					$\delta$ and $\epsilon$ are the parameters of bipartite non-signaling device $\mathrm{P_{HRW}}$ given in Eq. (\ref{eqn:RH-box}).}\label{table:dia_values}
			\end{center}
		\end{table}
		
		In all the four figures, the red dashed line represents the squashed mutual information $\widehat{\mathrm{I}}(A:B)_P$ between Alice and Bob. The optimal choices of the measurements by Alice and Bob in the squashing process  varies with $\delta$ and $\epsilon$. For Figs. \ref{fig:numerical-ub-all}(a) and \ref{fig:numerical-ub-all}(b), the optimum direct measurement choice is $(x = 0, y = 0)$ for $\delta < \varepsilon$, and any one of the other three input choices for $\delta \geq \varepsilon$. The measurement choice $(x = 0, y = 0)$ is optimal in the entire range of $\varepsilon$ for Fig. \ref{fig:numerical-ub-all}(c), and all measurements choices give the same mutual information for the choice of $\delta$ and $\epsilon$ in Fig.  \ref{fig:numerical-ub-all}(d).
		
		The nonlocality cost ${\cal N}_C(\mathrm{P_{HRW}})$ is plotted with the dashed-dot blue line in all the figures.

		Fig. \ref{fig:numerical-ub-all}(d) clearly shows that our measure, non-signaling squashed nonlocality ${\cal N}_{sq}$ is {\it not} a faithful measure of nonlocality. The orange curve is the upper bound on ${\cal N}_{sq}$, and we have found that the bound reaches to $0$ for $\varepsilon = 0.2$ (it remains equal 0 for $\varepsilon \in (0.2, 0.25]$ due to the convexity of the measure). It strongly suggests that there exists nonlocality which can not be turned into security. Indeed, for these devices, no protocol of distribution is known. Using wirings that is necessary for the key to be non-zero, imply that we enter to some extent the general scenario of $K_{DI}$ for which there is a wide class of attacks \cite{Salwey-Wolf}. Since our scenario is restricted, we can not postulate nonequivalence between nonlocality and secrecy in NSDI paradigm.

		\subsubsection{Method to obtain the upper bound on ${\cal N}_{sq}$}
		
		The non-signaling squashed nonlocality defined in Eq. (\ref{eqn:SN_full}), is the optimal
		conditional mutual information $\mathrm{I}(A:B|E')_{{\cal E}(\mathrm{P_{HRW}})}$, between Alice and Bob, when Eve holds the complete extension of the device $\mathrm{P_{HRW}}$.
		It involves a maximization over the measurement (input) choices of Alice and Bob. In our cryptographic protocol, we assume that Eve will perform an adaptive choice of measurements after learning Alice and Bob's measurements, followed by a post-processing channel. We also observed that an arbitrary adaptive measurement by Eve, direct or general,  with any post-processing channel, provides an upper bound on $N_{sq}$, which remains convex over $\varepsilon$, in the entire range of $\varepsilon$.

		We calculate the CE \cite{CE} of $\mathrm{P_{HRW}}$ numerically in the entire range of $\delta$ and $\epsilon$, where the device is non-local. 
		The most tighter upper bound we have obtained numerically, involve a direct measurement by Eve. This direct measurement is no doubt is a function of Alice and Bob's input choice, which is intended to reduce the correlation shared by them. This measurement on Eve's system creates the following minimal ensembles in part of Alice and Bob,  
		\ben
		&&  v = \left [ \frac 14 - \delta - 3\epsilon, \frac{1 + 4\epsilon}{8}, \frac{ 1+ 4\epsilon}{8},  \frac{ 1+ 4\epsilon}{8},\frac{ 1+ 4\epsilon}{8}, \frac{ 1+ 4\epsilon}{8}, \frac{ 1+ 4\epsilon}{8}, \frac{\delta}{2},  \frac{\delta}{2} \right ], \label{eq:min-prob}\\
		&& \mathcal{E}_{z_0} = [ \mathrm{B}_{000},  \mathrm{L}_{0000}, \mathrm{L}_{0010},\mathrm{L}_{0101}, \mathrm{L}_{0111}, \mathrm{L}_{1000}, \mathrm{L}_{1101}, \mathrm{L}_{1011}, \mathrm{L}_{1110}]. \label{eq:min-ens}
		\een 
		The same measurement leads us to the non-signaling squashed conditional  mutual information $\widehat{\mathrm{I}}(A:B|E)_{{\cal E}(P)}$ for all input choices of Alice and Bob, which we have plotted by the dashed-dotted blue line in all the figures of Fig.  \ref{fig:numerical-ub-all}. We have obtained that nonlocality cost of the shared device is $ {\cal N}_C(\mathrm{P_{HRW}}) =\widehat{\mathrm{I}}(A:B|E)_{{\cal E}(P)} $.

		The classical discrete post-processing channel $\Theta_{E'|E}$, that we have obtained is different for different input choice of Alice and Bob. And they are
		
		\ben
		\Theta_{E'|E}^{0,0} =
		\begin{array}{c|c|c|c|c|c|c|c|c|c}
			\mathrm{Device}  & \mathrm{B}_{000} &  \mathrm{L}_{0000} &  \mathrm{L}_{0010} &  \mathrm{L}_{0101} &  \mathrm{L}_{0111} &  \mathrm{L}_{1000} &  \mathrm{L}_{1101} &  \mathrm{L}_{1011} &  \mathrm{L}_{1110} \\
			\hline
			$\diagbox[width=1.8em, height=1.8em, innerrightsep=0pt]{$e'$}{$e$}$  & 0	& 1	&2  & 3 & 4 &5 & 6 & 7 & 8\\
			\hline 
			0 & \textbf{1} & 0 & 0 & 0 & 0 & 0 & 0 & \textbf{1} & \textbf{1} \\
			\hline
			1 & 0 & \textbf{1} & 0 & 0 & 0 & 0 & 0 & 0 & 0 \\
			\hline
			2 & 0 & 0 & \textbf{1} & 0 & 0 & 0 & 0 & 0 & 0 \\
			\hline
			3 & 0 & 0 & 0 & \textbf{1} & 0 & 0 & 0 & 0 & 0 \\
			\hline
			4 & 0 & 0 & 0 & 0 & \textbf{1} & 0 & 0 & 0 & 0 \\
			\hline
			5 & 0 & 0 & 0 & 0 & 0 & \textbf{1} & 0 & 0 & 0 \\
			\hline
			6 & 0 & 0 & 0 & 0 & 0 & 0 & \textbf{1} & 0 & 0 \\
			\hline
		\end{array},
		\een

		\ben
		\Theta_{E'|E}^{0,1} =
		\begin{array}{c|c|c|c|c|c|c|c|c|c}
			\mathrm{Box}  & \mathrm{B}_{000} &  \mathrm{L}_{0000} &  \mathrm{L}_{0010} &  \mathrm{L}_{0101} &  \mathrm{L}_{0111} &  \mathrm{L}_{1000} &  \mathrm{L}_{1101} &  \mathrm{L}_{1011} &  \mathrm{L}_{1110} \\
			\hline
			$\diagbox[width=1.8em, height=1.8em, innerrightsep=0pt]{$e'$}{$e$}$  & 0	& 1	&2  & 3 & 4 &5 & 6 & 7 & 8\\
			\hline 
			0 & \textbf{1} & 0 & 0 & 0 & 0 & \textbf{1} & \textbf{1} & 0 & 0\\
			\hline
			1 & 0 & \textbf{1} & 0 & 0 & 0 & 0 & 0 & 0 & 0 \\
			\hline
			2 & 0 & 0 & \textbf{1} & 0 & 0 & 0 & 0 & 0 & 0 \\
			\hline
			3 & 0 & 0 & 0 & \textbf{1} & 0 & 0 & 0 & 0 & 0 \\
			\hline
			4 & 0 & 0 & 0 & 0 & \textbf{1} & 0 & 0 & 0 & 0 \\
			\hline
			5 & 0 & 0 & 0 & 0 & 0 & 0 & 0 & \textbf{1} & 0 \\
			\hline
			6 & 0 & 0 & 0 & 0 & 0 & 0 & 0 & 0 & \textbf{1}  \\
			\hline
		\end{array},
		\een
		
		\ben
		\Theta_{E'|E}^{1,0} =
		\begin{array}{c|c|c|c|c|c|c|c|c|c}
			\mathrm{Device} & \mathrm{B}_{000} &  \mathrm{L}_{0000} &  \mathrm{L}_{0010} &  \mathrm{L}_{0101} &  \mathrm{L}_{0111} &  \mathrm{L}_{1000} &  \mathrm{L}_{1101} &  \mathrm{L}_{1011} &  \mathrm{L}_{1110} \\
			\hline
			$\diagbox[width=1.8em, height=1.8em, innerrightsep=0pt]{$e'$}{$e$}$  & 0	& 1	&2  & 3 & 4 &5 & 6 & 7 & 8\\
			\hline 
			0 & \textbf{1} & 0 &  \textbf{1} & 0 & \textbf{1} & 0 & 0 & 0 &0\\
			\hline
			1 & 0 & \textbf{1} & 0 & 0 & 0 & 0 & 0 & 0 & 0 \\
			\hline
			2 & 0 & 0 & 0 & \textbf{1} & 0 & 0 & 0 & 0 & 0 \\
			\hline
			3 & 0 & 0 & 0 & 0 & 0 & \textbf{1}  & 0 & 0 & 0 \\
			\hline
			4 & 0 & 0 & 0 & 0 & 0 & 0 & \textbf{1} & 0 & 0 \\
			\hline
			5 & 0 & 0 & 0 & 0 & 0 & 0 & 0 & \textbf{1} & 0 \\
			\hline
			6 & 0 & 0 & 0 & 0 & 0 & 0 & 0 & 0 & \textbf{1}  \\
			\hline
		\end{array},
		\een
		
		\begin{align}
		\Theta_{E'|E}^{1,1} =
		\begin{array}{c|c|c|c|c|c|c|c|c|c}
		\mathrm{Device}  & \mathrm{B}_{000} &  \mathrm{L}_{0000} &  \mathrm{L}_{0010} &  \mathrm{L}_{0101} &  \mathrm{L}_{0111} &  \mathrm{L}_{1000} &  \mathrm{L}_{1101} &  \mathrm{L}_{1011} &  \mathrm{L}_{1110} \\
		\hline
		$\diagbox[width=1.8em, height=1.8em, innerrightsep=0pt]{$e'$}{$e$}$  & 0	& 1	&2  & 3 & 4 &5 & 6 & 7 & 8\\
		\hline 
		0 & \textbf{1} &  \textbf{1} & 0 & \textbf{1} & 0 & 0 & 0 & 0 &0\\
		\hline
		1 & 0 & 0 & \textbf{1}  & 0 & 0 & 0 & 0 & 0 & 0 \\
		\hline
		2 & 0 & 0 & 0 & 0 & \textbf{1} & 0 & 0 & 0 & 0 \\
		\hline
		3 & 0 & 0 & 0 & 0 & 0 & \textbf{1}  & 0 & 0 & 0 \\
		\hline
		4 & 0 & 0 & 0 & 0 & 0 & 0 & \textbf{1} & 0 & 0 \\
		\hline
		5 & 0 & 0 & 0 & 0 & 0 & 0 & 0 & \textbf{1} & 0 \\
		\hline
		6 & 0 & 0 & 0 & 0 & 0 & 0 & 0 & 0 & \textbf{1}  \\
		\hline
		\end{array}.
		\end{align}
		
		Hence, the upper bound on the key, according to our numerical findings is 
		\begin{align}
		K_{DI}^{(iid)} \le {\cal N}_{sq}(P) \le \mathrm{LCH} \left\{\widehat{\mathrm{I}}\left(A:B|E\right)_\mathrm{\mathcal{E}\left(P_{RH}\right)(ABE|XYZ)},\widehat{\mathrm{I}}\left(A:B|E\right)_\mathrm{Q(ABE|XYZ)} \right\},
		\label{eqn:TRW2}
		\end{align}
		where $Q(ABE|XYZ) = \Theta_{E|E'}^{X,Y}\left(\mathcal{E}\left(\mathrm{P_{HRW}}\right)(ABE'|XYZ)\right)$, 
		an arbitrary optimal extension, which is obtained from CE by applying the above post-processing channel.
		
		The plot of the r.h.s. of the above inequality is given by the solid orange curve in Fig. \ref{fig:numerical-ub-all}. The color shade is used to separate the two regions, where the optimal measurement choices of the honest parties are coming from two different inputs. The light blue shade in Fig. \ref{fig:numerical-ub-all}(a) and  \ref{fig:numerical-ub-all}(b) represents the choices of optimal inputs to be $(x = 0, y = 0)$, whereas the dark blue shade is for the other choices of input  (all of them give rise to the same value). In Fig. \ref{fig:numerical-ub-all}(c) the optimal input by the honest parties is $(x = 0, y = 0)$, and in Fig.  \ref{fig:numerical-ub-all}(d) all the other set of inputs are equally likely,
		and the color shed has been chosen to light blue.
		
		We compare our upper bound with the key rate $\mathcal{R}\left(\left.\mathcal{P}\right|_{\mathrm{P_{HRW}}}\right)$, generated by H{\"a}nggi, Renner and Wolf \cite{hanggi-2009}, which is the magenta dotted line in all the figures in Fig. \ref{fig:numerical-ub-all}. It lies below the solid orange line, as it represents the NSDI key rate for a particular protocol, and we provide the upper bound over all possible protocols. 
		
		Moreover, if we compare the bounds among the  sub-figures of Fig. \ref{fig:numerical-ub-all}, we observe that for a fixed $\varepsilon$,  the bound is almost decreasing if one goes from Fig. \ref{fig:numerical-ub-all}(a) to \ref{fig:numerical-ub-all}(d). This is because in Fig. \ref{fig:numerical-ub-all}(a), the choices of the parameters $\delta$ and $\epsilon$ are such that the  probability of not winning  the CHSH game is smaller for one choice of the input compared to the other input choices of the honest parties.
		In Fig. \ref{fig:numerical-ub-all}(d) all the distribution has the same error $\varepsilon$, depicting the lowest bound, i.e., all the inputs give rise to the same error, which leads to no specific choice of inputs.

		The non-faithfulness of our measure is visible from 	Fig. \ref{fig:numerical-ub-all}(d). We have found that the bound reaches to $0$ for $\varepsilon = 0.2$ (it remains equal 0 for $\varepsilon \in (0.2, 0.25]$ due to the convexity of the measure). It strongly suggests that there exists nonlocality which can not be turned into security. Indeed, for these devices, no protocol of distribution is known. Using wirings that is necessary for the key to be non-zero, imply that we enter to some extent the general scenario of $K_{DI}$ for which there is a wide class of attacks \cite{Salwey-Wolf}.
	}

		\begin{figure}[h]
    \begin{center}
        \includegraphics[width=0.6\textwidth]{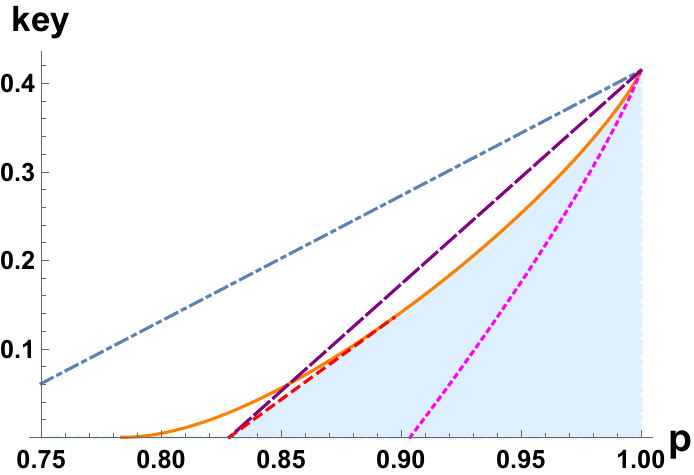}
    \end{center}
    \caption{ Plot of non-trivial upper bound on the non-signaling squashed nonlocality ${\cal N}_{sq}$, of $\mathrm{P}_\mathrm{AMP}\left(ab|xy\right)$ given in Eq. (\ref{eqn:AMP-box}), by the blue shaded region under the orange solid line and a red dashed line. The red dashed line is the (segment of) lower convex hull of the orange solid curve and the purple \textcolor{blue}{big-dashed} straight line. The solid orange line is obtained by the lower convex hull of several upper bounds of ${\cal N}_{sq}$, with the help of Eq. (339). Blue dashed-dotted line is the squashed conditional mutual information $\widehat{\mathrm{I}}(A:B|E)_{{\cal E}(\mathrm{P}_\mathrm{AMP})}$. The magenta dotted line is the lower bound on the key rate, whereas the purple \textcolor{blue}{big-}dashed line is the upper bound on intrinsic information of the eavesdropping strategy used in \cite{acin-2006-8}. We observe that the convexification technique resulting in the convex-hull bound allows to obtain tighter upper bound on ${\cal N}_{sq}$, and therefore the tightest known upper bound on the secret key-rate in the non-signaling scenario.   }
    \label{fig:numerical-ub-3222}
\end{figure}

	\subsection{Upper bound for the non-signaling device used by Ac\'in, Massar and Pironio}	
	
	In this section, we will find an upper bound on the non-signaling squashed nonlocality, for a device, which the honest parties Alice and Bob can obtain by performing quantum measurements on a shared bipartite quantum state, given in \cite{acin-2006-8}. The shared quantum state is the Werner state $\rho_{AB} = p \ket{\psi_+}\bra{\psi_+}_{AB} + \frac{1-p}{4}I_{AB}$, where $\ket{\psi_+}_{AB} = \frac{1}{\sqrt{2}}(\ket{0}_{A}\ket{0}_{B} + \ket{1}_{A}\ket{1}_{A})$, and $p \in [0,1]$. 
	One of the honest parties, Alice consider three possible measurement choices $x \in \{0,1,2\}$, whereas Bob chooses only two possible measurements $y \in  \{0,1\}$. Among those set of measurements when both the  measurement settings are $x = 0$ and $y = 0$,  the measurement bases coincides and only that choice of measurement has been used for the key distribution run. The other two measurements  $x \in \{1,2\}$, for Alice and two measurements $ y \in \{0,1\}$,for Bob, are for the test of non-local correlation present in the system i.e., for the violation of Bell inequality, of the shared state. 
	
	The shared probability distribution by both the parties, or the device obtained after the possible set of measurements is given by
	
	\ben 
		\mathrm{P}_\mathrm{AMP}\left(ab|xy\right)=
		\begin{array}{cc|cc|cc|cc}
			&\multicolumn{1}{c}{x} & \multicolumn{2}{c}{0}	& \multicolumn{2}{c}{1} & \multicolumn{2}{c}{2}\\
			y &$\diagbox[width=1.8em, height=1.8em, innerrightsep=0pt]{$b$}{$a$}$  & 0	& 1	&0  & 1 & 0 & 1\\
			\hline \\ [-0.9em]
			\multirow{2}{*}{0 }  & 0 & \frac{1 + p}{4}	& 	\frac{1 - p}{4}	& \frac{2 + \sqrt{2}p}{8}	& \frac{2 - \sqrt{2}p}{8} & \frac{2 + \sqrt{2}p}{8}	& \frac{2 - \sqrt{2}p}{8} \\[0.3em]
			& 1	& ~ \frac{1 - p}{4}~	& ~\frac{1+p}{4}~	& \frac{2 - \sqrt{2}p}{8} & \frac{2 + \sqrt{2}p}{8} & \frac{2 - \sqrt{2}p}{8}	& \frac{2 + \sqrt{2}p}{8}  \\ [0.2em]
			\hline  \\ [-0.9em]
			\multirow{2}{*}{1 }   &0 & \frac 14 &  \frac 14  &  \frac{2 + \sqrt{2}p}{8}	& \frac{2 - \sqrt{2}p}{8} & \frac{2 - \sqrt{2}p}{8}	& \frac{2 + \sqrt{2}p}{8} \\ [0.3em]
			& 1 & \frac 14	& \frac 14  &  \frac{2 - \sqrt{2}p}{8}	& \frac{2 + \sqrt{2}}{8} & \frac{2 + \sqrt{2}p}{8}	& \frac{2 - \sqrt{2}p}{8}
		\end{array}. ~~~~~~
		\label{eqn:AMP-box}
		\een

In the entire range of $p$, the device is a valid probability distribution but it  exhibit non-local correlation only for an small range of $p$. To compute the range of $p$, where  let us quantify the probability of not wining the CHSH game \cite{CHSH}, by the parameter $\varepsilon$,  which is 
		\begin{equation}\label{eq:CHSHerrorAMP}
		\varepsilon(\mathrm{P}_\mathrm{AMP}) = \text{Pr}(a \oplus b \neq (x- 1)\cdot y)_{\mathrm{P}_\mathrm{AMP}} =  \frac 14 \left( 2 - \sqrt{2}p\right).
		\end{equation}     
Note that for  $\mathrm{P}_\mathrm{AMP}$, Alice will use her inputs $x \in \{1,2\}$ for the detection of nonlocality. Now the device is non-local when $0 \leq \varepsilon < \frac 14$, hence the device may be useful for secure key agreement protocol in presence of non-signalling Eve in the range of  $\frac{1}{\sqrt{2}} < p \leq 1$.  

 To  make a rough estimation on the upper bound of ${\cal N}_{sq}$, of $\mathrm{P}_\mathrm{AMP}\left(ab|xy\right)$,  we first focus on  the  squashed conditional mutual information $\widehat{\mathrm{I}}(A:B|E)_{{\cal E}(\mathrm{P}_\mathrm{AMP})}$, where ${\cal E}(\mathrm{P}_\mathrm{AMP})$ is the complete extension of the  given quantum device.  In general, obtaining the complete extension of a given box, in this new $(3,2,2,2)$ polytope is an extremely difficult task and hence, 
we have found here only one exemplary minimal ensemble which up to our numerical search is an optimal eavesdropping strategy, i.e., achieving the $\min_z$, (see eqs. (\ref{eq:squashed-fn}) and Sec. \ref{entropies} for the definition of  $\widehat{\mathrm{I}}(A:B|E)_{{\cal E}(\mathrm{P}_\mathrm{AMP})}$), for the chosen values of the measurement setup by the honest parties for key sharing  $x = y = 0$. The minimal ensemble is
\ben
		 v &=& \left [\frac{p}{\sqrt{2}} -  \frac 12, \frac{p}{\sqrt{2}} -  \frac 12, \frac{2 - \sqrt{2}p}{8}, \frac{2 - \sqrt{2}p}{8}, \frac{2 - \sqrt{2}p}{8},\frac{2 - \sqrt{2}p}{8},\frac{2 - \sqrt{2}p}{8},\frac{2 - \sqrt{2}p}{8}, \frac{1 - p}{4}, \frac{1 - p}{4},  \right. \nonumber \\
	&& 	\hspace{4.5in} \left. \frac{(2 - \sqrt{2})p}{8},  \frac{(2 - \sqrt{2})p}{8} \right ],~~~~~~~ \label{eq:min-prob}\\
		 \mathcal{E}_{z_0} &=& [ ~~~~\mathrm{B}_{0}~~~,  ~~~~\mathrm{B}_{1}~~~~,  ~~~~\mathrm{L}_{0}~~~~, ~~~~\mathrm{L}_{1}~~~~,~~~~\mathrm{L}_{2}~~~~,~~~~\mathrm{L}_{3}~~~~,~~~~\mathrm{L}_{4}~~~~,~~~\mathrm{L}_{5}~~~~, ~~~\mathrm{L}_{6}~~,~~~\mathrm{L}_{7}~~, \nonumber \\
		&& 	\hspace{4.5in}	 ~~~~~~\mathrm{L}_{8}~~~~~,~~~~~\mathrm{L}_{9}~~~~]. \label{eq:min-ens-AMP}
		\een 		
	where $B_0, ~B_1$ are the two non-local extremal devices and $L_0, \ldots, L_9$ are the local deterministic devices (extremal), in the polytope of the devices where $\mathrm{P}_\mathrm{AMP})$ lies, and they are given by \cite{Jones_Masanes},

\ben 
		\mathrm{B}_0\left(ab|xy\right)=
		\begin{array}{cc|cc|cc|cc}
			&\multicolumn{1}{c}{x} & \multicolumn{2}{c}{0}	& \multicolumn{2}{c}{1} & \multicolumn{2}{c}{2}\\
			y &$\diagbox[width=1.8em, height=1.8em, innerrightsep=0pt]{$b$}{$a$}$  & 0	& 1	&0  & 1 & 0 & 1\\
			\hline \\ [-0.9em]
			\multirow{2}{*}{0 }  & 0 & \frac{1}{2}	& 0	& \frac{1}{2}	& 0 & \frac{1}{2}	& 0 \\[0.3em]
			& 1	&  0 & \frac{1}{2}	& 0	& \frac{1}{2}	& 0 & \frac{1}{2}  \\ [0.2em]
			\hline  \\ [-0.9em]
			\multirow{2}{*}{1 }   &0 &  \frac{1}{2}	& 0 & \frac{1}{2}	& 0 & 0 & \frac{1}{2} \\ [0.3em]
			& 1 & 0& \frac{1}{2}	& 0	& \frac{1}{2} & \frac{1}{2} & 0
		\end{array}, ~~~~~~
		\mathrm{B}_1\left(ab|xy\right)=
		\begin{array}{cc|cc|cc|cc}
			&\multicolumn{1}{c}{x} & \multicolumn{2}{c}{0}	& \multicolumn{2}{c}{1} & \multicolumn{2}{c}{2}\\
			y &$\diagbox[width=1.8em, height=1.8em, innerrightsep=0pt]{$b$}{$a$}$  & 0	& 1	&0  & 1 & 0 & 1\\
			\hline \\ [-0.9em]
			\multirow{2}{*}{0 }  & 0 & \frac{1}{2}	& 0	& \frac{1}{2}	& 0 & \frac{1}{2}	& 0 \\[0.3em]
			& 1	&  0 & \frac{1}{2}	& 0	& \frac{1}{2}	& 0 & \frac{1}{2}  \\ [0.2em]
			\hline  \\ [-0.9em]
			\multirow{2}{*}{1 }   &0 & 0& \frac{1}{2} & \frac{1}{2}	& 0 & 0 & \frac{1}{2} \\ [0.3em]
			& 1 & \frac{1}{2}	& 0 & 0	& \frac{1}{2} & \frac{1}{2} & 0	
		\end{array}. ~~~~~~
		\label{eqn:32B1}
		\een

			\ben 
		\mathrm{L}_0=
		\begin{array}{cc|cc|cc|cc}
			&\multicolumn{1}{c}{x} & \multicolumn{2}{c}{0}	& \multicolumn{2}{c}{1} & \multicolumn{2}{c}{2}\\
			y &$\diagbox[width=1.8em, height=1.8em, innerrightsep=0pt]{$b$}{$a$}$  & 0	& 1	&0  & 1 & 0 & 1\\
			\hline \\ [-0.9em]
			\multirow{2}{*}{0 }  & 0 & 1 &0  & 1 & 0 & 1	& 0 \\[0.3em]
			& 1	&  0 & 0 & 0  & 0 & 0 & 0  \\ [0.2em]
			\hline  \\ [-0.9em]
			\multirow{2}{*}{1 }  & 0 &  1 & 0 & 1& 0 & 1 & 0\\[0.3em]
			& 1	&  0 & 0 & 0  & 0 & 0 & 0	
 		\end{array}, ~~~
		\mathrm{L}_1=
		\begin{array}{cc|cc|cc|cc}
			&\multicolumn{1}{c}{x} & \multicolumn{2}{c}{0}	& \multicolumn{2}{c}{1} & \multicolumn{2}{c}{2}\\
			y &$\diagbox[width=1.8em, height=1.8em, innerrightsep=0pt]{$b$}{$a$}$  & 0	& 1	&0  & 1 & 0 & 1\\
			\hline \\ [-0.9em]
			\multirow{2}{*}{0 }  & 0 & 1 &0  & 1 & 0 & 0	& 1 \\[0.3em]
			& 1	&  0 & 0 & 0  & 0 & 0 & 0  \\ [0.2em]
			\hline  \\ [-0.9em]
			\multirow{2}{*}{1 }  & 0 &  1 & 0 & 1& 0 & 0 & 1\\[0.3em]
			& 1	&  0 & 0 & 0  & 0 & 0 & 0	
		\end{array}, ~~~
		\mathrm{L}_2=
		\begin{array}{cc|cc|cc|cc}
			&\multicolumn{1}{c}{x} & \multicolumn{2}{c}{0}	& \multicolumn{2}{c}{1} & \multicolumn{2}{c}{2}\\
			y &$\diagbox[width=1.8em, height=1.8em, innerrightsep=0pt]{$b$}{$a$}$  & 0	& 1	&0  & 1 & 0 & 1\\
			\hline \\ [-0.9em]
			\multirow{2}{*}{0 }  & 0 & 0 & 0 & 0  & 0 & 0 & 0 \\[0.3em]
			& 1	& 0 & 1 &0  & 1 & 0 & 1 \\ [0.2em]
			\hline  \\ [-0.9em]
			\multirow{2}{*}{1 }  & 0 & 0 & 0 & 0  & 0 & 0 & 0 \\[0.3em]
			& 1	& 0 & 1 &0  & 1 & 0 & 1	
		\end{array}.~~~
		\label{eqn:32L2}
		\een	
		
			\ben 
		\mathrm{L}_3 = ~
		\begin{array}{cc|cc|cc|cc}
			&\multicolumn{1}{c}{x} & \multicolumn{2}{c}{0}	& \multicolumn{2}{c}{1} & \multicolumn{2}{c}{2}\\
			y &$\diagbox[width=1.8em, height=1.8em, innerrightsep=0pt]{$b$}{$a$}$  & 0	& 1	&0  & 1 & 0 & 1\\
			\hline \\ [-0.9em]
			\multirow{2}{*}{0 }  & 0 & 0 & 0 & 0  & 0 & 0 & 0 \\[0.3em]
			& 1	& 0 & 1 &0  & 1 & 0 & 1 \\ [0.2em]
			\hline  \\ [-0.9em]
			\multirow{2}{*}{1 }  & 0 & 0 & 1 &0  & 1 & 0 & 1 \\[0.3em]
			& 1		& 0 & 0 & 0  & 0 & 0 & 0
		\end{array}, ~~~
		\mathrm{L}_4=
		\begin{array}{cc|cc|cc|cc}
			&\multicolumn{1}{c}{x} & \multicolumn{2}{c}{0}	& \multicolumn{2}{c}{1} & \multicolumn{2}{c}{2}\\
			y &$\diagbox[width=1.8em, height=1.8em, innerrightsep=0pt]{$b$}{$a$}$  & 0	& 1	&0  & 1 & 0 & 1\\
			\hline \\ [-0.9em]
			\multirow{2}{*}{0 }  & 0 & 0 & 0 & 0  & 0 & 0 & 0 \\[0.3em]
			& 1	& 0 & 1 &1  & 0 & 0 & 1 \\ [0.2em]
			\hline  \\ [-0.9em]
			\multirow{2}{*}{1 }  & 0 & 0 & 1 &1  & 0 & 0 & 1 \\[0.3em]
			& 1		& 0 & 0 & 0  & 0 & 0 & 0
		\end{array}, ~~~~
\mathrm{L}_5 = ~
			\begin{array}{cc|cc|cc|cc}
			&\multicolumn{1}{c}{x} & \multicolumn{2}{c}{0}	& \multicolumn{2}{c}{1} & \multicolumn{2}{c}{2}\\
			y &$\diagbox[width=1.8em, height=1.8em, innerrightsep=0pt]{$b$}{$a$}$  & 0	& 1	&0  & 1 & 0 & 1\\
			\hline \\ [-0.9em]
			\multirow{2}{*}{0 }  & 0 & 1 & 0 & 0  & 1 & 1 & 0 \\[0.3em]
			& 1	& 0 & 0 &0  & 0 & 0 & 0 \\ [0.2em]
			\hline  \\ [-0.9em]
			\multirow{2}{*}{1 }  & 0 & 0 & 0 & 0  & 0 & 0 & 0 \\[0.3em]
			& 1	& 1 & 0 &0  & 1 & 1 & 0	
		\end{array}. ~~~
		\label{eqn:32L5}
		\een	
		
			\ben 
			\mathrm{L}_6=
		\begin{array}{cc|cc|cc|cc}
			&\multicolumn{1}{c}{x} & \multicolumn{2}{c}{0}	& \multicolumn{2}{c}{1} & \multicolumn{2}{c}{2}\\
			y &$\diagbox[width=1.8em, height=1.8em, innerrightsep=0pt]{$b$}{$a$}$  & 0	& 1	&0  & 1 & 0 & 1\\
			\hline \\ [-0.9em]
			\multirow{2}{*}{0 }  & 0 & 0 & 0 & 0  & 0 & 0 & 0 \\[0.3em]
			& 1	& 1 & 0 &0  & 1 & 1 & 0 \\ [0.2em]
			\hline  \\ [-0.9em]
			\multirow{2}{*}{1 }  & 0 & 0 & 0 & 0  & 0 & 0 & 0 \\[0.3em]
			& 1	& 1 & 0 &0  & 1 & 1 & 0	
		\end{array}, ~~~~
		\mathrm{L}_7 = ~
		\begin{array}{cc|cc|cc|cc}
			&\multicolumn{1}{c}{x} & \multicolumn{2}{c}{0}	& \multicolumn{2}{c}{1} & \multicolumn{2}{c}{2}\\
			y &$\diagbox[width=1.8em, height=1.8em, innerrightsep=0pt]{$b$}{$a$}$  & 0	& 1	&0  & 1 & 0 & 1\\
			\hline \\ [-0.9em]
			\multirow{2}{*}{0 }  & 0 & 0 & 1 & 1  & 0 & 1 & 0 \\[0.3em]
			& 1	& 0 & 0 &0  & 0 & 0 & 0 \\ [0.2em]
			\hline  \\ [-0.9em]
			\multirow{2}{*}{1 }  & 0 & 0 & 0 &0  & 0 & 0 & 0 \\[0.3em]
			& 1		& 0 & 1 & 1  & 0 & 1 & 0
		\end{array}, ~~~
		\mathrm{L}_8=
		\begin{array}{cc|cc|cc|cc}
			&\multicolumn{1}{c}{x} & \multicolumn{2}{c}{0}	& \multicolumn{2}{c}{1} & \multicolumn{2}{c}{2}\\
			y &$\diagbox[width=1.8em, height=1.8em, innerrightsep=0pt]{$b$}{$a$}$  & 0	& 1	&0  & 1 & 0 & 1\\
			\hline \\ [-0.9em]
			\multirow{2}{*}{0 }  & 0 & 1 & 0 & 1  & 0 & 1 & 0 \\[0.3em]
			& 1	& 0 & 0 &0  & 0 & 0 & 0 \\ [0.2em]
			\hline  \\ [-0.9em]
			\multirow{2}{*}{1 }  & 0 & 0 & 0 & 0  & 0 & 0 & 0 \\[0.3em]
			& 1	& 1 & 0 &1  & 0 & 1 & 0	
		\end{array}. ~~~~~~
		\label{eqn:32L5}
		\een

			\ben 
		\mathrm{L}_{9}(ab|xy)=
		\begin{array}{cc|cc|cc|cc}
			&\multicolumn{1}{c}{x} & \multicolumn{2}{c}{0}	& \multicolumn{2}{c}{1} & \multicolumn{2}{c}{2}\\
			y &$\diagbox[width=1.8em, height=1.8em, innerrightsep=0pt]{$b$}{$a$}$  & 0	& 1	&0  & 1 & 0 & 1\\
			\hline \\ [-0.9em]
			\multirow{2}{*}{0 }  & 0 & 0 & 0 & 0  & 0 & 0 & 0 \\[0.3em]
			& 1	& 0 & 1 &0  & 1 & 1 & 0 \\ [0.2em]
			\hline  \\ [-0.9em]
			\multirow{2}{*}{1 }  & 0 & 0 & 0 & 0  & 0 & 0 & 0 \\[0.3em]
			& 1		& 0 & 1 & 0  & 1 & 1 & 0
		\end{array}, ~~~~
		\een

For the given decomposition of the device $\mathrm{P}_\mathrm{AMP}\left(ab|xy\right)$, the  squashed conditional mutual information reduces to
$\widehat{\mathrm{I}}(A:B|E)_{{\cal E}(\mathrm{P}_\mathrm{AMP})} = \sqrt{2}p -1$,  which is equal to the
nonlocality cost of the shared device i.e.,  $ {\cal N}_C(\mathrm{P_{AMP}})$. It
reaches to $\sqrt{2} - 1$, for $p = 1$, i.e., when the Bell state is shared. 

To obtained the upper bound on ${\cal N}_{sq}(\mathrm{P}_\mathrm{AMP})$, we will again apply some post-processing channel $\Theta_{E|E'}$, on the output of Eve $E$, and apply the procedure of getting the lower convex hull, by the relation
\begin{align}
    {\cal N}_{sq}(\mathrm{P}_\mathrm{AMP}) \le \mathrm{LCH} \left\{\widehat{\mathrm{I}}\left(A:B|E\right)_\mathrm{\mathcal{E}\left(\mathrm{P}_\mathrm{AMP}\right)(ABE|XYZ)},\widehat{\mathrm{I}}\left(A:B|E\right)_\mathrm{Q_{AMP}(ABE|XYZ)} \right\},  \label{eqn:AMPLCH}
\end{align}
		where $\mathrm{Q_{AMP}}(ABE|XYZ) = \Theta_{E|E'}\left(\mathcal{E}\left(\mathrm{P_{AMP}}\right)(ABE'|XYZ)\right)$, is 
		an arbitrary extension of $\mathrm{P_{AMP}}$, upon applying the post-processing channel $\Theta_{E|E'}$, given by 
	\begin{align}
		\Theta_{E|E'} =
		\begin{array}{c|c|c|c|c|c|c|c|c|c|c|c|c}
			\mathrm{Device}  & \mathrm{B}_{0} & \mathrm{B}_{1} & \mathrm{L}_{0} &  \mathrm{L}_{1} &  \mathrm{L}_{2} &  \mathrm{L}_{3} &  \mathrm{L}_{4} &  \mathrm{L}_{5} &  \mathrm{L}_{6} &  \mathrm{L}_{7} &  \mathrm{L}_{8} &  \mathrm{L}_{9} \\
			\hline
			$\diagbox[width=1.8em, height=1.8em, innerrightsep=0pt]{$e$}{$e'$}$  & 0	& 1	&2  & 3 & 4 &5 & 6 & 7 & 8 & 9 & 10 & 11\\
			\hline 
			0 & \textbf{1} & \textbf{1} & 0 & 0 & 0 & 0 & \textbf{1} & \textbf{1} & 0 & 0 & 0 & 0\\
			\hline
			1 & 0 & 0 & \textbf{1} & 0 & 0 & 0 & 0 & 0 & 0 & 0 & 0 & 0\\
			\hline
			2 & 0 & 0 & 0 & \textbf{1} & 0 & 0 & 0 & 0 & 0 & 0 & 0 & 0\\
			\hline
			3 & 0 & 0 & 0 & 0 & \textbf{1} & 0 & 0 & 0 & 0 & 0 & 0 & 0\\
			\hline
			4 & 0 & 0 & 0 & 0 & 0 & \textbf{1} & 0 & 0 & 0 & 0 & 0 & 0\\
			\hline
			5 & 0 & 0 & 0 & 0 & 0 & 0 & 0 & 0 & \textbf{1} & 0 & 0 & 0\\
			\hline
			6 & 0 & 0 & 0 & 0 & 0 & 0 & 0 & 0 & 0 &\textbf{1} & 0 & 0 \\
			\hline
			7 & 0 & 0 & 0 & 0 & 0 & 0 & 0 & 0 & 0 & 0 & \textbf{1} & 0 \\
			\hline
			8 & 0 & 0 & 0 & 0 & 0 & 0 & 0 & 0 & 0 & 0 & 0 & \textbf{1} \\
			\hline
		\end{array},
		\end{align}
Note that here we need only one post-processing channel, because in the squashing procedure unlike Sec. \ref{sec:HRW_bound}, Eve's know which outcomes of Alice and Bob are used for the key generation run.

The upper bound on ${\cal N}_{sq}(\mathrm{P}_\mathrm{AMP})$, i.e., the right hand side of (339), has been plotted in figure \ref{fig:numerical-ub-3222}, by the orange line, which vanishes for $p \approx 0.783$, and from the procedure of lower convex hull we will consider it $0$, for all $p < 0.783$. The magenta dotted line is the lower bound on the key rate of \cite{acin-2006-8}, whereas the violate dashed line is the upper bound on the intrinsic information $I(A:B\downarrow E)$, of \cite{acin-2006-8}, for a particular eavesdropping strategy. We have  found that our bound on ${\cal N}_{sq}$ is better than the bound on $I(A:B\downarrow E)$, by \cite{acin-2006-8}, for  $p > 0.853$.


\end{document}